\documentclass[aps,prx,twocolumn,superscriptaddress,longbibliography]{revtex4-2}
\usepackage{color}
\usepackage{dcolumn}
\usepackage{graphicx}
\usepackage{amsmath}
\usepackage{amssymb}
\usepackage{bm}
\usepackage{braket}
\usepackage[hidelinks]{hyperref}
\usepackage{multirow}

\usepackage{soul}
\usepackage{xcolor}

\begin{document}

\title{Arbitrary controlled-phase gate on fluxonium qubits using differential ac-Stark shifts}

\author{Haonan Xiong}
\thanks{Equal contribution author}
\affiliation{Department of Physics, Joint Quantum Institute, and Center for Nanophysics and Advanced Materials, University of Maryland, College Park, MD 20742, USA}

\author{Quentin Ficheux}
\thanks{Equal contribution author}
\affiliation{Department of Physics, Joint Quantum Institute, and Center for Nanophysics and Advanced Materials, University of Maryland, College Park, MD 20742, USA}

\author{Aaron Somoroff}
\affiliation{Department of Physics, Joint Quantum Institute, and Center for Nanophysics and Advanced Materials, University of Maryland, College Park, MD 20742, USA}

\author{Long B. Nguyen}
\affiliation{Department of Physics, Joint Quantum Institute, and Center for Nanophysics and Advanced Materials, University of Maryland, College Park, MD 20742, USA}

\author{Ebru Dogan}
\affiliation{Department of Physics, University of Massachusetts-Amherst, Amherst, MA, 01003, USA}

\author{Dario Rosenstock}
\affiliation{Department of Physics, University of Massachusetts-Amherst, Amherst, MA, 01003, USA}

\author{Chen Wang}
\affiliation{Department of Physics, University of Massachusetts-Amherst, Amherst, MA, 01003, USA}

\author{Konstantin N. Nesterov}
\affiliation{Department of Physics and  Wisconsin Quantum Institute, University of Wisconsin - Madison, Madison, WI 53706, USA}

\author{Maxim G. Vavilov}
\affiliation{Department of Physics and  Wisconsin Quantum Institute, University of Wisconsin - Madison, Madison, WI 53706, USA}

\author{Vladimir E. Manucharyan}
\affiliation{Department of Physics, Joint Quantum Institute, and Center for Nanophysics and Advanced Materials, University of Maryland, College Park, MD 20742, USA}

\date{\today}

\begin{abstract}

Large scale quantum computing motivates the invention of two-qubit gate schemes that not only maximize the gate fidelity but also draw minimal resources. In the case of superconducting qubits, the weak anharmonicity of transmons imposes profound constraints on the gate design, leading to increased complexity of devices and control protocols. Here we demonstrate a resource-efficient control over the interaction of strongly-anharmonic fluxonium qubits. Namely, applying an off-resonant drive to non-computational transitions in a pair of capacitively-coupled fluxoniums induces a $\textrm{ZZ}$-interaction due to unequal ac-Stark shifts of the computational levels.
With a continuous choice of frequency and amplitude, the drive can either cancel the static $\textrm{ZZ}$-term or increase it by an order of magnitude to enable a controlled-phase (CP) gate with an arbitrary programmed phase shift. The cross-entropy benchmarking of these non-Clifford operations yields a sub $1\%$ error, limited solely by incoherent processes. Our result demonstrates the advantages of strongly-anharmonic circuits over transmons in designing the next generation of quantum processors.

\end{abstract}

\maketitle

Wiring up a pair of superconducting qubits creates an unintentional $\textrm{ZZ}$-term in the two-qubit Hamiltonian, where Z is the single-qubit Pauli $\sigma_z$-operator.  This $\textrm{ZZ}$-interaction arises from the repulsion between computational and non-computational energy levels of the coupled circuit; the effect would be absent for purely two-level systems. On one hand, such an interaction realizes a controlled-phase logical operation, inducing a phase shift on one qubit depending on the state of the other one~\cite{Strauch2003,Dicarlo2009}. On the other hand, a small but non-zero $\textrm{ZZ}$-term would induce coherent errors during single-qubit operations and lead to quantum cross-talk across the qubit register~\cite{Gambetta2012,McKay2019,Arute2019,Rudinger2019,Krinner2020a,McKay2020,Morvan2020}. Therefore, any high-fidelity multiqubit system must either achieve an in situ control over the $\textrm{ZZ}$-term or have it permanently eliminated. The former approach invokes flux-tunable qubits and coupler circuits~\cite{Ku2020,Zhao2020,Arute2019,Foxen2020,Negirneac2020,Sung2020}, which comes at the price of limiting the qubit coherence time by the 1/f flux noise and introducing a new error channel due to the leakage of quantum information into the coupler degrees of freedom \cite{Sung2020,Collodo2020}. The latter approach critically relies on fine-tuning the circuit parameters at the nanofabrication stage, which both reduces the device yield and narrows the parameter space available for performance optimization~\cite{Chow2011,Sheldon2016,Tripathi2019,Jurcevic2020,Kandala2020}.

A technologically attractive control scheme would be to tune the $\textrm{ZZ}$-term in a fixed-parameter circuit using a microwave drive \cite{Chow2013a,Paik2016,Krinner2020}. Indeed, as long as the drive does not resonantly excite the circuit transitions, its effect would be reduced to ac-Stark shifts, which generally modify the level repulsion structure and hence the magnitude of the $\textrm{ZZ}$-term. Unfortunately, such ideas proved impractical for transmon qubits, largely due to their weak anharmonicity, responsible for leakage of quantum information outside the computational subspace. Recently, a complete suppression of the static $\textrm{ZZ}$-term was demonstrated in a weakly-anharmonic capacitively-shunted flux qubit thanks to breaking the parity selection rule with an external flux-bias \cite{Noguchi2020}. However, the qubit is unavoidably exposed to the first-order 1/f flux noise, which limits the qubit coherence and hence the gate error. 

In this work we develop the ideas from Refs. \cite{Chow2013a,Paik2016,Krinner2020,Noguchi2020} for strongly-anharmonic qubits -- fluxoniums~\cite{Manucharyan2009, Manucharyan2012}. We consider fluxoniums that are coupled by a fixed capacitance, which is compatible with transmon-based processors, and biased at the half-integer flux quantum, practically eliminating the flux-noise decoherence channel. Previously, we have demonstrated a controlled-Z (CZ) gate in such a system by applying a near-resonant drive to the transition between first and second excited states of one of the qubits~\cite{Ficheux2020}. In that experiment, the $\textrm{ZZ}$-term was naturally suppressed by the special choice of relatively low qubit frequencies (about $100~\textrm{MHz}$). Here we (i) explore a much broader range of qubit frequencies (up to $1.3~\textrm{GHz}$), for which the static $\textrm{ZZ}$-term is significant, and (ii) use off-resonant driving, the effect of which can be understood in terms of radiation-dressing and light-shifts. Our key new result is that the static $\textrm{ZZ}$-term can be either suppressed to zero or enhanced to about $10~\textrm{MHz}$ on demand, resulting in a resource-efficient controlled-phase (CP) gate free of appreciable spurious effects, such as coherent state leakage. 

\begin{figure*}[t]
    \centering
    \includegraphics[width=0.9\linewidth]{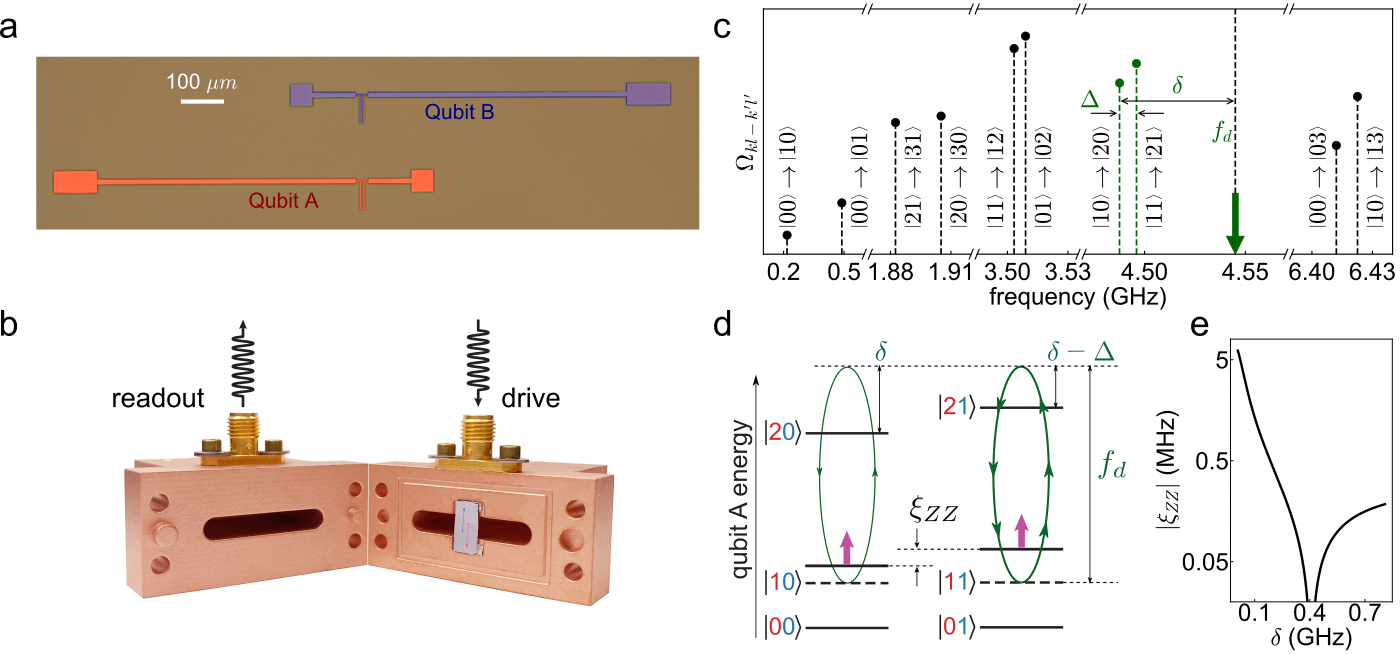}
    \caption{\textbf{Differential ac-Stark shift in a two-fluxonium circuit.}
  (a) False color picture of the two-fluxonium device. The device is similar to that reported in Ref. \cite{Ficheux2020}. (b) Photograph of the cavity that has the same design as the one used in this experiment. All the microwave drives share the same drive port. The transmitted signal at the cavity frequency is collected from the readout port. The two halves of the cavity are sealed with indium [not shown]. The size of the fluxoniums in this picture is exaggerated for better visibility. (c) Spectrum of allowed transitions of the two-fluxonium circuit extracted from spectroscopy data. The simulated quantity $\Omega_{kl-k'l'}$ defines an on-resonance Rabi frequency that a drive field would induce, assuming the field amplitude is frequency-independent. Note that only transitions between states with different parity are allowed. The green arrow indicates the drive frequency $f_d$ used to induce the differential ac-Stark shift. (d) Schematic of the energy level diagram for qubit B in the ground state (left) and qubit B in the excited state (right). A drive at frequency $f_d$ pulls the energy level $|11\rangle$ more than the energy level $|10\rangle$, without affecting levels $|00\rangle$ and $|01\rangle$, which is equivalent to a differential ac-Stark shift $\xi_{ZZ}^{\textrm{drive}} >0$. (e) Calculated total $\textrm{ZZ}$-interaction rate $\xi_{ZZ} = -|\xi_{ZZ}^\mathrm{static}| + \xi_{ZZ}^\mathrm{drive}$ for a fixed drive amplitude ($\Omega_{11-21}=52$ MHz). Note, at an appropriate detuning $\delta$, the total qubit-qubit interaction is switched off, we get $\xi_{ZZ} =0$.}
    \label{fig:fig1}
\end{figure*}

The practicality of our scheme is illustrated by the performance of the controlled-phase (CP) gate with an arbitrary programmed phase, implemented by rapidly turning the $\textrm{ZZ}$-term on and off. We used the cross-entropy benchmarking technique~\cite{Neill2018,Arute2019,Barends2019,Foxen2020} to measure the CP gate error for a set of 16 equally spaced phase values, obtaining an error of about $3 \times 10^{-3}$ per radian, while the single qubit error reaches $1\times 10^{-3}$. Notably, both numbers are limited by a manifestly sub-optimal coherence in the current experimental setup, yet, they are already close to the state-of-the-art for transmons. In fact, our phase-averaged CP gate error of $8 \times 10^{-3}$ is only few times larger than the best result recently demonstrated on a flux-controlled processor~\cite{Foxen2020,Negirneac2020}. This direct access to the complete family of CP gates is important since they can reduce the required circuit depth in a number of useful algorithms~\cite{Barends2016,OMalley2016,Kandala2017,Ganzhorn2019,Foxen2020,Aleiner2020} such as the quantum approximate optimization algorithms (QAOA) and variational quantum eigensolvers (VQE) ~\cite{Kitaev1995,Whitfield2011,Lacroix2020}. The minimalism of our qubit-qubit interaction control scheme, enabled in large part by the strong anharmonicity of fluxoniums, can provide a significant advantage for constructing large-scale quantum processors.\\

\noindent\textbf{Results}\\

\noindent We describe experiments on two devices with significantly distinct spectral properties. In the main article, we focus on the first device, while the Supplementary Note 7 supports our conclusions with the data obtained with the second device. The device geometry  (shown in Fig.~\ref{fig:fig1}a) and the measurement setup are similar to those previously reported in Ref.~\cite{Ficheux2020}. We conventionally label the coupled energy eigenstates as $|k l \rangle$, where the $k$ and $l$ indices are the uncoupled eigenstates of qubits A and B, respectively. For example, qubit A's (B's) computational transition is labeled $|00\rangle - |10\rangle$ ($|00\rangle - |01\rangle$) and has a frequency $f_A = 217.2 \mathrm{~MHz}$ ($f_B = 488.9 \mathrm{~MHz}$). While the qubit lifetimes are above $100~\mu\textrm{s}$, the Ramsey coherence times are in the $10-14~\mu s$ interval, limited by insufficient thermalization of the measurement lines. The relevant part of the measured two-qubit spectrum is shown in Fig.~\ref{fig:fig1}c. The simulated quantity $\Omega_{kl-k'l'}$ attached to every transition is the Rabi frequency that would be induced by a resonant drive with the same amplitude for all transitions. This frequency scale quantifies the effective coupling of the external drive to the circuit transitions, reflecting both the values of matrix elements of fluxonium charge operators and the asymmetric coupling of the drive field to each qubit (we use only one port to drive both qubits at shown in Fig.~\ref{fig:fig1}b). Details of spectroscopic and time-domain characterization of our device are provided in the Supplementary Note 2.\\

\noindent\textbf{Differential ac-Stark shift.} The computational states $|00\rangle$, $|10\rangle$, $|01\rangle$, and $|11\rangle$ are separated in energy from non-computational states by at least a few $\textrm{GHz}$, which exceeds the qubit frequencies by almost an order of magnitude. Within the computational subspace, the two-qubit dynamics obeys the Hamiltonian
\begin{equation}
    \frac{\hat{H}}{h} = f_A \frac{ZI}{2} + f_B\frac{IZ}{2} + \xi_{ZZ} \frac{ZZ}{4},
    \label{eq:ZZHamiltonian}
\end{equation}
where Z is the corresponding Pauli matrix and $\xi_{ZZ} = \xi_{ZZ}^{\textrm{static}} = -357~\textrm{kHz}$ is the $\textrm{ZZ}$-interaction strength in the present device. The quantity $\xi_{ZZ}^{\textrm{static}}$ has a negative sign because the non-computational levels push stronger on level $|11\rangle$ than on the other computational levels.

\begin{figure}
    \centering
    \includegraphics[width=86mm]{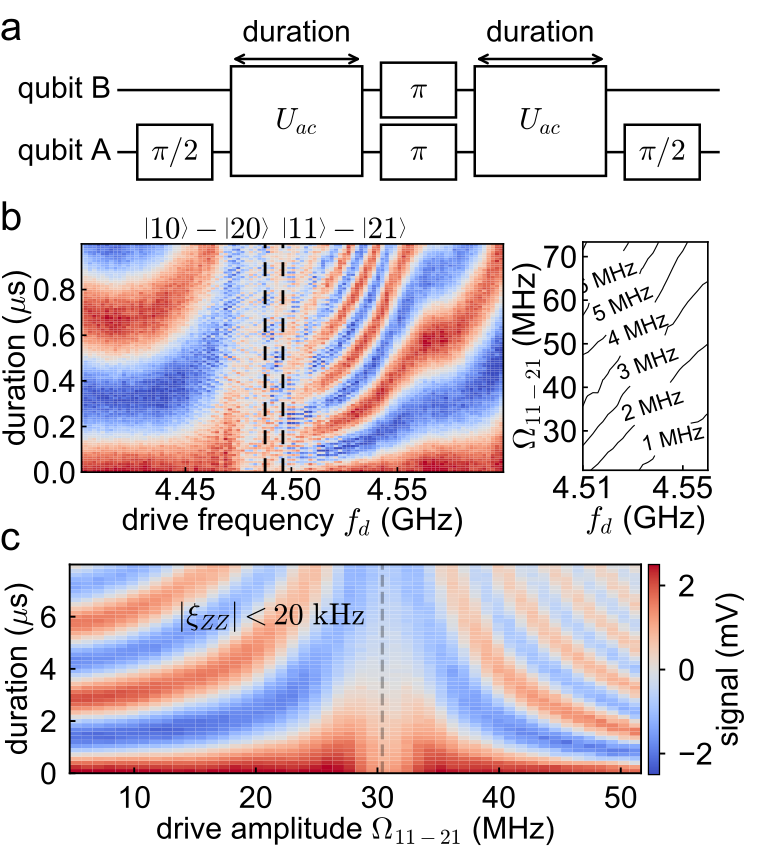}
    \caption{\textbf{Tuning the $\textrm{ZZ}$-interaction.} (a) Pulse sequence used to measure the interaction rate $\xi_{ZZ}$. The qubits evolve under $U_{ac}$ that corresponds to the $\textrm{ZZ}$-term in Eq.~\ref{eq:ZZHamiltonian}. We use a refocusing pulse on each qubit in the middle of the sequence to cancel single-qubit $Z$ rotations. (b) Induced $\textrm{ZZ}$-interactions (a two-qubit Ramsey-type fringe) as a function of drive frequency and amplitude around the $|10 \rangle-| 20 \rangle$ and $|11 \rangle - |21 \rangle$ transitions. The color scale is proportional to $\left<ZI\right>$. The oscillation rate expectedly increases on approaching a resonance condition (left panel). The induced interaction rate reaches about $\xi_{ZZ} = 5 \mathrm{~MHz}$ (right panel). (c) Cancellation of the qubit-qubit interaction. When applying a drive at $4.65 \mathrm{~GHz}$, the $\textrm{ZZ}$-oscillations slow down and speed back up as the drive amplitude passes though the value $\Omega_{11-21} = 30.4 \mathrm{~MHz}$. At this point, we can extract that $\xi_{ZZ} < 20~\textrm{kHz}$.}
    \label{fig:fig2}
\end{figure}

Tuning the magnitude of $\xi_{ZZ}$ by an externally applied microwave drive can be understood as follows. First, let us note that capacitive coupling splits the otherwise degenerate transitions $|10\rangle - |20\rangle$ and $|11\rangle - |21\rangle$ by $\Delta = 8.47~\textrm{MHz}$. The same is true for other transition pairs, e.g. $|11\rangle - |12\rangle$, $|01\rangle - |02\rangle$, or $|00\rangle - |03\rangle$, $|10\rangle - |13\rangle$, although the splitting may vary. Consider driving the circuit at the frequency $f_d$, blue-detuned from the $|10\rangle - |20\rangle$ resonance by an amount $\delta \gg \Delta$ (see Fig.~\ref{fig:fig1}d). Except for the paired transition $|11\rangle - |21\rangle$, there are no other circuit transitions in the $\textrm{GHz}$-vicinity, which reduces the effect of the off-resonant drive to creating ac-Stark shifts. Specifically, if qubit B is in the ground state, there is a positive ac-Stark shift $\delta f^{\textrm{Stark}}(\Omega,\delta) =(\sqrt{\Omega^2+\delta^2}-\delta)/2$ on the qubit A frequency, which can be thought of as pulling level $|10\rangle$ towards level $|20\rangle$ by the $\delta$-detuned drive. However, this shift is larger when qubit B is in the excited state, because the detuning $(\delta-\Delta)$ is smaller and the effective drive amplitude $\Omega$ is larger. Therefore, each qubit acquires a differential ac-Stark shift, which is equivalent to modifying the $\textrm{ZZ}$-term in Eq.~\eqref{eq:ZZHamiltonian} as $\xi _{ZZ}= \xi_{ZZ}^{\textrm{static}}+ \xi_{ZZ}^{\textrm{drive}}$, where $\xi_{ZZ}^{\textrm{drive}} = \delta f^{\textrm{Stark}}(\Omega_{11-21},\delta-\Delta)-\delta f^{\textrm{Stark}}(\Omega_{10-20},\delta)$. Because  $\xi_{ZZ}^{\textrm{static}} <0$ and $\xi_{ZZ}^{\textrm{drive}} >0$ for typical fluxonium parameters, the total qubit-qubit interaction $\xi_{ZZ}$ in Eq.~\eqref{eq:ZZHamiltonian} can be tuned through zero or increased by about an order of magnitude compared to the static value by adjusting the drive frequency and amplitude (see Fig.~\ref{fig:fig1}e and Fig.~\ref{fig:fig2}b, right panel).\\

\noindent\textbf{Tomography of drive-tuned $\mathrm{ZZ}$-interaction.} We verify that the qubit-qubit interaction indeed takes the form of Eq.~\eqref{eq:ZZHamiltonian} using a tomography protocol depicted in Fig.~\ref{fig:fig2}a \cite{Chow2013}. The pulse sequence shown results in the observation of Ramsey-type fringes oscillating at the frequency $\xi_{ZZ}$. Figure~\ref{fig:fig2}b shows the measured oscillations by sweeping the driving frequency around $f_d \approx 4.5~\textrm{GHz}$ and fixing the amplitude such that $\Omega_\mathrm{11-21}=52~\mathrm{MHz}$. As the drive frequency approaches either the $| 10 \rangle - |20 \rangle$ or $| 11 \rangle - | 21 \rangle$ transitions (marked by the black dashed lines at $4.488 \mathrm{~GHz}$ and $4.496\mathrm{~GHz}$, respectively), the $\textrm{ZZ}$-interaction rate reaches about $6 \mathrm{~MHz}$ before the off-resonant ac-Stark shift picture breaks down (Fig.~\ref{fig:fig2}c. b, right inset). Beyond that point, the drive field and the two non-computational transitions undergo a coherent energy exchange, witnessed by the detection of rapidly oscillating ripple features in Fig.~\ref{fig:fig2}b. The $\textrm{ZZ}$-interaction can also be controlled by off-resonantly driving the pair of doublets of transitions $|00\rangle - |03\rangle$, $|10\rangle - |13\rangle$, $|01\rangle - |31\rangle$, $|00\rangle - |30\rangle$, near the frequency $f_d' \approx 6.5~\textrm{GHz}$, where we obtained $\xi_{ZZ} > 10~\textrm{MHz}$ (Supplementary Note 6 and Supplementary Fig.~6). Furthermore, with a more optimal structure of the non-computational transitions, we observed $\xi_{ZZ} > 23~\textrm{MHz}$ in the second device (Supplementary Note 7).

\begin{figure*}
    \centering
    \includegraphics[width=0.9\linewidth]{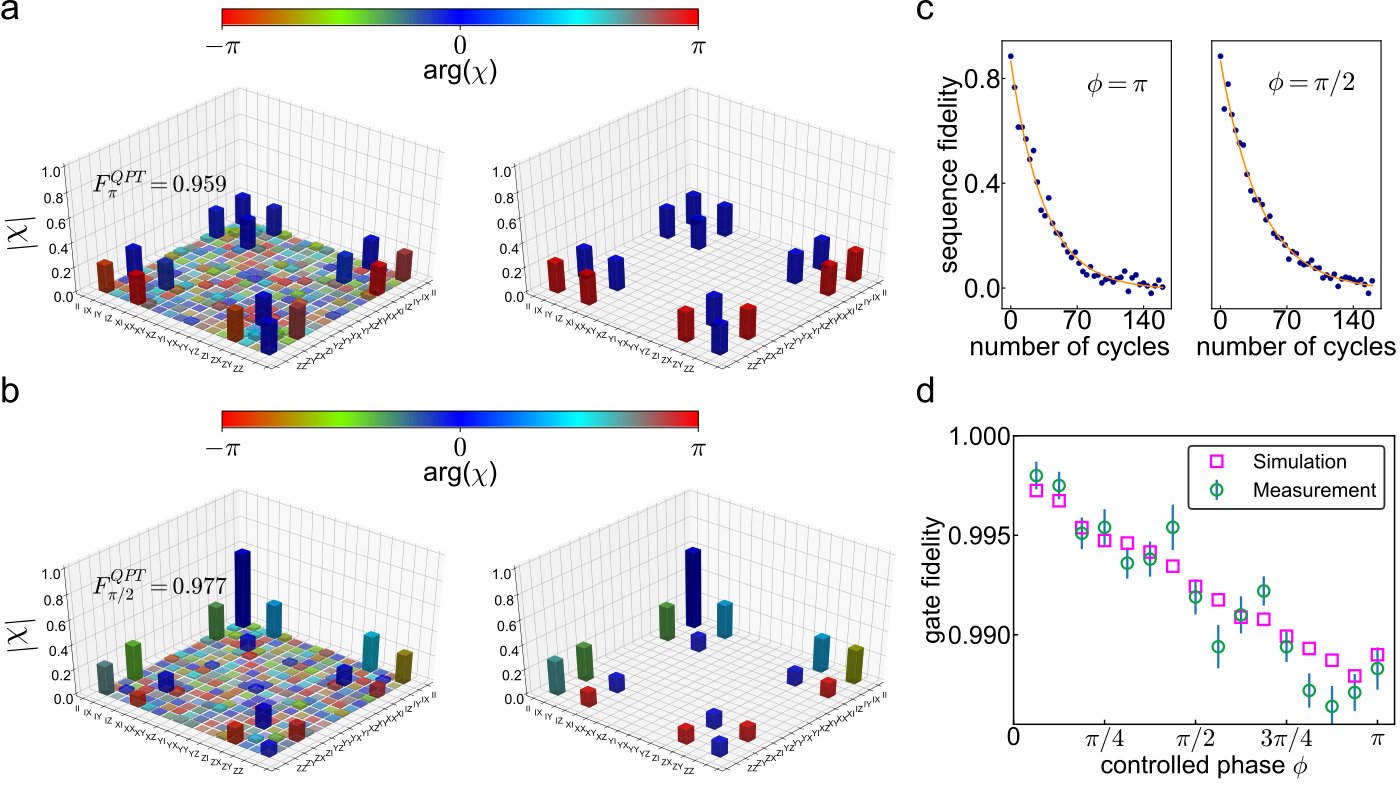}
    \caption{\textbf{Benchmarking of arbitrary CP gates.} (a,b) Examples of the quantum process tomography (QPT) for the two values of the controlled phase $\phi = \pi$ (top) and $\phi = \pi/2$ (bottom). Left (right) panels indicate experimental (theoretical) values. The process tomography $\chi$-matrix reproduces the ideal process with a fidelity $F^{QPT}_{\phi =\pi} = 0.959$ and $F^{QPT}_{\phi =\pi/2} = 0.977$. (c) Cross-entropy benchmarking (XEB) for the phases $\phi= \pi/2~ \mathrm{and}~\pi$ corresponding to the process tomography shown in (a,b). We extract the Pauli and gate errors from the decay of the XEB sequence fidelity with the number of cycles. The gate fidelity at $\phi=\pi/2$ reaches $(99.2\pm 0.1) \%$. (d) Fidelity of the controlled-phase gate family. The green circles are obtained by cross-entropy benchmarking and the magenta squares are obtained by master equation simulations using the average $T_1$ and $T_2^E$ shown in Table~\ref{table:T1T2} (see Supplementary Note 8). The gate fidelity averaged over the complete family exceeds $99.2 \%$.}
    \label{fig:fig4}
\end{figure*}

The immediate application of our differential ac-Stark shift phenomenon is to permanently cancel the static $\textrm{ZZ}$-interaction between qubits. This is achieved by applying a microwave tone at the frequency $f_d = 4.65~\mathrm{GHz}$ and the amplitude such that $\Omega_{11-21}=30~\mathrm{MHz}$ (see Fig.~\ref{fig:fig2} c). As we fine-tune the drive amplitude, the two-qubit Ramsey fringe slows down from the static value of $357~\textrm{kHz}$ to the value $\xi_{ZZ} < 20~\textrm{kHz}$, beyond which the oscillations cannot be resolved due to the qubit's finite coherence time. We characterized single-qubit operations, in the presence of the $\textrm{ZZ}$-canceling drive, using conventional randomized benchmarking sequences~\cite{McKay2017}. In the case of individual bechmarking, the average single-qubit gate fidelity is $0.9969$ ($45~\textrm{ns}$ long pulses) for qubit A and $0.9991$ for qubit B ($26~\textrm{ns}$ long pulses). The simultaneous qubit benchmarking requires longer pulse sequences, which reduces the fidelity to $0.9963\pm 0.0001$ for qubit A and $0.9957\pm 0.0001$ for qubit B (See Supplementary Note 4 for details).

\noindent\textbf{Arbitrary controlled-phase (CP) gate.} The $\textrm{ZZ}$-term in Eq.~\eqref{eq:ZZHamiltonian} can be switched on and off on a time scale of about $10~\textrm{ns}$ using Gaussian-edge pulses supplemented with the commonly used derivative removal (DRAG) distortion \cite{Motzoi2009,Gambetta2011}. During the time interval of $\xi_{ZZ}(t)\neq 0$, the states $|10 \rangle$ and $| 11 \rangle$ accumulate different phases $\phi_{10}$ and $\phi_{11}$, which is equivalent to the action of a unitary evolution operator $U = \mathrm{diag}(1,e^{-i \phi_{10}},1,e^{-i \phi_{11}})$ in the computational subspace. Using virtual Z rotations \cite{McKay2017}, the accumulated phase can be entirely assigned to any state such as $| 11 \rangle$ to implement a controlled-phase operation $U_{\mathrm{CP}}(\phi)=\mathrm{diag}(1,1,1,e^{-i\phi})$ where $\phi=\phi_{11}-\phi_{10}$. In principle, one can modulate both the drive frequency and amplitude during the gate pulse, but for simplicity we fixed the drive frequency and tried two different values, $f_d = 4.545~\textrm{GHz}$ (detuned by $49~\textrm{MHz}$ from transition $|11\rangle - |21\rangle$) and $f_d' = 6.665~\textrm{GHz}$ (detuned by $55~\textrm{MHz}$ from $|00\rangle - |30\rangle$, see Supplementary Note 6).

We start the CP gate characterization by performing quantum process tomography (QPT). The process tomography matrix (the $\chi$-matrix) \cite{Nielsen2000,Corcoles2013} is obtained by preparing $16$ independent input states, applying the CP gate, and performing the state tomography of the final quantum state (see Supplementary Note 5). By comparing the measured $\chi$-matrix to the theoretical one, we perform the initial tune-up of the pulse parameters required to implement the CP gate with the given phase $\phi$: gate duration, drive amplitude, DRAG coefficient, and the phases of the virtual Z-rotations. Optimized QPT examples for $\phi = \pi$ and $\phi = \pi/2$, with the single-qubit Z-rotations adjusted to exhibit only the $\mathrm{ZZ}$ evolution, are shown in Figs.~\ref{fig:fig4}a,b. The $\chi$-matrix fidelity \cite{Chow2009} reached $0.959$ for $\phi=\pi$ and $0.977$ for $\phi = \pi/2$, likely limited by the state preparation and measurement (SPAM) errors.

At $\phi = \pi$ the CP gate belongs to the Clifford group, i.e. it becomes the controlled-Z gate. Hence, it can be characterized using randomized benchmarking (RB), which evades SPAM limitations. Using procedures similar to those described in Ref.~\cite{Ficheux2020}, we optimized the gate pulses and obtained a CZ gate fidelity of $0.989 \pm 0.001$ at $f_d = 4.545~\textrm{GHz}$ and $0.991\pm0.001$ at $f_d' = 6.665~\textrm{GHz}$. Next, we characterize the CP gate at $\phi = \pi$ using the cross-entropy (XEB) benchmarking technique, which is applicable to non-Clifford operations, and also evades the SPAM errors~\cite{Arute2019, Barends2019, Foxen2020}. The XEB procedure consists of a succession of cycles, each composed of one randomly chosen single-qubit Clifford gate on each qubit and a given CP gate. The sequence fidelity is calculated from the cross-entropy between the measured and expected qubit state distribution \cite{Barends2019}, which decays exponentially with the number of cycles (see Methods). The optimal gate parameters are found by optimizing the sequence fidelity at a fixed number of cycles with the Nelder-Mead algorithm. The XEB procedure applied to the CZ gate at $f_d = 4.545~\textrm{GHz}$ yields a gate fidelity of $0.988\pm 0.001$ which agrees with the results of randomized benchmarking and hence validates the use of XEB for other values $\phi$ of the controlled phase.

Finally, we apply the XEB procedure to a family of CP gates with the value of $\phi$ equally spaced by $\pi/16$. The extracted gate error grows approximately linearly in $\phi$ with a slope of about $3\times10^{-3}$ per radian (Fig.~\ref{fig:fig4}d, circular markers). For $\phi = \pi/16$, the CP gate error reaches $2\times 10^{-3}$, which is close to the experimental resolution limit. In order to understand the origin of the gate error, we performed detailed master-equation simulations of the driven two-fluxonium system (see Supplementary Note 8). Our numerical model closely reproduces the data while relying only on experimentally measured parameters (Fig.~\ref{fig:fig4}d, square markers). According to the model, the error is entirely due to incoherent processes, while the coherent error is absent down to the $10^{-4}$ level thanks to the strong anharmonicity of fluxonium's non-computational transitions.\\

\noindent\textbf{Discussion}\\

\noindent The off-resonant drive in our experiment essentially replaces the original (undriven) qubit states with the ``dressed" ones, e.g. $|10\rangle \rightarrow |10\rangle + (\lambda/2) |20\rangle$ and $|11\rangle \rightarrow |11\rangle + (\lambda/2) |21\rangle$ in the case shown in Fig.~\ref{fig:fig1}d, where $\lambda \approx \Omega/\delta$. It is this dressing that leads to the tunable interaction rate $\xi_{ZZ}$ in Eq.~\ref{eq:ZZHamiltonian}. One can further show that in the presence of energy relaxation between the undressed single-qubit states $|20\rangle$ and $|10\rangle$ at a rate $1/T_1^{(2\rightarrow 1)}$, the dressed qubit transition $|00\rangle - |10\rangle$ inherits a pure dephasing rate $\lambda^2/8T_1^{(2\rightarrow 1)}$. Remarkably, dressing due to the $\textrm{ZZ}$-cancellation drive is so weak, $\lambda^2 \approx 0.04$, that even for a low relaxation time $T_1^{(2\rightarrow 1)}\approx 5~\mu\textrm{s}$ in our present experiment, the extra dephasing becomes significant only at the $1~\textrm{ms}$ level. Thus, the highest measured single-dressed-qubit gate fidelity reached $0.999$, and it is limited by the undressed coherence time of about $10-15~\mu\textrm{s}$. In other words, our scheme for eliminating the $\textrm{ZZ}$-interaction has practically no adverse effects, and hence can be generally used to cancel the quantum cross-talk in fluxonium-based processors.

During the CP gate, both $\xi_{ZZ}$ and $\lambda$ temporarily grow stronger, and hence the energy relaxation of the non-computational transitions becomes an important contributor to the incoherent gate error. Let us note that the $|1\rangle - |2\rangle$ transition of fluxoniums has the frequency and charge matrix element of typical transmons, whose relaxation time usually belongs to the $10-100~\mu\textrm{s}$ range. Therefore, we expect the same range for $T_1^{(2\rightarrow 1)}$ in fluxoniums once the fabrication and thermalization procedures are properly optimized. In this case, a $\textrm{ZZ}$-interaction rate in the range $\xi_{ZZ} = 5-10~\textrm{MHz}$ can be induced with no decoherence exposure at the level of a few hundred microseconds. In fact, the numerical model projects a CP gate error (for $\phi = \pi$) well in the $10^{-4}$ range for the next generation of fluxonium devices.

The demonstrated all-microwave control over the $\textrm{ZZ}$-interaction is technologically attractive for scaling-up superconducting quantum processors. Our scheme does not require a close arrangement of qubit frequencies, in contrast to the case of the cross-resonance (CR) gate~\cite{Sheldon2016,Tripathi2019} for transmons, which would mitigate the spectral crowding issues~\cite{Sheldon2016, Krinner2020a} and enable multiplexing of the qubit control (for example, here we used only one input port for the entire experiment). Furthermore, the CP gate can be activated by a broad selection of drive frequencies in the $\textrm{GHz}$-range and the required drive power is comparable to single-qubit rotations~\cite{Krinner2020}. Most importantly, while certain circuit parameter combinations would further mitigate the errors, the CP gate requires no special parameter combinations and hence would be tolerant to fabrication variability. For example, the qubit frequency could be as low as $100~\textrm{MHz}$, as in our earlier demonstration of the CZ-gate~\cite{Ficheux2020}, but it can also be around $200-500~\textrm{MHz}$ (main device here) or even $700-1.3~\textrm{GHz}$ (second device here). 

Among the previously explored two-qubit gates with a high degree of flexibility, our scheme is most reminiscent of the resonator-induced phase (RIP) gate~\cite{Cross2015,Paik2016,Puri2016} for transmons. The RIP gate populates an auxiliary bus mode with off-resonant photons to induce a differential Stark shift. In comparison, our CP gate requires no auxiliary modes or complex pulse-shaping, has closely-confined driven dynamics, and already enables a higher fidelity in devices with very sub-optimal coherence, all thanks to the strong anharmonicity of fluxoniums. 

\bigskip

%\footnotesize

\noindent\textbf{Methods}\\

\noindent\textbf{Experimental setup.} We use the same readout and initialization procedures described in Ref.~\cite{Ficheux2020}. The device is embedded into a rectangular copper cavity resonator with a resonance frequency $f_C = 7.538 \mathrm{~GHz}$ and linewidth $\kappa/2 \pi = 5 \mathrm{~MHz}$, thermally anchored to the base plate of a dilution refrigerator at $14 \mathrm{~mK}$. External driving is provided through a single input port to the cavity and the transmission signal is monitored using a stronger coupled output port. Spectroscopy vs.~flux data was used to accurately extract circuit parameters and calculate transition matrix elements. The table of relevant coherence times is provided below. 

\begin{center}
\begin{table}[h!]
\begin{tabular}{ | c | c | c | c | } \hline
  & $T_1 \mathrm{~(\mu s)}$& $T_2^R \mathrm{~(\mu s)}$ & $T_2^E \mathrm{~(\mu s)}$\\ \hline
  $\left| 00 \right> - \left| 10 \right>$ & $158-207$ & $10-12$ & $14-15$ \\
  $\left| 00 \right> - \left| 01 \right>$ & $116-141$ & $13$ & $20-25$ \\ 
  $\left| 11 \right> - \left| 21 \right>$ & $4.9-6.2$ & $2.6-2.8$ & $3.3$ \\ \hline
\end{tabular}
\caption{\label{table:T1T2} Energy relaxation time $T_1$, Ramsey coherence time $T_2^R$, and spin echo coherence time $T_2^E$. The ranges corresponds to time fluctuations during the experiment.}
\end{table}
\end{center}

We perform a single-shot joint readout of the two qubit states \cite{Filipp2009} by preamplifying the readout signal with the Josephson Traveling Parametric Wave Amplifier (JTWPA) \cite{Macklin2015}. The equilibrium population of the two qubit states are obtained by fitting the single-shot histograms with 4 Gaussian distributions, and we compensate the readout error with an empirical model. Prior to each experiment, the qubits are intialized by populating the cavity with a large number of photons, which conveniently prepares the two qubits in a mixed state with the excited state populations of $69$ \% and $82$ \%, respectively. Such a degree of state initialization is sufficient to perform accurate gate error measurement, including the quantum process tomography. The details of experimental procedures are provided in the supplementary material.\\

\noindent \textbf{Cross-entropy benchmarking (XEB).} The corrected populations are used to calculate the cross-entropy between the experimental populations and the expected populations starting from the state after the initialization pulse. The cross-entropy yields the XEB sequence fidelity. By fitting the sequence fidelity versus number of cycles with $Ap^m+B$, the Pauli error per cycle $r_\mathrm{cycle}^P$ is given by
\begin{eqnarray}
    r_{\mathrm{cycle}}&=&\frac{N-1}{N}(1-p) \\
    r_{\mathrm{cycle}}^P&=&\frac{N+1}{N}r_{\mathrm{cycle}}
\label{eq:rP} 
\end{eqnarray}
where $N$ is the dimension of the system. Then we extract the CP gate Pauli error $r_{\mathrm{CP}}^P$ from the equation $(1-r_{\mathrm{cycle}}^P) = (1-r_{A}^P)(1-r_{B}^P)(1-r_{\mathrm{CP}}^P)$, where $r_{A}^P,~r_{B}^P$ are the average single-qubit Pauli error measured from the simultaneous RB experiment described in Supplementary Note 4. The arbitrary CP gate fidelity is calculated by converting Pauli error back to the gate error $r_{\mathrm{CP}}$ as shown in Fig.~\ref{fig:fig4}c and Fig.~\ref{fig:fig4}d.\\

\footnotesize

\noindent\textbf{Acknowledgements.}
This work was supported by the DOE and by the ARO-LPS HiPS program. V.E.M. and M.G.V acknowledge the Faculty Research Award from Google. We thank Lincoln Labs and IARPA for providing a Josephson Traveling Wave Parametric Amplifier.

\footnotesize

%\noindent\textbf{Author contributions.}
%H.X. and Q.F. contributed equally to this work.
%H.X., Q.F., and L.B.N. measured the device which A.S. designed and fabricated. K.N.N. and M.G.V. assisted by L.B.N provided theory support. L.B.N. also contributed to designing the sample, the measurement setup, and to performing initial device characterization. E.D., D.R., and C.W. provided additional data supporting our conclusions in the supplementary note 7 and contributed to discussions of the described gate scheme. H.X., Q.F., M.G.V, and V.E.M. wrote the manuscript with input from all authors. V.E.M. managed the project.

%\footnotesize

%\noindent \textbf{Competing interests.}
%The authors declare no competing financial interests.

%\bibliography{references2}

\begin{thebibliography}{48}%
\makeatletter
\providecommand \@ifxundefined [1]{%
 \@ifx{#1\undefined}
}%
\providecommand \@ifnum [1]{%
 \ifnum #1\expandafter \@firstoftwo
 \else \expandafter \@secondoftwo
 \fi
}%
\providecommand \@ifx [1]{%
 \ifx #1\expandafter \@firstoftwo
 \else \expandafter \@secondoftwo
 \fi
}%
\providecommand \natexlab [1]{#1}%
\providecommand \enquote  [1]{``#1''}%
\providecommand \bibnamefont  [1]{#1}%
\providecommand \bibfnamefont [1]{#1}%
\providecommand \citenamefont [1]{#1}%
\providecommand \href@noop [0]{\@secondoftwo}%
\providecommand \href [0]{\begingroup \@sanitize@url \@href}%
\providecommand \@href[1]{\@@startlink{#1}\@@href}%
\providecommand \@@href[1]{\endgroup#1\@@endlink}%
\providecommand \@sanitize@url [0]{\catcode `\\12\catcode `\$12\catcode
  `\&12\catcode `\#12\catcode `\^12\catcode `\_12\catcode `\%12\relax}%
\providecommand \@@startlink[1]{}%
\providecommand \@@endlink[0]{}%
\providecommand \url  [0]{\begingroup\@sanitize@url \@url }%
\providecommand \@url [1]{\endgroup\@href {#1}{\urlprefix }}%
\providecommand \urlprefix  [0]{URL }%
\providecommand \Eprint [0]{\href }%
\providecommand \doibase [0]{https://doi.org/}%
\providecommand \selectlanguage [0]{\@gobble}%
\providecommand \bibinfo  [0]{\@secondoftwo}%
\providecommand \bibfield  [0]{\@secondoftwo}%
\providecommand \translation [1]{[#1]}%
\providecommand \BibitemOpen [0]{}%
\providecommand \bibitemStop [0]{}%
\providecommand \bibitemNoStop [0]{.\EOS\space}%
\providecommand \EOS [0]{\spacefactor3000\relax}%
\providecommand \BibitemShut  [1]{\csname bibitem#1\endcsname}%
\let\auto@bib@innerbib\@empty
%</preamble>
\bibitem [{\citenamefont {Strauch}\ \emph {et~al.}(2003)\citenamefont
  {Strauch}, \citenamefont {Johnson}, \citenamefont {Dragt}, \citenamefont
  {Lobb}, \citenamefont {Anderson},\ and\ \citenamefont
  {Wellstood}}]{Strauch2003}%
  \BibitemOpen
  \bibfield  {author} {\bibinfo {author} {\bibfnamefont {F.~W.}\ \bibnamefont
  {Strauch}}, \bibinfo {author} {\bibfnamefont {P.~R.}\ \bibnamefont
  {Johnson}}, \bibinfo {author} {\bibfnamefont {A.~J.}\ \bibnamefont {Dragt}},
  \bibinfo {author} {\bibfnamefont {C.~J.}\ \bibnamefont {Lobb}}, \bibinfo
  {author} {\bibfnamefont {J.~R.}\ \bibnamefont {Anderson}},\ and\ \bibinfo
  {author} {\bibfnamefont {F.~C.}\ \bibnamefont {Wellstood}},\ }\bibfield
  {title} {\bibinfo {title} {{Quantum logic gates for coupled superconducting
  phase qubits}},\ }\href {https://doi.org/10.1103/PhysRevLett.91.167005}
  {\bibfield  {journal} {\bibinfo  {journal} {Physical Review Letters}\
  }\textbf {\bibinfo {volume} {91}},\ \bibinfo {pages} {167005} (\bibinfo
  {year} {2003})}\BibitemShut {NoStop}%
\bibitem [{\citenamefont {Dicarlo}\ \emph {et~al.}(2009)\citenamefont
  {Dicarlo}, \citenamefont {Chow}, \citenamefont {Gambetta}, \citenamefont
  {Bishop}, \citenamefont {Johnson}, \citenamefont {Schuster}, \citenamefont
  {Majer}, \citenamefont {Blais}, \citenamefont {Frunzio}, \citenamefont
  {Girvin},\ and\ \citenamefont {Schoelkopf}}]{Dicarlo2009}%
  \BibitemOpen
  \bibfield  {author} {\bibinfo {author} {\bibfnamefont {L.}~\bibnamefont
  {Dicarlo}}, \bibinfo {author} {\bibfnamefont {J.~M.}\ \bibnamefont {Chow}},
  \bibinfo {author} {\bibfnamefont {J.~M.}\ \bibnamefont {Gambetta}}, \bibinfo
  {author} {\bibfnamefont {L.~S.}\ \bibnamefont {Bishop}}, \bibinfo {author}
  {\bibfnamefont {B.~R.}\ \bibnamefont {Johnson}}, \bibinfo {author}
  {\bibfnamefont {D.~I.}\ \bibnamefont {Schuster}}, \bibinfo {author}
  {\bibfnamefont {J.}~\bibnamefont {Majer}}, \bibinfo {author} {\bibfnamefont
  {A.}~\bibnamefont {Blais}}, \bibinfo {author} {\bibfnamefont
  {L.}~\bibnamefont {Frunzio}}, \bibinfo {author} {\bibfnamefont {S.~M.}\
  \bibnamefont {Girvin}},\ and\ \bibinfo {author} {\bibfnamefont {R.~J.}\
  \bibnamefont {Schoelkopf}},\ }\bibfield  {title} {\bibinfo {title}
  {{Demonstration of two-qubit algorithms with a superconducting quantum
  processor}},\ }\href {https://doi.org/10.1038/nature08121} {\bibfield
  {journal} {\bibinfo  {journal} {Nature}\ }\textbf {\bibinfo {volume} {460}},\
  \bibinfo {pages} {240} (\bibinfo {year} {2009})}\BibitemShut {NoStop}%
\bibitem [{\citenamefont {Gambetta}\ \emph {et~al.}(2012)\citenamefont
  {Gambetta}, \citenamefont {C{\'{o}}rcoles}, \citenamefont {Merkel},
  \citenamefont {Johnson}, \citenamefont {Smolin}, \citenamefont {Chow},
  \citenamefont {Ryan}, \citenamefont {Rigetti}, \citenamefont {Poletto},
  \citenamefont {Ohki}, \citenamefont {Ketchen},\ and\ \citenamefont
  {Steffen}}]{Gambetta2012}%
  \BibitemOpen
  \bibfield  {author} {\bibinfo {author} {\bibfnamefont {J.~M.}\ \bibnamefont
  {Gambetta}}, \bibinfo {author} {\bibfnamefont {A.~D.}\ \bibnamefont
  {C{\'{o}}rcoles}}, \bibinfo {author} {\bibfnamefont {S.~T.}\ \bibnamefont
  {Merkel}}, \bibinfo {author} {\bibfnamefont {B.~R.}\ \bibnamefont {Johnson}},
  \bibinfo {author} {\bibfnamefont {J.~A.}\ \bibnamefont {Smolin}}, \bibinfo
  {author} {\bibfnamefont {J.~M.}\ \bibnamefont {Chow}}, \bibinfo {author}
  {\bibfnamefont {C.~A.}\ \bibnamefont {Ryan}}, \bibinfo {author}
  {\bibfnamefont {C.}~\bibnamefont {Rigetti}}, \bibinfo {author} {\bibfnamefont
  {S.}~\bibnamefont {Poletto}}, \bibinfo {author} {\bibfnamefont {T.~A.}\
  \bibnamefont {Ohki}}, \bibinfo {author} {\bibfnamefont {M.~B.}\ \bibnamefont
  {Ketchen}},\ and\ \bibinfo {author} {\bibfnamefont {M.}~\bibnamefont
  {Steffen}},\ }\bibfield  {title} {\bibinfo {title} {{Characterization of
  addressability by simultaneous randomized benchmarking}},\ }\href
  {https://doi.org/10.1103/PhysRevLett.109.240504} {\bibfield  {journal}
  {\bibinfo  {journal} {Physical Review Letters}\ }\textbf {\bibinfo {volume}
  {109}},\ \bibinfo {pages} {240504} (\bibinfo {year} {2012})}\BibitemShut
  {NoStop}%
\bibitem [{\citenamefont {McKay}\ \emph {et~al.}(2019)\citenamefont {McKay},
  \citenamefont {Sheldon}, \citenamefont {Smolin}, \citenamefont {Chow},\ and\
  \citenamefont {Gambetta}}]{McKay2019}%
  \BibitemOpen
  \bibfield  {author} {\bibinfo {author} {\bibfnamefont {D.~C.}\ \bibnamefont
  {McKay}}, \bibinfo {author} {\bibfnamefont {S.}~\bibnamefont {Sheldon}},
  \bibinfo {author} {\bibfnamefont {J.~A.}\ \bibnamefont {Smolin}}, \bibinfo
  {author} {\bibfnamefont {J.~M.}\ \bibnamefont {Chow}},\ and\ \bibinfo
  {author} {\bibfnamefont {J.~M.}\ \bibnamefont {Gambetta}},\ }\bibfield
  {title} {\bibinfo {title} {{Three-Qubit Randomized Benchmarking}},\ }\href
  {https://doi.org/10.1103/PhysRevLett.122.200502} {\bibfield  {journal}
  {\bibinfo  {journal} {Physical Review Letters}\ }\textbf {\bibinfo {volume}
  {122}},\ \bibinfo {pages} {200502} (\bibinfo {year} {2019})}\BibitemShut
  {NoStop}%
\bibitem [{\citenamefont {Arute}\ \emph {et~al.}(2019)\citenamefont {Arute},
  \citenamefont {Arya}, \citenamefont {Babbush}, \citenamefont {Bacon},
  \citenamefont {Bardin}, \citenamefont {Barends}, \citenamefont {Biswas},
  \citenamefont {Boixo}, \citenamefont {Brandao}, \citenamefont {Buell},
  \citenamefont {Burkett}, \citenamefont {Chen}, \citenamefont {Chen},
  \citenamefont {Chiaro}, \citenamefont {Collins}, \citenamefont {Courtney},
  \citenamefont {Dunsworth}, \citenamefont {Farhi}, \citenamefont {Foxen},
  \citenamefont {Fowler}, \citenamefont {Gidney}, \citenamefont {Giustina},
  \citenamefont {Graff}, \citenamefont {Guerin}, \citenamefont {Habegger},
  \citenamefont {Harrigan}, \citenamefont {Hartmann}, \citenamefont {Ho},
  \citenamefont {Hoffmann}, \citenamefont {Huang}, \citenamefont {Humble},
  \citenamefont {Isakov}, \citenamefont {Jeffrey}, \citenamefont {Jiang},
  \citenamefont {Kafri}, \citenamefont {Kechedzhi}, \citenamefont {Kelly},
  \citenamefont {Klimov}, \citenamefont {Knysh}, \citenamefont {Korotkov},
  \citenamefont {Kostritsa}, \citenamefont {Landhuis}, \citenamefont
  {Lindmark}, \citenamefont {Lucero}, \citenamefont {Lyakh}, \citenamefont
  {Mandr{\`{a}}}, \citenamefont {McClean}, \citenamefont {McEwen},
  \citenamefont {Megrant}, \citenamefont {Mi}, \citenamefont {Michielsen},
  \citenamefont {Mohseni}, \citenamefont {Mutus}, \citenamefont {Naaman},
  \citenamefont {Neeley}, \citenamefont {Neill}, \citenamefont {Niu},
  \citenamefont {Ostby}, \citenamefont {Petukhov}, \citenamefont {Platt},
  \citenamefont {Quintana}, \citenamefont {Rieffel}, \citenamefont {Roushan},
  \citenamefont {Rubin}, \citenamefont {Sank}, \citenamefont {Satzinger},
  \citenamefont {Smelyanskiy}, \citenamefont {Sung}, \citenamefont
  {Trevithick}, \citenamefont {Vainsencher}, \citenamefont {Villalonga},
  \citenamefont {White}, \citenamefont {Yao}, \citenamefont {Yeh},
  \citenamefont {Zalcman}, \citenamefont {Neven},\ and\ \citenamefont
  {Martinis}}]{Arute2019}%
  \BibitemOpen
  \bibfield  {author} {\bibinfo {author} {\bibfnamefont {F.}~\bibnamefont
  {Arute}}, \bibinfo {author} {\bibfnamefont {K.}~\bibnamefont {Arya}},
  \bibinfo {author} {\bibfnamefont {R.}~\bibnamefont {Babbush}}, \bibinfo
  {author} {\bibfnamefont {D.}~\bibnamefont {Bacon}}, \bibinfo {author}
  {\bibfnamefont {J.~C.}\ \bibnamefont {Bardin}}, \bibinfo {author}
  {\bibfnamefont {R.}~\bibnamefont {Barends}}, \bibinfo {author} {\bibfnamefont
  {R.}~\bibnamefont {Biswas}}, \bibinfo {author} {\bibfnamefont
  {S.}~\bibnamefont {Boixo}}, \bibinfo {author} {\bibfnamefont {F.~G.}\
  \bibnamefont {Brandao}}, \bibinfo {author} {\bibfnamefont {D.~A.}\
  \bibnamefont {Buell}}, \bibinfo {author} {\bibfnamefont {B.}~\bibnamefont
  {Burkett}}, \bibinfo {author} {\bibfnamefont {Y.}~\bibnamefont {Chen}},
  \bibinfo {author} {\bibfnamefont {Z.}~\bibnamefont {Chen}}, \bibinfo {author}
  {\bibfnamefont {B.}~\bibnamefont {Chiaro}}, \bibinfo {author} {\bibfnamefont
  {R.}~\bibnamefont {Collins}}, \bibinfo {author} {\bibfnamefont
  {W.}~\bibnamefont {Courtney}}, \bibinfo {author} {\bibfnamefont
  {A.}~\bibnamefont {Dunsworth}}, \bibinfo {author} {\bibfnamefont
  {E.}~\bibnamefont {Farhi}}, \bibinfo {author} {\bibfnamefont
  {B.}~\bibnamefont {Foxen}}, \bibinfo {author} {\bibfnamefont
  {A.}~\bibnamefont {Fowler}}, \bibinfo {author} {\bibfnamefont
  {C.}~\bibnamefont {Gidney}}, \bibinfo {author} {\bibfnamefont
  {M.}~\bibnamefont {Giustina}}, \bibinfo {author} {\bibfnamefont
  {R.}~\bibnamefont {Graff}}, \bibinfo {author} {\bibfnamefont
  {K.}~\bibnamefont {Guerin}}, \bibinfo {author} {\bibfnamefont
  {S.}~\bibnamefont {Habegger}}, \bibinfo {author} {\bibfnamefont {M.~P.}\
  \bibnamefont {Harrigan}}, \bibinfo {author} {\bibfnamefont {M.~J.}\
  \bibnamefont {Hartmann}}, \bibinfo {author} {\bibfnamefont {A.}~\bibnamefont
  {Ho}}, \bibinfo {author} {\bibfnamefont {M.}~\bibnamefont {Hoffmann}},
  \bibinfo {author} {\bibfnamefont {T.}~\bibnamefont {Huang}}, \bibinfo
  {author} {\bibfnamefont {T.~S.}\ \bibnamefont {Humble}}, \bibinfo {author}
  {\bibfnamefont {S.~V.}\ \bibnamefont {Isakov}}, \bibinfo {author}
  {\bibfnamefont {E.}~\bibnamefont {Jeffrey}}, \bibinfo {author} {\bibfnamefont
  {Z.}~\bibnamefont {Jiang}}, \bibinfo {author} {\bibfnamefont
  {D.}~\bibnamefont {Kafri}}, \bibinfo {author} {\bibfnamefont
  {K.}~\bibnamefont {Kechedzhi}}, \bibinfo {author} {\bibfnamefont
  {J.}~\bibnamefont {Kelly}}, \bibinfo {author} {\bibfnamefont {P.~V.}\
  \bibnamefont {Klimov}}, \bibinfo {author} {\bibfnamefont {S.}~\bibnamefont
  {Knysh}}, \bibinfo {author} {\bibfnamefont {A.}~\bibnamefont {Korotkov}},
  \bibinfo {author} {\bibfnamefont {F.}~\bibnamefont {Kostritsa}}, \bibinfo
  {author} {\bibfnamefont {D.}~\bibnamefont {Landhuis}}, \bibinfo {author}
  {\bibfnamefont {M.}~\bibnamefont {Lindmark}}, \bibinfo {author}
  {\bibfnamefont {E.}~\bibnamefont {Lucero}}, \bibinfo {author} {\bibfnamefont
  {D.}~\bibnamefont {Lyakh}}, \bibinfo {author} {\bibfnamefont
  {S.}~\bibnamefont {Mandr{\`{a}}}}, \bibinfo {author} {\bibfnamefont {J.~R.}\
  \bibnamefont {McClean}}, \bibinfo {author} {\bibfnamefont {M.}~\bibnamefont
  {McEwen}}, \bibinfo {author} {\bibfnamefont {A.}~\bibnamefont {Megrant}},
  \bibinfo {author} {\bibfnamefont {X.}~\bibnamefont {Mi}}, \bibinfo {author}
  {\bibfnamefont {K.}~\bibnamefont {Michielsen}}, \bibinfo {author}
  {\bibfnamefont {M.}~\bibnamefont {Mohseni}}, \bibinfo {author} {\bibfnamefont
  {J.}~\bibnamefont {Mutus}}, \bibinfo {author} {\bibfnamefont
  {O.}~\bibnamefont {Naaman}}, \bibinfo {author} {\bibfnamefont
  {M.}~\bibnamefont {Neeley}}, \bibinfo {author} {\bibfnamefont
  {C.}~\bibnamefont {Neill}}, \bibinfo {author} {\bibfnamefont {M.~Y.}\
  \bibnamefont {Niu}}, \bibinfo {author} {\bibfnamefont {E.}~\bibnamefont
  {Ostby}}, \bibinfo {author} {\bibfnamefont {A.}~\bibnamefont {Petukhov}},
  \bibinfo {author} {\bibfnamefont {J.~C.}\ \bibnamefont {Platt}}, \bibinfo
  {author} {\bibfnamefont {C.}~\bibnamefont {Quintana}}, \bibinfo {author}
  {\bibfnamefont {E.~G.}\ \bibnamefont {Rieffel}}, \bibinfo {author}
  {\bibfnamefont {P.}~\bibnamefont {Roushan}}, \bibinfo {author} {\bibfnamefont
  {N.~C.}\ \bibnamefont {Rubin}}, \bibinfo {author} {\bibfnamefont
  {D.}~\bibnamefont {Sank}}, \bibinfo {author} {\bibfnamefont {K.~J.}\
  \bibnamefont {Satzinger}}, \bibinfo {author} {\bibfnamefont {V.}~\bibnamefont
  {Smelyanskiy}}, \bibinfo {author} {\bibfnamefont {K.~J.}\ \bibnamefont
  {Sung}}, \bibinfo {author} {\bibfnamefont {M.~D.}\ \bibnamefont
  {Trevithick}}, \bibinfo {author} {\bibfnamefont {A.}~\bibnamefont
  {Vainsencher}}, \bibinfo {author} {\bibfnamefont {B.}~\bibnamefont
  {Villalonga}}, \bibinfo {author} {\bibfnamefont {T.}~\bibnamefont {White}},
  \bibinfo {author} {\bibfnamefont {Z.~J.}\ \bibnamefont {Yao}}, \bibinfo
  {author} {\bibfnamefont {P.}~\bibnamefont {Yeh}}, \bibinfo {author}
  {\bibfnamefont {A.}~\bibnamefont {Zalcman}}, \bibinfo {author} {\bibfnamefont
  {H.}~\bibnamefont {Neven}},\ and\ \bibinfo {author} {\bibfnamefont {J.~M.}\
  \bibnamefont {Martinis}},\ }\bibfield  {title} {\bibinfo {title} {{Quantum
  supremacy using a programmable superconducting processor}},\ }\href
  {https://doi.org/10.1038/s41586-019-1666-5} {\bibfield  {journal} {\bibinfo
  {journal} {Nature}\ }\textbf {\bibinfo {volume} {574}},\ \bibinfo {pages}
  {505} (\bibinfo {year} {2019})}\BibitemShut {NoStop}%
\bibitem [{\citenamefont {Rudinger}\ \emph {et~al.}(2019)\citenamefont
  {Rudinger}, \citenamefont {Proctor}, \citenamefont {Langharst}, \citenamefont
  {Sarovar}, \citenamefont {Young},\ and\ \citenamefont
  {Blume-Kohout}}]{Rudinger2019}%
  \BibitemOpen
  \bibfield  {author} {\bibinfo {author} {\bibfnamefont {K.}~\bibnamefont
  {Rudinger}}, \bibinfo {author} {\bibfnamefont {T.}~\bibnamefont {Proctor}},
  \bibinfo {author} {\bibfnamefont {D.}~\bibnamefont {Langharst}}, \bibinfo
  {author} {\bibfnamefont {M.}~\bibnamefont {Sarovar}}, \bibinfo {author}
  {\bibfnamefont {K.}~\bibnamefont {Young}},\ and\ \bibinfo {author}
  {\bibfnamefont {R.}~\bibnamefont {Blume-Kohout}},\ }\bibfield  {title}
  {\bibinfo {title} {{Probing Context-Dependent Errors in Quantum
  Processors}},\ }\href@noop {} {\bibfield  {journal} {\bibinfo  {journal}
  {Physical Review X}\ }\textbf {\bibinfo {volume} {9}} (\bibinfo {year}
  {2019})}\BibitemShut {NoStop}%
\bibitem [{\citenamefont {Krinner}\ \emph
  {et~al.}(2020{\natexlab{a}})\citenamefont {Krinner}, \citenamefont {Lazar},
  \citenamefont {Remm}, \citenamefont {Andersen}, \citenamefont {Lacroix},
  \citenamefont {Norris}, \citenamefont {Hellings}, \citenamefont {Gabureac},
  \citenamefont {Eichler},\ and\ \citenamefont {Wallraff}}]{Krinner2020a}%
  \BibitemOpen
  \bibfield  {author} {\bibinfo {author} {\bibfnamefont {S.}~\bibnamefont
  {Krinner}}, \bibinfo {author} {\bibfnamefont {S.}~\bibnamefont {Lazar}},
  \bibinfo {author} {\bibfnamefont {A.}~\bibnamefont {Remm}}, \bibinfo {author}
  {\bibfnamefont {C.~K.}\ \bibnamefont {Andersen}}, \bibinfo {author}
  {\bibfnamefont {N.}~\bibnamefont {Lacroix}}, \bibinfo {author} {\bibfnamefont
  {G.~J.}\ \bibnamefont {Norris}}, \bibinfo {author} {\bibfnamefont
  {C.}~\bibnamefont {Hellings}}, \bibinfo {author} {\bibfnamefont
  {M.}~\bibnamefont {Gabureac}}, \bibinfo {author} {\bibfnamefont
  {C.}~\bibnamefont {Eichler}},\ and\ \bibinfo {author} {\bibfnamefont
  {A.}~\bibnamefont {Wallraff}},\ }\bibfield  {title} {\bibinfo {title}
  {{Benchmarking coherent errors in controlled-phase gates due to spectator
  qubits}},\ }\href@noop {} {\bibfield  {journal} {\bibinfo  {journal}
  {Physical Review Applied}\ }\textbf {\bibinfo {volume} {14}} (\bibinfo {year}
  {2020}{\natexlab{a}})}\BibitemShut {NoStop}%
\bibitem [{\citenamefont {McKay}\ \emph {et~al.}(2020)\citenamefont {McKay},
  \citenamefont {Cross}, \citenamefont {Wood},\ and\ \citenamefont
  {Gambetta}}]{McKay2020}%
  \BibitemOpen
  \bibfield  {author} {\bibinfo {author} {\bibfnamefont {D.~C.}\ \bibnamefont
  {McKay}}, \bibinfo {author} {\bibfnamefont {A.~W.}\ \bibnamefont {Cross}},
  \bibinfo {author} {\bibfnamefont {C.~J.}\ \bibnamefont {Wood}},\ and\
  \bibinfo {author} {\bibfnamefont {J.~M.}\ \bibnamefont {Gambetta}},\
  }\bibfield  {title} {\bibinfo {title} {{Correlated randomized
  benchmarking}},\ }\href@noop {} {\bibfield  {journal} {\bibinfo  {journal}
  {arXiv:2003.02354}\ } (\bibinfo {year} {2020})}\BibitemShut {NoStop}%
\bibitem [{\citenamefont {Morvan}\ \emph {et~al.}(2020)\citenamefont {Morvan},
  \citenamefont {Ramasesh}, \citenamefont {Blok}, \citenamefont {Kreikebaum},
  \citenamefont {O’Brien}, \citenamefont {Chen}, \citenamefont {Mitchell},
  \citenamefont {Naik}, \citenamefont {Santiago},\ and\ \citenamefont
  {Siddiqi}}]{Morvan2020}%
  \BibitemOpen
  \bibfield  {author} {\bibinfo {author} {\bibfnamefont {A.}~\bibnamefont
  {Morvan}}, \bibinfo {author} {\bibfnamefont {V.~V.}\ \bibnamefont
  {Ramasesh}}, \bibinfo {author} {\bibfnamefont {M.~S.}\ \bibnamefont {Blok}},
  \bibinfo {author} {\bibfnamefont {J.~M.}\ \bibnamefont {Kreikebaum}},
  \bibinfo {author} {\bibfnamefont {K.}~\bibnamefont {O’Brien}}, \bibinfo
  {author} {\bibfnamefont {L.}~\bibnamefont {Chen}}, \bibinfo {author}
  {\bibfnamefont {B.~K.}\ \bibnamefont {Mitchell}}, \bibinfo {author}
  {\bibfnamefont {R.~K.}\ \bibnamefont {Naik}}, \bibinfo {author}
  {\bibfnamefont {D.~I.}\ \bibnamefont {Santiago}},\ and\ \bibinfo {author}
  {\bibfnamefont {I.}~\bibnamefont {Siddiqi}},\ }\bibfield  {title} {\bibinfo
  {title} {{Qutrit randomized benchmarking}},\ }\href@noop {} {\bibfield
  {journal} {\bibinfo  {journal} {arXiv:2008.09134}\ } (\bibinfo {year}
  {2020})}\BibitemShut {NoStop}%
\bibitem [{\citenamefont {Ku}\ \emph {et~al.}(2020)\citenamefont {Ku},
  \citenamefont {Xu}, \citenamefont {Brink}, \citenamefont {McKay},
  \citenamefont {Hertzberg}, \citenamefont {Ansari},\ and\ \citenamefont
  {Plourde}}]{Ku2020}%
  \BibitemOpen
  \bibfield  {author} {\bibinfo {author} {\bibfnamefont {J.}~\bibnamefont
  {Ku}}, \bibinfo {author} {\bibfnamefont {X.}~\bibnamefont {Xu}}, \bibinfo
  {author} {\bibfnamefont {M.}~\bibnamefont {Brink}}, \bibinfo {author}
  {\bibfnamefont {D.~C.}\ \bibnamefont {McKay}}, \bibinfo {author}
  {\bibfnamefont {J.~B.}\ \bibnamefont {Hertzberg}}, \bibinfo {author}
  {\bibfnamefont {M.~H.}\ \bibnamefont {Ansari}},\ and\ \bibinfo {author}
  {\bibfnamefont {B.~L.}\ \bibnamefont {Plourde}},\ }\bibfield  {title}
  {\bibinfo {title} {{Suppression of Unwanted ZZ Interactions in a Hybrid
  Two-Qubit System}},\ }\href {https://doi.org/10.1103/PhysRevLett.125.200504}
  {\bibfield  {journal} {\bibinfo  {journal} {Physical review letters}\
  }\textbf {\bibinfo {volume} {125}},\ \bibinfo {pages} {200504} (\bibinfo
  {year} {2020})}\BibitemShut {NoStop}%
\bibitem [{\citenamefont {Zhao}\ \emph {et~al.}(2020)\citenamefont {Zhao},
  \citenamefont {Xu}, \citenamefont {Lan}, \citenamefont {Chu}, \citenamefont
  {Tan}, \citenamefont {Yu},\ and\ \citenamefont {Yu}}]{Zhao2020}%
  \BibitemOpen
  \bibfield  {author} {\bibinfo {author} {\bibfnamefont {P.}~\bibnamefont
  {Zhao}}, \bibinfo {author} {\bibfnamefont {P.}~\bibnamefont {Xu}}, \bibinfo
  {author} {\bibfnamefont {D.}~\bibnamefont {Lan}}, \bibinfo {author}
  {\bibfnamefont {J.}~\bibnamefont {Chu}}, \bibinfo {author} {\bibfnamefont
  {X.}~\bibnamefont {Tan}}, \bibinfo {author} {\bibfnamefont {H.}~\bibnamefont
  {Yu}},\ and\ \bibinfo {author} {\bibfnamefont {Y.}~\bibnamefont {Yu}},\
  }\bibfield  {title} {\bibinfo {title} {{High-contrast ZZ interaction using
  superconducting qubits with opposite-sign anharmonicity}},\ }\href
  {http://arxiv.org/abs/2002.07560} {\bibfield  {journal} {\bibinfo  {journal}
  {arXiv:2002.07560}\ } (\bibinfo {year} {2020})}\BibitemShut {NoStop}%
\bibitem [{\citenamefont {Foxen}\ \emph {et~al.}(2020)\citenamefont {Foxen},
  \citenamefont {Neill}, \citenamefont {Dunsworth}, \citenamefont {Roushan},
  \citenamefont {Chiaro}, \citenamefont {Megrant}, \citenamefont {Kelly},
  \citenamefont {Chen}, \citenamefont {Satzinger}, \citenamefont {Barends},
  \citenamefont {Arute}, \citenamefont {Arya}, \citenamefont {Babbush},
  \citenamefont {Bacon}, \citenamefont {Bardin}, \citenamefont {Boixo},
  \citenamefont {Buell}, \citenamefont {Burkett}, \citenamefont {Chen},
  \citenamefont {Collins}, \citenamefont {Farhi}, \citenamefont {Fowler},
  \citenamefont {Gidney}, \citenamefont {Giustina}, \citenamefont {Graff},
  \citenamefont {Harrigan}, \citenamefont {Huang}, \citenamefont {Isakov},
  \citenamefont {Jeffrey}, \citenamefont {Jiang}, \citenamefont {Kafri},
  \citenamefont {Kechedzhi}, \citenamefont {Klimov}, \citenamefont {Korotkov},
  \citenamefont {Kostritsa}, \citenamefont {Landhuis}, \citenamefont {Lucero},
  \citenamefont {Mcclean}, \citenamefont {Mcewen}, \citenamefont {Mi},
  \citenamefont {Mohseni}, \citenamefont {Mutus}, \citenamefont {Naaman},
  \citenamefont {Neeley}, \citenamefont {Niu}, \citenamefont {Petukhov},
  \citenamefont {Quintana}, \citenamefont {Rubin}, \citenamefont {Sank},
  \citenamefont {Smelyanskiy}, \citenamefont {Vainsencher}, \citenamefont
  {White}, \citenamefont {Yao}, \citenamefont {Yeh}, \citenamefont {Zalcman},
  \citenamefont {Neven},\ and\ \citenamefont {Martinis}}]{Foxen2020}%
  \BibitemOpen
  \bibfield  {author} {\bibinfo {author} {\bibfnamefont {B.}~\bibnamefont
  {Foxen}}, \bibinfo {author} {\bibfnamefont {C.}~\bibnamefont {Neill}},
  \bibinfo {author} {\bibfnamefont {A.}~\bibnamefont {Dunsworth}}, \bibinfo
  {author} {\bibfnamefont {P.}~\bibnamefont {Roushan}}, \bibinfo {author}
  {\bibfnamefont {B.}~\bibnamefont {Chiaro}}, \bibinfo {author} {\bibfnamefont
  {A.}~\bibnamefont {Megrant}}, \bibinfo {author} {\bibfnamefont
  {J.}~\bibnamefont {Kelly}}, \bibinfo {author} {\bibfnamefont
  {Z.}~\bibnamefont {Chen}}, \bibinfo {author} {\bibfnamefont {K.}~\bibnamefont
  {Satzinger}}, \bibinfo {author} {\bibfnamefont {R.}~\bibnamefont {Barends}},
  \bibinfo {author} {\bibfnamefont {F.}~\bibnamefont {Arute}}, \bibinfo
  {author} {\bibfnamefont {K.}~\bibnamefont {Arya}}, \bibinfo {author}
  {\bibfnamefont {R.}~\bibnamefont {Babbush}}, \bibinfo {author} {\bibfnamefont
  {D.}~\bibnamefont {Bacon}}, \bibinfo {author} {\bibfnamefont {J.~C.}\
  \bibnamefont {Bardin}}, \bibinfo {author} {\bibfnamefont {S.}~\bibnamefont
  {Boixo}}, \bibinfo {author} {\bibfnamefont {D.}~\bibnamefont {Buell}},
  \bibinfo {author} {\bibfnamefont {B.}~\bibnamefont {Burkett}}, \bibinfo
  {author} {\bibfnamefont {Y.}~\bibnamefont {Chen}}, \bibinfo {author}
  {\bibfnamefont {R.}~\bibnamefont {Collins}}, \bibinfo {author} {\bibfnamefont
  {E.}~\bibnamefont {Farhi}}, \bibinfo {author} {\bibfnamefont
  {A.}~\bibnamefont {Fowler}}, \bibinfo {author} {\bibfnamefont
  {C.}~\bibnamefont {Gidney}}, \bibinfo {author} {\bibfnamefont
  {M.}~\bibnamefont {Giustina}}, \bibinfo {author} {\bibfnamefont
  {R.}~\bibnamefont {Graff}}, \bibinfo {author} {\bibfnamefont
  {M.}~\bibnamefont {Harrigan}}, \bibinfo {author} {\bibfnamefont
  {T.}~\bibnamefont {Huang}}, \bibinfo {author} {\bibfnamefont {S.~V.}\
  \bibnamefont {Isakov}}, \bibinfo {author} {\bibfnamefont {E.}~\bibnamefont
  {Jeffrey}}, \bibinfo {author} {\bibfnamefont {Z.}~\bibnamefont {Jiang}},
  \bibinfo {author} {\bibfnamefont {D.}~\bibnamefont {Kafri}}, \bibinfo
  {author} {\bibfnamefont {K.}~\bibnamefont {Kechedzhi}}, \bibinfo {author}
  {\bibfnamefont {P.}~\bibnamefont {Klimov}}, \bibinfo {author} {\bibfnamefont
  {A.}~\bibnamefont {Korotkov}}, \bibinfo {author} {\bibfnamefont
  {F.}~\bibnamefont {Kostritsa}}, \bibinfo {author} {\bibfnamefont
  {D.}~\bibnamefont {Landhuis}}, \bibinfo {author} {\bibfnamefont
  {E.}~\bibnamefont {Lucero}}, \bibinfo {author} {\bibfnamefont
  {J.}~\bibnamefont {Mcclean}}, \bibinfo {author} {\bibfnamefont
  {M.}~\bibnamefont {Mcewen}}, \bibinfo {author} {\bibfnamefont
  {X.}~\bibnamefont {Mi}}, \bibinfo {author} {\bibfnamefont {M.}~\bibnamefont
  {Mohseni}}, \bibinfo {author} {\bibfnamefont {J.~Y.}\ \bibnamefont {Mutus}},
  \bibinfo {author} {\bibfnamefont {O.}~\bibnamefont {Naaman}}, \bibinfo
  {author} {\bibfnamefont {M.}~\bibnamefont {Neeley}}, \bibinfo {author}
  {\bibfnamefont {M.}~\bibnamefont {Niu}}, \bibinfo {author} {\bibfnamefont
  {A.}~\bibnamefont {Petukhov}}, \bibinfo {author} {\bibfnamefont
  {C.}~\bibnamefont {Quintana}}, \bibinfo {author} {\bibfnamefont
  {N.}~\bibnamefont {Rubin}}, \bibinfo {author} {\bibfnamefont
  {D.}~\bibnamefont {Sank}}, \bibinfo {author} {\bibfnamefont {V.}~\bibnamefont
  {Smelyanskiy}}, \bibinfo {author} {\bibfnamefont {A.}~\bibnamefont
  {Vainsencher}}, \bibinfo {author} {\bibfnamefont {T.~C.}\ \bibnamefont
  {White}}, \bibinfo {author} {\bibfnamefont {Z.}~\bibnamefont {Yao}}, \bibinfo
  {author} {\bibfnamefont {P.}~\bibnamefont {Yeh}}, \bibinfo {author}
  {\bibfnamefont {A.}~\bibnamefont {Zalcman}}, \bibinfo {author} {\bibfnamefont
  {H.}~\bibnamefont {Neven}},\ and\ \bibinfo {author} {\bibfnamefont {J.~M.}\
  \bibnamefont {Martinis}},\ }\bibfield  {title} {\bibinfo {title}
  {{Demonstrating a Continuous Set of Two-Qubit Gates for Near-Term Quantum
  Algorithms}},\ }\href {http://arxiv.org/abs/2001.08343} {\bibfield  {journal}
  {\bibinfo  {journal} {Physical Review Letters}\ }\textbf {\bibinfo {volume}
  {125}} (\bibinfo {year} {2020})}\BibitemShut {NoStop}%
\bibitem [{\citenamefont {Neg{\^{i}}rneac}\ \emph {et~al.}(2020)\citenamefont
  {Neg{\^{i}}rneac}, \citenamefont {Ali}, \citenamefont {Muthusubramanian},
  \citenamefont {Battistel}, \citenamefont {Sagastizabal}, \citenamefont
  {Moreira}, \citenamefont {Marques}, \citenamefont {Vlothuizen}, \citenamefont
  {Beekman}, \citenamefont {Haider}, \citenamefont {Bruno},\ and\ \citenamefont
  {DiCarlo}}]{Negirneac2020}%
  \BibitemOpen
  \bibfield  {author} {\bibinfo {author} {\bibfnamefont {V.}~\bibnamefont
  {Neg{\^{i}}rneac}}, \bibinfo {author} {\bibfnamefont {H.}~\bibnamefont
  {Ali}}, \bibinfo {author} {\bibfnamefont {N.}~\bibnamefont
  {Muthusubramanian}}, \bibinfo {author} {\bibfnamefont {F.}~\bibnamefont
  {Battistel}}, \bibinfo {author} {\bibfnamefont {R.}~\bibnamefont
  {Sagastizabal}}, \bibinfo {author} {\bibfnamefont {M.~S.}\ \bibnamefont
  {Moreira}}, \bibinfo {author} {\bibfnamefont {J.~F.}\ \bibnamefont
  {Marques}}, \bibinfo {author} {\bibfnamefont {W.}~\bibnamefont {Vlothuizen}},
  \bibinfo {author} {\bibfnamefont {M.}~\bibnamefont {Beekman}}, \bibinfo
  {author} {\bibfnamefont {N.}~\bibnamefont {Haider}}, \bibinfo {author}
  {\bibfnamefont {A.}~\bibnamefont {Bruno}},\ and\ \bibinfo {author}
  {\bibfnamefont {L.}~\bibnamefont {DiCarlo}},\ }\bibfield  {title} {\bibinfo
  {title} {{High-fidelity controlled-Z gate with maximal intermediate leakage
  operating at the speed limit in a superconducting quantum processor}},\
  }\href {http://arxiv.org/abs/2008.07411} {\bibfield  {journal} {\bibinfo
  {journal} {arXiv:2008.07411}\ } (\bibinfo {year} {2020})}\BibitemShut
  {NoStop}%
\bibitem [{\citenamefont {Sung}\ \emph {et~al.}(2020)\citenamefont {Sung},
  \citenamefont {Ding}, \citenamefont {Braum{\"{u}}ller}, \citenamefont
  {Veps{\"{a}}l{\"{a}}inen}, \citenamefont {Kannan}, \citenamefont
  {Kjaergaard}, \citenamefont {Greene}, \citenamefont {Samach}, \citenamefont
  {McNally}, \citenamefont {Kim}, \citenamefont {Melville}, \citenamefont
  {Niedzielski}, \citenamefont {Schwartz}, \citenamefont {Yoder}, \citenamefont
  {Orlando}, \citenamefont {Gustavsson},\ and\ \citenamefont
  {Oliver}}]{Sung2020}%
  \BibitemOpen
  \bibfield  {author} {\bibinfo {author} {\bibfnamefont {Y.}~\bibnamefont
  {Sung}}, \bibinfo {author} {\bibfnamefont {L.}~\bibnamefont {Ding}}, \bibinfo
  {author} {\bibfnamefont {J.}~\bibnamefont {Braum{\"{u}}ller}}, \bibinfo
  {author} {\bibfnamefont {A.}~\bibnamefont {Veps{\"{a}}l{\"{a}}inen}},
  \bibinfo {author} {\bibfnamefont {B.}~\bibnamefont {Kannan}}, \bibinfo
  {author} {\bibfnamefont {M.}~\bibnamefont {Kjaergaard}}, \bibinfo {author}
  {\bibfnamefont {A.}~\bibnamefont {Greene}}, \bibinfo {author} {\bibfnamefont
  {G.~O.}\ \bibnamefont {Samach}}, \bibinfo {author} {\bibfnamefont
  {C.}~\bibnamefont {McNally}}, \bibinfo {author} {\bibfnamefont
  {D.}~\bibnamefont {Kim}}, \bibinfo {author} {\bibfnamefont {A.}~\bibnamefont
  {Melville}}, \bibinfo {author} {\bibfnamefont {B.~M.}\ \bibnamefont
  {Niedzielski}}, \bibinfo {author} {\bibfnamefont {M.~E.}\ \bibnamefont
  {Schwartz}}, \bibinfo {author} {\bibfnamefont {J.~L.}\ \bibnamefont {Yoder}},
  \bibinfo {author} {\bibfnamefont {T.~P.}\ \bibnamefont {Orlando}}, \bibinfo
  {author} {\bibfnamefont {S.}~\bibnamefont {Gustavsson}},\ and\ \bibinfo
  {author} {\bibfnamefont {W.~D.}\ \bibnamefont {Oliver}},\ }\bibfield  {title}
  {\bibinfo {title} {{Realization of high-fidelity CZ and ZZ-free iSWAP gates
  with a tunable coupler}},\ }\href {http://arxiv.org/abs/2011.01261}
  {\bibfield  {journal} {\bibinfo  {journal} {arXiv:2011.01261}\ } (\bibinfo
  {year} {2020})}\BibitemShut {NoStop}%
\bibitem [{\citenamefont {Collodo}\ \emph {et~al.}(2020)\citenamefont
  {Collodo}, \citenamefont {Herrmann}, \citenamefont {Lacroix}, \citenamefont
  {Andersen}, \citenamefont {Remm}, \citenamefont {Lazar}, \citenamefont
  {Besse}, \citenamefont {Walter}, \citenamefont {Wallraff},\ and\
  \citenamefont {Eichler}}]{Collodo2020}%
  \BibitemOpen
  \bibfield  {author} {\bibinfo {author} {\bibfnamefont {M.~C.}\ \bibnamefont
  {Collodo}}, \bibinfo {author} {\bibfnamefont {J.}~\bibnamefont {Herrmann}},
  \bibinfo {author} {\bibfnamefont {N.}~\bibnamefont {Lacroix}}, \bibinfo
  {author} {\bibfnamefont {C.~K.}\ \bibnamefont {Andersen}}, \bibinfo {author}
  {\bibfnamefont {A.}~\bibnamefont {Remm}}, \bibinfo {author} {\bibfnamefont
  {S.}~\bibnamefont {Lazar}}, \bibinfo {author} {\bibfnamefont {J.~C.}\
  \bibnamefont {Besse}}, \bibinfo {author} {\bibfnamefont {T.}~\bibnamefont
  {Walter}}, \bibinfo {author} {\bibfnamefont {A.}~\bibnamefont {Wallraff}},\
  and\ \bibinfo {author} {\bibfnamefont {C.}~\bibnamefont {Eichler}},\
  }\bibfield  {title} {\bibinfo {title} {{Implementation of Conditional Phase
  Gates Based on Tunable ZZ Interactions}},\ }\href
  {http://arxiv.org/abs/2005.08863} {\bibfield  {journal} {\bibinfo  {journal}
  {Physical Review Letters}\ }\textbf {\bibinfo {volume} {125}} (\bibinfo
  {year} {2020})}\BibitemShut {NoStop}%
\bibitem [{\citenamefont {Chow}\ \emph {et~al.}(2011)\citenamefont {Chow},
  \citenamefont {C{\'{o}}rcoles}, \citenamefont {Gambetta}, \citenamefont
  {Rigetti}, \citenamefont {Johnson}, \citenamefont {Smolin}, \citenamefont
  {Rozen}, \citenamefont {Keefe}, \citenamefont {Rothwell}, \citenamefont
  {Ketchen},\ and\ \citenamefont {Steffen}}]{Chow2011}%
  \BibitemOpen
  \bibfield  {author} {\bibinfo {author} {\bibfnamefont {J.~M.}\ \bibnamefont
  {Chow}}, \bibinfo {author} {\bibfnamefont {A.~D.}\ \bibnamefont
  {C{\'{o}}rcoles}}, \bibinfo {author} {\bibfnamefont {J.~M.}\ \bibnamefont
  {Gambetta}}, \bibinfo {author} {\bibfnamefont {C.}~\bibnamefont {Rigetti}},
  \bibinfo {author} {\bibfnamefont {B.~R.}\ \bibnamefont {Johnson}}, \bibinfo
  {author} {\bibfnamefont {J.~A.}\ \bibnamefont {Smolin}}, \bibinfo {author}
  {\bibfnamefont {J.~R.}\ \bibnamefont {Rozen}}, \bibinfo {author}
  {\bibfnamefont {G.~A.}\ \bibnamefont {Keefe}}, \bibinfo {author}
  {\bibfnamefont {M.~B.}\ \bibnamefont {Rothwell}}, \bibinfo {author}
  {\bibfnamefont {M.~B.}\ \bibnamefont {Ketchen}},\ and\ \bibinfo {author}
  {\bibfnamefont {M.}~\bibnamefont {Steffen}},\ }\bibfield  {title} {\bibinfo
  {title} {{Simple all-microwave entangling gate for fixed-frequency
  superconducting qubits}},\ }\href
  {https://doi.org/10.1103/PhysRevLett.107.080502} {\bibfield  {journal}
  {\bibinfo  {journal} {Physical Review Letters}\ }\textbf {\bibinfo {volume}
  {107}},\ \bibinfo {pages} {080502} (\bibinfo {year} {2011})}\BibitemShut
  {NoStop}%
\bibitem [{\citenamefont {Sheldon}\ \emph {et~al.}(2016)\citenamefont
  {Sheldon}, \citenamefont {Magesan}, \citenamefont {Chow},\ and\ \citenamefont
  {Gambetta}}]{Sheldon2016}%
  \BibitemOpen
  \bibfield  {author} {\bibinfo {author} {\bibfnamefont {S.}~\bibnamefont
  {Sheldon}}, \bibinfo {author} {\bibfnamefont {E.}~\bibnamefont {Magesan}},
  \bibinfo {author} {\bibfnamefont {J.~M.}\ \bibnamefont {Chow}},\ and\
  \bibinfo {author} {\bibfnamefont {J.~M.}\ \bibnamefont {Gambetta}},\
  }\bibfield  {title} {\bibinfo {title} {{Procedure for systematically tuning
  up cross-talk in the cross-resonance gate}},\ }\href
  {https://doi.org/10.1103/PhysRevA.93.060302} {\bibfield  {journal} {\bibinfo
  {journal} {Physical Review A}\ }\textbf {\bibinfo {volume} {93}},\ \bibinfo
  {pages} {060302} (\bibinfo {year} {2016})}\BibitemShut {NoStop}%
\bibitem [{\citenamefont {Tripathi}\ \emph {et~al.}(2019)\citenamefont
  {Tripathi}, \citenamefont {Khezri},\ and\ \citenamefont
  {Korotkov}}]{Tripathi2019}%
  \BibitemOpen
  \bibfield  {author} {\bibinfo {author} {\bibfnamefont {V.}~\bibnamefont
  {Tripathi}}, \bibinfo {author} {\bibfnamefont {M.}~\bibnamefont {Khezri}},\
  and\ \bibinfo {author} {\bibfnamefont {A.~N.}\ \bibnamefont {Korotkov}},\
  }\bibfield  {title} {\bibinfo {title} {{Operation and intrinsic error budget
  of a two-qubit cross-resonance gate}},\ }\href
  {https://doi.org/10.1103/PhysRevA.100.012301} {\bibfield  {journal} {\bibinfo
   {journal} {Physical Review A}\ }\textbf {\bibinfo {volume} {100}},\ \bibinfo
  {pages} {012301} (\bibinfo {year} {2019})}\BibitemShut {NoStop}%
\bibitem [{\citenamefont {Jurcevic}\ \emph {et~al.}(2020)\citenamefont
  {Jurcevic}, \citenamefont {Javadi-Abhari}, \citenamefont {Bishop},
  \citenamefont {Lauer}, \citenamefont {Bogorin}, \citenamefont {Brink},
  \citenamefont {Capelluto}, \citenamefont {G{\"{u}}nl{\"{u}}k}, \citenamefont
  {Itoko}, \citenamefont {Kanazawa}, \citenamefont {Kandala}, \citenamefont
  {Keefe}, \citenamefont {Krsulich}, \citenamefont {Landers}, \citenamefont
  {Lewandowski}, \citenamefont {McClure}, \citenamefont {Nannicini},
  \citenamefont {Narasgond}, \citenamefont {Nayfeh}, \citenamefont {Pritchett},
  \citenamefont {Rothwell}, \citenamefont {Srinivasan}, \citenamefont
  {Sundaresan}, \citenamefont {Wang}, \citenamefont {Wei}, \citenamefont
  {Wood}, \citenamefont {Yau}, \citenamefont {Zhang}, \citenamefont {Dial},
  \citenamefont {Chow},\ and\ \citenamefont {Gambetta}}]{Jurcevic2020}%
  \BibitemOpen
  \bibfield  {author} {\bibinfo {author} {\bibfnamefont {P.}~\bibnamefont
  {Jurcevic}}, \bibinfo {author} {\bibfnamefont {A.}~\bibnamefont
  {Javadi-Abhari}}, \bibinfo {author} {\bibfnamefont {L.~S.}\ \bibnamefont
  {Bishop}}, \bibinfo {author} {\bibfnamefont {I.}~\bibnamefont {Lauer}},
  \bibinfo {author} {\bibfnamefont {D.~F.}\ \bibnamefont {Bogorin}}, \bibinfo
  {author} {\bibfnamefont {M.}~\bibnamefont {Brink}}, \bibinfo {author}
  {\bibfnamefont {L.}~\bibnamefont {Capelluto}}, \bibinfo {author}
  {\bibfnamefont {O.}~\bibnamefont {G{\"{u}}nl{\"{u}}k}}, \bibinfo {author}
  {\bibfnamefont {T.}~\bibnamefont {Itoko}}, \bibinfo {author} {\bibfnamefont
  {N.}~\bibnamefont {Kanazawa}}, \bibinfo {author} {\bibfnamefont
  {A.}~\bibnamefont {Kandala}}, \bibinfo {author} {\bibfnamefont {G.~A.}\
  \bibnamefont {Keefe}}, \bibinfo {author} {\bibfnamefont {K.}~\bibnamefont
  {Krsulich}}, \bibinfo {author} {\bibfnamefont {W.}~\bibnamefont {Landers}},
  \bibinfo {author} {\bibfnamefont {E.~P.}\ \bibnamefont {Lewandowski}},
  \bibinfo {author} {\bibfnamefont {D.~T.}\ \bibnamefont {McClure}}, \bibinfo
  {author} {\bibfnamefont {G.}~\bibnamefont {Nannicini}}, \bibinfo {author}
  {\bibfnamefont {A.}~\bibnamefont {Narasgond}}, \bibinfo {author}
  {\bibfnamefont {H.~M.}\ \bibnamefont {Nayfeh}}, \bibinfo {author}
  {\bibfnamefont {E.}~\bibnamefont {Pritchett}}, \bibinfo {author}
  {\bibfnamefont {M.~B.}\ \bibnamefont {Rothwell}}, \bibinfo {author}
  {\bibfnamefont {S.}~\bibnamefont {Srinivasan}}, \bibinfo {author}
  {\bibfnamefont {N.}~\bibnamefont {Sundaresan}}, \bibinfo {author}
  {\bibfnamefont {C.}~\bibnamefont {Wang}}, \bibinfo {author} {\bibfnamefont
  {K.~X.}\ \bibnamefont {Wei}}, \bibinfo {author} {\bibfnamefont {C.~J.}\
  \bibnamefont {Wood}}, \bibinfo {author} {\bibfnamefont {J.-B.}\ \bibnamefont
  {Yau}}, \bibinfo {author} {\bibfnamefont {E.~J.}\ \bibnamefont {Zhang}},
  \bibinfo {author} {\bibfnamefont {O.~E.}\ \bibnamefont {Dial}}, \bibinfo
  {author} {\bibfnamefont {J.~M.}\ \bibnamefont {Chow}},\ and\ \bibinfo
  {author} {\bibfnamefont {J.~M.}\ \bibnamefont {Gambetta}},\ }\bibfield
  {title} {\bibinfo {title} {{Demonstration of quantum volume 64 on a
  superconducting quantum computing system}},\ }\href
  {http://arxiv.org/abs/2008.08571} {\bibfield  {journal} {\bibinfo  {journal}
  {arXiv:2008.08571}\ } (\bibinfo {year} {2020})}\BibitemShut {NoStop}%
\bibitem [{\citenamefont {Kandala}\ \emph {et~al.}(2020)\citenamefont
  {Kandala}, \citenamefont {Wei}, \citenamefont {Srinivasan}, \citenamefont
  {Magesan}, \citenamefont {Carnevale}, \citenamefont {Keefe}, \citenamefont
  {Klaus}, \citenamefont {Dial},\ and\ \citenamefont {McKay}}]{Kandala2020}%
  \BibitemOpen
  \bibfield  {author} {\bibinfo {author} {\bibfnamefont {A.}~\bibnamefont
  {Kandala}}, \bibinfo {author} {\bibfnamefont {K.~X.}\ \bibnamefont {Wei}},
  \bibinfo {author} {\bibfnamefont {S.}~\bibnamefont {Srinivasan}}, \bibinfo
  {author} {\bibfnamefont {E.}~\bibnamefont {Magesan}}, \bibinfo {author}
  {\bibfnamefont {S.}~\bibnamefont {Carnevale}}, \bibinfo {author}
  {\bibfnamefont {G.~A.}\ \bibnamefont {Keefe}}, \bibinfo {author}
  {\bibfnamefont {D.}~\bibnamefont {Klaus}}, \bibinfo {author} {\bibfnamefont
  {O.}~\bibnamefont {Dial}},\ and\ \bibinfo {author} {\bibfnamefont {D.~C.}\
  \bibnamefont {McKay}},\ }\bibfield  {title} {\bibinfo {title} {{Demonstration
  of a High-Fidelity CNOT for Fixed-Frequency Transmons with Engineered ZZ
  Suppression}},\ }\href {https://arxiv.org/abs/2011.07050
  http://arxiv.org/abs/2011.07050} {\bibfield  {journal} {\bibinfo  {journal}
  {arXiv:2011.07050}\ } (\bibinfo {year} {2020})}\BibitemShut {NoStop}%
\bibitem [{\citenamefont {Chow}\ \emph {et~al.}(2013)\citenamefont {Chow},
  \citenamefont {Gambetta}, \citenamefont {Cross}, \citenamefont {Merkel},
  \citenamefont {Rigetti},\ and\ \citenamefont {Steffen}}]{Chow2013a}%
  \BibitemOpen
  \bibfield  {author} {\bibinfo {author} {\bibfnamefont {J.~M.}\ \bibnamefont
  {Chow}}, \bibinfo {author} {\bibfnamefont {J.~M.}\ \bibnamefont {Gambetta}},
  \bibinfo {author} {\bibfnamefont {A.~W.}\ \bibnamefont {Cross}}, \bibinfo
  {author} {\bibfnamefont {S.~T.}\ \bibnamefont {Merkel}}, \bibinfo {author}
  {\bibfnamefont {C.}~\bibnamefont {Rigetti}},\ and\ \bibinfo {author}
  {\bibfnamefont {M.}~\bibnamefont {Steffen}},\ }\bibfield  {title} {\bibinfo
  {title} {{Microwave-activated conditional-phase gate for superconducting
  qubits}},\ }\href {https://doi.org/10.1088/1367-2630/15/11/115012} {\bibfield
   {journal} {\bibinfo  {journal} {New Journal of Physics}\ }\textbf {\bibinfo
  {volume} {15}},\ \bibinfo {pages} {115012} (\bibinfo {year}
  {2013})}\BibitemShut {NoStop}%
\bibitem [{\citenamefont {Paik}\ \emph {et~al.}(2016)\citenamefont {Paik},
  \citenamefont {Mezzacapo}, \citenamefont {Sandberg}, \citenamefont {McClure},
  \citenamefont {Abdo}, \citenamefont {C{\'{o}}rcoles}, \citenamefont {Dial},
  \citenamefont {Bogorin}, \citenamefont {Plourde}, \citenamefont {Steffen},
  \citenamefont {Cross}, \citenamefont {Gambetta},\ and\ \citenamefont
  {Chow}}]{Paik2016}%
  \BibitemOpen
  \bibfield  {author} {\bibinfo {author} {\bibfnamefont {H.}~\bibnamefont
  {Paik}}, \bibinfo {author} {\bibfnamefont {A.}~\bibnamefont {Mezzacapo}},
  \bibinfo {author} {\bibfnamefont {M.}~\bibnamefont {Sandberg}}, \bibinfo
  {author} {\bibfnamefont {D.~T.}\ \bibnamefont {McClure}}, \bibinfo {author}
  {\bibfnamefont {B.}~\bibnamefont {Abdo}}, \bibinfo {author} {\bibfnamefont
  {A.~D.}\ \bibnamefont {C{\'{o}}rcoles}}, \bibinfo {author} {\bibfnamefont
  {O.}~\bibnamefont {Dial}}, \bibinfo {author} {\bibfnamefont {D.~F.}\
  \bibnamefont {Bogorin}}, \bibinfo {author} {\bibfnamefont {B.~L.}\
  \bibnamefont {Plourde}}, \bibinfo {author} {\bibfnamefont {M.}~\bibnamefont
  {Steffen}}, \bibinfo {author} {\bibfnamefont {A.~W.}\ \bibnamefont {Cross}},
  \bibinfo {author} {\bibfnamefont {J.~M.}\ \bibnamefont {Gambetta}},\ and\
  \bibinfo {author} {\bibfnamefont {J.~M.}\ \bibnamefont {Chow}},\ }\bibfield
  {title} {\bibinfo {title} {{Experimental Demonstration of a Resonator-Induced
  Phase Gate in a Multiqubit Circuit-QED System}},\ }\href@noop {} {\bibfield
  {journal} {\bibinfo  {journal} {Physical Review Letters}\ }\textbf {\bibinfo
  {volume} {117}} (\bibinfo {year} {2016})}\BibitemShut {NoStop}%
\bibitem [{\citenamefont {Krinner}\ \emph
  {et~al.}(2020{\natexlab{b}})\citenamefont {Krinner}, \citenamefont
  {Kurpiers}, \citenamefont {Royer}, \citenamefont {Magnard}, \citenamefont
  {Tsitsilin}, \citenamefont {Besse}, \citenamefont {Remm}, \citenamefont
  {Blais},\ and\ \citenamefont {Wallraff}}]{Krinner2020}%
  \BibitemOpen
  \bibfield  {author} {\bibinfo {author} {\bibfnamefont {S.}~\bibnamefont
  {Krinner}}, \bibinfo {author} {\bibfnamefont {P.}~\bibnamefont {Kurpiers}},
  \bibinfo {author} {\bibfnamefont {B.}~\bibnamefont {Royer}}, \bibinfo
  {author} {\bibfnamefont {P.}~\bibnamefont {Magnard}}, \bibinfo {author}
  {\bibfnamefont {I.}~\bibnamefont {Tsitsilin}}, \bibinfo {author}
  {\bibfnamefont {J.~C.}\ \bibnamefont {Besse}}, \bibinfo {author}
  {\bibfnamefont {A.}~\bibnamefont {Remm}}, \bibinfo {author} {\bibfnamefont
  {A.}~\bibnamefont {Blais}},\ and\ \bibinfo {author} {\bibfnamefont
  {A.}~\bibnamefont {Wallraff}},\ }\bibfield  {title} {\bibinfo {title}
  {{Demonstration of an All-Microwave Controlled-Phase Gate between Far-Detuned
  Qubits}},\ }\href {http://arxiv.org/abs/2006.10639} {\bibfield  {journal}
  {\bibinfo  {journal} {Physical Review Applied}\ }\textbf {\bibinfo {volume}
  {14}} (\bibinfo {year} {2020}{\natexlab{b}})}\BibitemShut {NoStop}%
\bibitem [{\citenamefont {Noguchi}\ \emph {et~al.}(2020)\citenamefont
  {Noguchi}, \citenamefont {Osada}, \citenamefont {Masuda}, \citenamefont
  {Kono}, \citenamefont {Heya}, \citenamefont {Wolski}, \citenamefont
  {Takahashi}, \citenamefont {Sugiyama}, \citenamefont {Lachance-Quirion},\
  and\ \citenamefont {Nakamura}}]{Noguchi2020}%
  \BibitemOpen
  \bibfield  {author} {\bibinfo {author} {\bibfnamefont {A.}~\bibnamefont
  {Noguchi}}, \bibinfo {author} {\bibfnamefont {A.}~\bibnamefont {Osada}},
  \bibinfo {author} {\bibfnamefont {S.}~\bibnamefont {Masuda}}, \bibinfo
  {author} {\bibfnamefont {S.}~\bibnamefont {Kono}}, \bibinfo {author}
  {\bibfnamefont {K.}~\bibnamefont {Heya}}, \bibinfo {author} {\bibfnamefont
  {S.~P.}\ \bibnamefont {Wolski}}, \bibinfo {author} {\bibfnamefont
  {H.}~\bibnamefont {Takahashi}}, \bibinfo {author} {\bibfnamefont
  {T.}~\bibnamefont {Sugiyama}}, \bibinfo {author} {\bibfnamefont
  {D.}~\bibnamefont {Lachance-Quirion}},\ and\ \bibinfo {author} {\bibfnamefont
  {Y.}~\bibnamefont {Nakamura}},\ }\bibfield  {title} {\bibinfo {title} {{Fast
  parametric two-qubit gates with suppressed residual interaction using the
  second-order nonlinearity of a cubic transmon}},\ }\href
  {http://arxiv.org/abs/2005.02630} {\bibfield  {journal} {\bibinfo  {journal}
  {Physical Review A}\ }\textbf {\bibinfo {volume} {102}} (\bibinfo {year}
  {2020})}\BibitemShut {NoStop}%
\bibitem [{\citenamefont {Manucharyan}\ \emph {et~al.}(2009)\citenamefont
  {Manucharyan}, \citenamefont {Koch}, \citenamefont {Glazman},\ and\
  \citenamefont {Devoret}}]{Manucharyan2009}%
  \BibitemOpen
  \bibfield  {author} {\bibinfo {author} {\bibfnamefont {V.~E.}\ \bibnamefont
  {Manucharyan}}, \bibinfo {author} {\bibfnamefont {J.}~\bibnamefont {Koch}},
  \bibinfo {author} {\bibfnamefont {L.~I.}\ \bibnamefont {Glazman}},\ and\
  \bibinfo {author} {\bibfnamefont {M.~H.}\ \bibnamefont {Devoret}},\
  }\bibfield  {title} {\bibinfo {title} {{Fluxonium: Single cooper-pair circuit
  free of charge offsets}},\ }\href {https://doi.org/10.1126/science.1175552}
  {\bibfield  {journal} {\bibinfo  {journal} {Science}\ }\textbf {\bibinfo
  {volume} {326}},\ \bibinfo {pages} {113} (\bibinfo {year}
  {2009})}\BibitemShut {NoStop}%
\bibitem [{\citenamefont {Manucharyan}\ \emph {et~al.}(2012)\citenamefont
  {Manucharyan}, \citenamefont {Masluk}, \citenamefont {Kamal}, \citenamefont
  {Koch}, \citenamefont {Glazman},\ and\ \citenamefont
  {Devoret}}]{Manucharyan2012}%
  \BibitemOpen
  \bibfield  {author} {\bibinfo {author} {\bibfnamefont {V.~E.}\ \bibnamefont
  {Manucharyan}}, \bibinfo {author} {\bibfnamefont {N.~A.}\ \bibnamefont
  {Masluk}}, \bibinfo {author} {\bibfnamefont {A.}~\bibnamefont {Kamal}},
  \bibinfo {author} {\bibfnamefont {J.}~\bibnamefont {Koch}}, \bibinfo {author}
  {\bibfnamefont {L.~I.}\ \bibnamefont {Glazman}},\ and\ \bibinfo {author}
  {\bibfnamefont {M.~H.}\ \bibnamefont {Devoret}},\ }\bibfield  {title}
  {\bibinfo {title} {{Evidence for coherent quantum phase slips across a
  Josephson junction array}},\ }\href
  {https://doi.org/10.1103/PhysRevB.85.024521} {\bibfield  {journal} {\bibinfo
  {journal} {Physical Review B - Condensed Matter and Materials Physics}\
  }\textbf {\bibinfo {volume} {85}},\ \bibinfo {pages} {24521} (\bibinfo {year}
  {2012})}\BibitemShut {NoStop}%
\bibitem [{\citenamefont {Ficheux}\ \emph {et~al.}(2020)\citenamefont
  {Ficheux}, \citenamefont {Nguyen}, \citenamefont {Somoroff}, \citenamefont
  {Xiong}, \citenamefont {Nesterov}, \citenamefont {Vavilov},\ and\
  \citenamefont {Manucharyan}}]{Ficheux2020}%
  \BibitemOpen
  \bibfield  {author} {\bibinfo {author} {\bibfnamefont {Q.}~\bibnamefont
  {Ficheux}}, \bibinfo {author} {\bibfnamefont {L.~B.}\ \bibnamefont {Nguyen}},
  \bibinfo {author} {\bibfnamefont {A.}~\bibnamefont {Somoroff}}, \bibinfo
  {author} {\bibfnamefont {H.}~\bibnamefont {Xiong}}, \bibinfo {author}
  {\bibfnamefont {K.~N.}\ \bibnamefont {Nesterov}}, \bibinfo {author}
  {\bibfnamefont {M.~G.}\ \bibnamefont {Vavilov}},\ and\ \bibinfo {author}
  {\bibfnamefont {V.~E.}\ \bibnamefont {Manucharyan}},\ }\bibfield  {title}
  {\bibinfo {title} {{Fast logic with slow qubits: microwave-activated
  controlled-Z gate on low-frequency fluxoniums}},\ }\href
  {http://arxiv.org/abs/2011.02634} {\bibfield  {journal} {\bibinfo  {journal}
  {arXiv:2011.02634}\ } (\bibinfo {year} {2020})}\BibitemShut {NoStop}%
\bibitem [{\citenamefont {Neill}\ \emph {et~al.}(2018)\citenamefont {Neill},
  \citenamefont {Roushan}, \citenamefont {Kechedzhi}, \citenamefont {Boixo},
  \citenamefont {Isakov}, \citenamefont {Smelyanskiy}, \citenamefont {Megrant},
  \citenamefont {Chiaro}, \citenamefont {Dunsworth}, \citenamefont {Arya},
  \citenamefont {Barends}, \citenamefont {Burkett}, \citenamefont {Chen},
  \citenamefont {Chen}, \citenamefont {Fowler}, \citenamefont {Foxen},
  \citenamefont {Giustina}, \citenamefont {Graff}, \citenamefont {Jeffrey},
  \citenamefont {Huang}, \citenamefont {Kelly}, \citenamefont {Klimov},
  \citenamefont {Lucero}, \citenamefont {Mutus}, \citenamefont {Neeley},
  \citenamefont {Quintana}, \citenamefont {Sank}, \citenamefont {Vainsencher},
  \citenamefont {Wenner}, \citenamefont {White}, \citenamefont {Neven},\ and\
  \citenamefont {Martinis}}]{Neill2018}%
  \BibitemOpen
  \bibfield  {author} {\bibinfo {author} {\bibfnamefont {C.}~\bibnamefont
  {Neill}}, \bibinfo {author} {\bibfnamefont {P.}~\bibnamefont {Roushan}},
  \bibinfo {author} {\bibfnamefont {K.}~\bibnamefont {Kechedzhi}}, \bibinfo
  {author} {\bibfnamefont {S.}~\bibnamefont {Boixo}}, \bibinfo {author}
  {\bibfnamefont {S.~V.}\ \bibnamefont {Isakov}}, \bibinfo {author}
  {\bibfnamefont {V.}~\bibnamefont {Smelyanskiy}}, \bibinfo {author}
  {\bibfnamefont {A.}~\bibnamefont {Megrant}}, \bibinfo {author} {\bibfnamefont
  {B.}~\bibnamefont {Chiaro}}, \bibinfo {author} {\bibfnamefont
  {A.}~\bibnamefont {Dunsworth}}, \bibinfo {author} {\bibfnamefont
  {K.}~\bibnamefont {Arya}}, \bibinfo {author} {\bibfnamefont {R.}~\bibnamefont
  {Barends}}, \bibinfo {author} {\bibfnamefont {B.}~\bibnamefont {Burkett}},
  \bibinfo {author} {\bibfnamefont {Y.}~\bibnamefont {Chen}}, \bibinfo {author}
  {\bibfnamefont {Z.}~\bibnamefont {Chen}}, \bibinfo {author} {\bibfnamefont
  {A.}~\bibnamefont {Fowler}}, \bibinfo {author} {\bibfnamefont
  {B.}~\bibnamefont {Foxen}}, \bibinfo {author} {\bibfnamefont
  {M.}~\bibnamefont {Giustina}}, \bibinfo {author} {\bibfnamefont
  {R.}~\bibnamefont {Graff}}, \bibinfo {author} {\bibfnamefont
  {E.}~\bibnamefont {Jeffrey}}, \bibinfo {author} {\bibfnamefont
  {T.}~\bibnamefont {Huang}}, \bibinfo {author} {\bibfnamefont
  {J.}~\bibnamefont {Kelly}}, \bibinfo {author} {\bibfnamefont
  {P.}~\bibnamefont {Klimov}}, \bibinfo {author} {\bibfnamefont
  {E.}~\bibnamefont {Lucero}}, \bibinfo {author} {\bibfnamefont
  {J.}~\bibnamefont {Mutus}}, \bibinfo {author} {\bibfnamefont
  {M.}~\bibnamefont {Neeley}}, \bibinfo {author} {\bibfnamefont
  {C.}~\bibnamefont {Quintana}}, \bibinfo {author} {\bibfnamefont
  {D.}~\bibnamefont {Sank}}, \bibinfo {author} {\bibfnamefont {A.}~\bibnamefont
  {Vainsencher}}, \bibinfo {author} {\bibfnamefont {J.}~\bibnamefont {Wenner}},
  \bibinfo {author} {\bibfnamefont {T.~C.}\ \bibnamefont {White}}, \bibinfo
  {author} {\bibfnamefont {H.}~\bibnamefont {Neven}},\ and\ \bibinfo {author}
  {\bibfnamefont {J.~M.}\ \bibnamefont {Martinis}},\ }\bibfield  {title}
  {\bibinfo {title} {{A blueprint for demonstrating quantum supremacy with
  superconducting qubits}},\ }\href {https://doi.org/10.1126/science.aao4309}
  {\bibfield  {journal} {\bibinfo  {journal} {Science}\ }\textbf {\bibinfo
  {volume} {360}},\ \bibinfo {pages} {195} (\bibinfo {year}
  {2018})}\BibitemShut {NoStop}%
\bibitem [{\citenamefont {Barends}\ \emph {et~al.}(2019)\citenamefont
  {Barends}, \citenamefont {Quintana}, \citenamefont {Petukhov}, \citenamefont
  {Chen}, \citenamefont {Kafri}, \citenamefont {Kechedzhi}, \citenamefont
  {Collins}, \citenamefont {Naaman}, \citenamefont {Boixo}, \citenamefont
  {Arute}, \citenamefont {Arya}, \citenamefont {Buell}, \citenamefont
  {Burkett}, \citenamefont {Chen}, \citenamefont {Chiaro}, \citenamefont
  {Dunsworth}, \citenamefont {Foxen}, \citenamefont {Fowler}, \citenamefont
  {Gidney}, \citenamefont {Giustina}, \citenamefont {Graff}, \citenamefont
  {Huang}, \citenamefont {Jeffrey}, \citenamefont {Kelly}, \citenamefont
  {Klimov}, \citenamefont {Kostritsa}, \citenamefont {Landhuis}, \citenamefont
  {Lucero}, \citenamefont {McEwen}, \citenamefont {Megrant}, \citenamefont
  {Mi}, \citenamefont {Mutus}, \citenamefont {Neeley}, \citenamefont {Neill},
  \citenamefont {Ostby}, \citenamefont {Roushan}, \citenamefont {Sank},
  \citenamefont {Satzinger}, \citenamefont {Vainsencher}, \citenamefont
  {White}, \citenamefont {Yao}, \citenamefont {Yeh}, \citenamefont {Zalcman},
  \citenamefont {Neven}, \citenamefont {Smelyanskiy},\ and\ \citenamefont
  {Martinis}}]{Barends2019}%
  \BibitemOpen
  \bibfield  {author} {\bibinfo {author} {\bibfnamefont {R.}~\bibnamefont
  {Barends}}, \bibinfo {author} {\bibfnamefont {C.~M.}\ \bibnamefont
  {Quintana}}, \bibinfo {author} {\bibfnamefont {A.~G.}\ \bibnamefont
  {Petukhov}}, \bibinfo {author} {\bibfnamefont {Y.}~\bibnamefont {Chen}},
  \bibinfo {author} {\bibfnamefont {D.}~\bibnamefont {Kafri}}, \bibinfo
  {author} {\bibfnamefont {K.}~\bibnamefont {Kechedzhi}}, \bibinfo {author}
  {\bibfnamefont {R.}~\bibnamefont {Collins}}, \bibinfo {author} {\bibfnamefont
  {O.}~\bibnamefont {Naaman}}, \bibinfo {author} {\bibfnamefont
  {S.}~\bibnamefont {Boixo}}, \bibinfo {author} {\bibfnamefont
  {F.}~\bibnamefont {Arute}}, \bibinfo {author} {\bibfnamefont
  {K.}~\bibnamefont {Arya}}, \bibinfo {author} {\bibfnamefont {D.}~\bibnamefont
  {Buell}}, \bibinfo {author} {\bibfnamefont {B.}~\bibnamefont {Burkett}},
  \bibinfo {author} {\bibfnamefont {Z.}~\bibnamefont {Chen}}, \bibinfo {author}
  {\bibfnamefont {B.}~\bibnamefont {Chiaro}}, \bibinfo {author} {\bibfnamefont
  {A.}~\bibnamefont {Dunsworth}}, \bibinfo {author} {\bibfnamefont
  {B.}~\bibnamefont {Foxen}}, \bibinfo {author} {\bibfnamefont
  {A.}~\bibnamefont {Fowler}}, \bibinfo {author} {\bibfnamefont
  {C.}~\bibnamefont {Gidney}}, \bibinfo {author} {\bibfnamefont
  {M.}~\bibnamefont {Giustina}}, \bibinfo {author} {\bibfnamefont
  {R.}~\bibnamefont {Graff}}, \bibinfo {author} {\bibfnamefont
  {T.}~\bibnamefont {Huang}}, \bibinfo {author} {\bibfnamefont
  {E.}~\bibnamefont {Jeffrey}}, \bibinfo {author} {\bibfnamefont
  {J.}~\bibnamefont {Kelly}}, \bibinfo {author} {\bibfnamefont {P.~V.}\
  \bibnamefont {Klimov}}, \bibinfo {author} {\bibfnamefont {F.}~\bibnamefont
  {Kostritsa}}, \bibinfo {author} {\bibfnamefont {D.}~\bibnamefont {Landhuis}},
  \bibinfo {author} {\bibfnamefont {E.}~\bibnamefont {Lucero}}, \bibinfo
  {author} {\bibfnamefont {M.}~\bibnamefont {McEwen}}, \bibinfo {author}
  {\bibfnamefont {A.}~\bibnamefont {Megrant}}, \bibinfo {author} {\bibfnamefont
  {X.}~\bibnamefont {Mi}}, \bibinfo {author} {\bibfnamefont {J.}~\bibnamefont
  {Mutus}}, \bibinfo {author} {\bibfnamefont {M.}~\bibnamefont {Neeley}},
  \bibinfo {author} {\bibfnamefont {C.}~\bibnamefont {Neill}}, \bibinfo
  {author} {\bibfnamefont {E.}~\bibnamefont {Ostby}}, \bibinfo {author}
  {\bibfnamefont {P.}~\bibnamefont {Roushan}}, \bibinfo {author} {\bibfnamefont
  {D.}~\bibnamefont {Sank}}, \bibinfo {author} {\bibfnamefont {K.~J.}\
  \bibnamefont {Satzinger}}, \bibinfo {author} {\bibfnamefont {A.}~\bibnamefont
  {Vainsencher}}, \bibinfo {author} {\bibfnamefont {T.}~\bibnamefont {White}},
  \bibinfo {author} {\bibfnamefont {J.}~\bibnamefont {Yao}}, \bibinfo {author}
  {\bibfnamefont {P.}~\bibnamefont {Yeh}}, \bibinfo {author} {\bibfnamefont
  {A.}~\bibnamefont {Zalcman}}, \bibinfo {author} {\bibfnamefont
  {H.}~\bibnamefont {Neven}}, \bibinfo {author} {\bibfnamefont {V.~N.}\
  \bibnamefont {Smelyanskiy}},\ and\ \bibinfo {author} {\bibfnamefont {J.~M.}\
  \bibnamefont {Martinis}},\ }\bibfield  {title} {\bibinfo {title} {{Diabatic
  Gates for Frequency-Tunable Superconducting Qubits}},\ }\href
  {https://doi.org/10.1103/PhysRevLett.123.210501} {\bibfield  {journal}
  {\bibinfo  {journal} {Physical Review Letters}\ }\textbf {\bibinfo {volume}
  {123}},\ \bibinfo {pages} {210501} (\bibinfo {year} {2019})}\BibitemShut
  {NoStop}%
\bibitem [{\citenamefont {Barends}\ \emph {et~al.}(2016)\citenamefont
  {Barends}, \citenamefont {Shabani}, \citenamefont {Lamata}, \citenamefont
  {Kelly}, \citenamefont {Mezzacapo}, \citenamefont {Heras}, \citenamefont
  {Babbush}, \citenamefont {Fowler}, \citenamefont {Campbell}, \citenamefont
  {Chen}, \citenamefont {Chen}, \citenamefont {Chiaro}, \citenamefont
  {Dunsworth}, \citenamefont {Jeffrey}, \citenamefont {Lucero}, \citenamefont
  {Megrant}, \citenamefont {Mutus}, \citenamefont {Neeley}, \citenamefont
  {Neill}, \citenamefont {O'Malley}, \citenamefont {Quintana}, \citenamefont
  {Roushan}, \citenamefont {Sank}, \citenamefont {Vainsencher}, \citenamefont
  {Wenner}, \citenamefont {White}, \citenamefont {Solano}, \citenamefont
  {Neven},\ and\ \citenamefont {Martinis}}]{Barends2016}%
  \BibitemOpen
  \bibfield  {author} {\bibinfo {author} {\bibfnamefont {R.}~\bibnamefont
  {Barends}}, \bibinfo {author} {\bibfnamefont {A.}~\bibnamefont {Shabani}},
  \bibinfo {author} {\bibfnamefont {L.}~\bibnamefont {Lamata}}, \bibinfo
  {author} {\bibfnamefont {J.}~\bibnamefont {Kelly}}, \bibinfo {author}
  {\bibfnamefont {A.}~\bibnamefont {Mezzacapo}}, \bibinfo {author}
  {\bibfnamefont {U.~L.}\ \bibnamefont {Heras}}, \bibinfo {author}
  {\bibfnamefont {R.}~\bibnamefont {Babbush}}, \bibinfo {author} {\bibfnamefont
  {A.~G.}\ \bibnamefont {Fowler}}, \bibinfo {author} {\bibfnamefont
  {B.}~\bibnamefont {Campbell}}, \bibinfo {author} {\bibfnamefont
  {Y.}~\bibnamefont {Chen}}, \bibinfo {author} {\bibfnamefont {Z.}~\bibnamefont
  {Chen}}, \bibinfo {author} {\bibfnamefont {B.}~\bibnamefont {Chiaro}},
  \bibinfo {author} {\bibfnamefont {A.}~\bibnamefont {Dunsworth}}, \bibinfo
  {author} {\bibfnamefont {E.}~\bibnamefont {Jeffrey}}, \bibinfo {author}
  {\bibfnamefont {E.}~\bibnamefont {Lucero}}, \bibinfo {author} {\bibfnamefont
  {A.}~\bibnamefont {Megrant}}, \bibinfo {author} {\bibfnamefont {J.~Y.}\
  \bibnamefont {Mutus}}, \bibinfo {author} {\bibfnamefont {M.}~\bibnamefont
  {Neeley}}, \bibinfo {author} {\bibfnamefont {C.}~\bibnamefont {Neill}},
  \bibinfo {author} {\bibfnamefont {P.~J.}\ \bibnamefont {O'Malley}}, \bibinfo
  {author} {\bibfnamefont {C.}~\bibnamefont {Quintana}}, \bibinfo {author}
  {\bibfnamefont {P.}~\bibnamefont {Roushan}}, \bibinfo {author} {\bibfnamefont
  {D.}~\bibnamefont {Sank}}, \bibinfo {author} {\bibfnamefont {A.}~\bibnamefont
  {Vainsencher}}, \bibinfo {author} {\bibfnamefont {J.}~\bibnamefont {Wenner}},
  \bibinfo {author} {\bibfnamefont {T.~C.}\ \bibnamefont {White}}, \bibinfo
  {author} {\bibfnamefont {E.}~\bibnamefont {Solano}}, \bibinfo {author}
  {\bibfnamefont {H.}~\bibnamefont {Neven}},\ and\ \bibinfo {author}
  {\bibfnamefont {J.~M.}\ \bibnamefont {Martinis}},\ }\bibfield  {title}
  {\bibinfo {title} {{Digitized adiabatic quantum computing with a
  superconducting circuit}},\ }\href {https://doi.org/10.1038/nature17658}
  {\bibfield  {journal} {\bibinfo  {journal} {Nature}\ }\textbf {\bibinfo
  {volume} {534}},\ \bibinfo {pages} {222} (\bibinfo {year}
  {2016})}\BibitemShut {NoStop}%
\bibitem [{\citenamefont {O'Malley}\ \emph {et~al.}(2016)\citenamefont
  {O'Malley}, \citenamefont {Babbush}, \citenamefont {Kivlichan}, \citenamefont
  {Romero}, \citenamefont {McClean}, \citenamefont {Barends}, \citenamefont
  {Kelly}, \citenamefont {Roushan}, \citenamefont {Tranter}, \citenamefont
  {Ding}, \citenamefont {Campbell}, \citenamefont {Chen}, \citenamefont {Chen},
  \citenamefont {Chiaro}, \citenamefont {Dunsworth}, \citenamefont {Fowler},
  \citenamefont {Jeffrey}, \citenamefont {Lucero}, \citenamefont {Megrant},
  \citenamefont {Mutus}, \citenamefont {Neeley}, \citenamefont {Neill},
  \citenamefont {Quintana}, \citenamefont {Sank}, \citenamefont {Vainsencher},
  \citenamefont {Wenner}, \citenamefont {White}, \citenamefont {Coveney},
  \citenamefont {Love}, \citenamefont {Neven}, \citenamefont {Aspuru-Guzik},\
  and\ \citenamefont {Martinis}}]{OMalley2016}%
  \BibitemOpen
  \bibfield  {author} {\bibinfo {author} {\bibfnamefont {P.~J.}\ \bibnamefont
  {O'Malley}}, \bibinfo {author} {\bibfnamefont {R.}~\bibnamefont {Babbush}},
  \bibinfo {author} {\bibfnamefont {I.~D.}\ \bibnamefont {Kivlichan}}, \bibinfo
  {author} {\bibfnamefont {J.}~\bibnamefont {Romero}}, \bibinfo {author}
  {\bibfnamefont {J.~R.}\ \bibnamefont {McClean}}, \bibinfo {author}
  {\bibfnamefont {R.}~\bibnamefont {Barends}}, \bibinfo {author} {\bibfnamefont
  {J.}~\bibnamefont {Kelly}}, \bibinfo {author} {\bibfnamefont
  {P.}~\bibnamefont {Roushan}}, \bibinfo {author} {\bibfnamefont
  {A.}~\bibnamefont {Tranter}}, \bibinfo {author} {\bibfnamefont
  {N.}~\bibnamefont {Ding}}, \bibinfo {author} {\bibfnamefont {B.}~\bibnamefont
  {Campbell}}, \bibinfo {author} {\bibfnamefont {Y.}~\bibnamefont {Chen}},
  \bibinfo {author} {\bibfnamefont {Z.}~\bibnamefont {Chen}}, \bibinfo {author}
  {\bibfnamefont {B.}~\bibnamefont {Chiaro}}, \bibinfo {author} {\bibfnamefont
  {A.}~\bibnamefont {Dunsworth}}, \bibinfo {author} {\bibfnamefont {A.~G.}\
  \bibnamefont {Fowler}}, \bibinfo {author} {\bibfnamefont {E.}~\bibnamefont
  {Jeffrey}}, \bibinfo {author} {\bibfnamefont {E.}~\bibnamefont {Lucero}},
  \bibinfo {author} {\bibfnamefont {A.}~\bibnamefont {Megrant}}, \bibinfo
  {author} {\bibfnamefont {J.~Y.}\ \bibnamefont {Mutus}}, \bibinfo {author}
  {\bibfnamefont {M.}~\bibnamefont {Neeley}}, \bibinfo {author} {\bibfnamefont
  {C.}~\bibnamefont {Neill}}, \bibinfo {author} {\bibfnamefont
  {C.}~\bibnamefont {Quintana}}, \bibinfo {author} {\bibfnamefont
  {D.}~\bibnamefont {Sank}}, \bibinfo {author} {\bibfnamefont {A.}~\bibnamefont
  {Vainsencher}}, \bibinfo {author} {\bibfnamefont {J.}~\bibnamefont {Wenner}},
  \bibinfo {author} {\bibfnamefont {T.~C.}\ \bibnamefont {White}}, \bibinfo
  {author} {\bibfnamefont {P.~V.}\ \bibnamefont {Coveney}}, \bibinfo {author}
  {\bibfnamefont {P.~J.}\ \bibnamefont {Love}}, \bibinfo {author}
  {\bibfnamefont {H.}~\bibnamefont {Neven}}, \bibinfo {author} {\bibfnamefont
  {A.}~\bibnamefont {Aspuru-Guzik}},\ and\ \bibinfo {author} {\bibfnamefont
  {J.~M.}\ \bibnamefont {Martinis}},\ }\bibfield  {title} {\bibinfo {title}
  {{Scalable quantum simulation of molecular energies}},\ }\href@noop {}
  {\bibfield  {journal} {\bibinfo  {journal} {Physical Review X}\ }\textbf
  {\bibinfo {volume} {6}} (\bibinfo {year} {2016})}\BibitemShut {NoStop}%
\bibitem [{\citenamefont {Kandala}\ \emph {et~al.}(2017)\citenamefont
  {Kandala}, \citenamefont {Mezzacapo}, \citenamefont {Temme}, \citenamefont
  {Takita}, \citenamefont {Brink}, \citenamefont {Chow},\ and\ \citenamefont
  {Gambetta}}]{Kandala2017}%
  \BibitemOpen
  \bibfield  {author} {\bibinfo {author} {\bibfnamefont {A.}~\bibnamefont
  {Kandala}}, \bibinfo {author} {\bibfnamefont {A.}~\bibnamefont {Mezzacapo}},
  \bibinfo {author} {\bibfnamefont {K.}~\bibnamefont {Temme}}, \bibinfo
  {author} {\bibfnamefont {M.}~\bibnamefont {Takita}}, \bibinfo {author}
  {\bibfnamefont {M.}~\bibnamefont {Brink}}, \bibinfo {author} {\bibfnamefont
  {J.~M.}\ \bibnamefont {Chow}},\ and\ \bibinfo {author} {\bibfnamefont
  {J.~M.}\ \bibnamefont {Gambetta}},\ }\bibfield  {title} {\bibinfo {title}
  {{Hardware-efficient variational quantum eigensolver for small molecules and
  quantum magnets}},\ }\href {https://doi.org/10.1038/nature23879} {\bibfield
  {journal} {\bibinfo  {journal} {Nature}\ }\textbf {\bibinfo {volume} {549}},\
  \bibinfo {pages} {242} (\bibinfo {year} {2017})}\BibitemShut {NoStop}%
\bibitem [{\citenamefont {Ganzhorn}\ \emph {et~al.}(2019)\citenamefont
  {Ganzhorn}, \citenamefont {Egger}, \citenamefont {Barkoutsos}, \citenamefont
  {Ollitrault}, \citenamefont {Salis}, \citenamefont {Moll}, \citenamefont
  {Roth}, \citenamefont {Fuhrer}, \citenamefont {Mueller}, \citenamefont
  {Woerner}, \citenamefont {Tavernelli},\ and\ \citenamefont
  {Filipp}}]{Ganzhorn2019}%
  \BibitemOpen
  \bibfield  {author} {\bibinfo {author} {\bibfnamefont {M.}~\bibnamefont
  {Ganzhorn}}, \bibinfo {author} {\bibfnamefont {D.~J.}\ \bibnamefont {Egger}},
  \bibinfo {author} {\bibfnamefont {P.}~\bibnamefont {Barkoutsos}}, \bibinfo
  {author} {\bibfnamefont {P.}~\bibnamefont {Ollitrault}}, \bibinfo {author}
  {\bibfnamefont {G.}~\bibnamefont {Salis}}, \bibinfo {author} {\bibfnamefont
  {N.}~\bibnamefont {Moll}}, \bibinfo {author} {\bibfnamefont {M.}~\bibnamefont
  {Roth}}, \bibinfo {author} {\bibfnamefont {A.}~\bibnamefont {Fuhrer}},
  \bibinfo {author} {\bibfnamefont {P.}~\bibnamefont {Mueller}}, \bibinfo
  {author} {\bibfnamefont {S.}~\bibnamefont {Woerner}}, \bibinfo {author}
  {\bibfnamefont {I.}~\bibnamefont {Tavernelli}},\ and\ \bibinfo {author}
  {\bibfnamefont {S.}~\bibnamefont {Filipp}},\ }\bibfield  {title} {\bibinfo
  {title} {{Gate-Efficient Simulation of Molecular Eigenstates on a Quantum
  Computer}},\ }\href@noop {} {\bibfield  {journal} {\bibinfo  {journal}
  {Physical Review Applied}\ }\textbf {\bibinfo {volume} {11}} (\bibinfo {year}
  {2019})}\BibitemShut {NoStop}%
\bibitem [{\citenamefont {Aleiner}\ \emph {et~al.}(2020)\citenamefont
  {Aleiner}, \citenamefont {Arute}, \citenamefont {Arya}, \citenamefont
  {Atalaya}, \citenamefont {Babbush}, \citenamefont {Bardin}, \citenamefont
  {Barends}, \citenamefont {Bengtsson}, \citenamefont {Boixo}, \citenamefont
  {Bourassa}, \citenamefont {Broughton}, \citenamefont {Buckley}, \citenamefont
  {Buell}, \citenamefont {Burkett}, \citenamefont {Bushnell}, \citenamefont
  {Chen}, \citenamefont {Chen}, \citenamefont {Chiaro}, \citenamefont
  {Collins}, \citenamefont {Courtney}, \citenamefont {Demura}, \citenamefont
  {Derk}, \citenamefont {Dunsworth}, \citenamefont {Eppens}, \citenamefont
  {Erickson}, \citenamefont {Farhi}, \citenamefont {Fowler}, \citenamefont
  {Foxen}, \citenamefont {Gidney}, \citenamefont {Giustina}, \citenamefont
  {Gross}, \citenamefont {Harrigan}, \citenamefont {Harrington}, \citenamefont
  {Hilton}, \citenamefont {Ho}, \citenamefont {Hong}, \citenamefont {Huang},
  \citenamefont {Huggins}, \citenamefont {Ioffe}, \citenamefont {Isakov},
  \citenamefont {Jeffrey}, \citenamefont {Jiang}, \citenamefont {Jones},
  \citenamefont {Kafri}, \citenamefont {Kechedzhi}, \citenamefont {Kelly},
  \citenamefont {Kim}, \citenamefont {Klimov}, \citenamefont {Korotkov},
  \citenamefont {Kostritsa}, \citenamefont {Landhuis}, \citenamefont {Laptev},
  \citenamefont {Lucero}, \citenamefont {Martin}, \citenamefont {McClean},
  \citenamefont {McCourt}, \citenamefont {McEwen}, \citenamefont {Megrant},
  \citenamefont {Mi}, \citenamefont {Miao}, \citenamefont {Mohseni},
  \citenamefont {Mruczkiewicz}, \citenamefont {Mutus}, \citenamefont {Naaman},
  \citenamefont {Neeley}, \citenamefont {Neill}, \citenamefont {Neven},
  \citenamefont {Newman}, \citenamefont {Niu}, \citenamefont {O'Brien},
  \citenamefont {Opremcak}, \citenamefont {Ostby}, \citenamefont {Pat{\'{o}}},
  \citenamefont {Petukhov}, \citenamefont {Quintana}, \citenamefont {Redd},
  \citenamefont {Roushan}, \citenamefont {Rubin}, \citenamefont {Sank},
  \citenamefont {Satzinger}, \citenamefont {Shvarts}, \citenamefont
  {Smelyanskiy}, \citenamefont {Strain}, \citenamefont {Szalay}, \citenamefont
  {Trevithick}, \citenamefont {Villalonga}, \citenamefont {White},
  \citenamefont {Yao}, \citenamefont {Yeh},\ and\ \citenamefont
  {Zalcman}}]{Aleiner2020}%
  \BibitemOpen
  \bibfield  {author} {\bibinfo {author} {\bibfnamefont {I.}~\bibnamefont
  {Aleiner}}, \bibinfo {author} {\bibfnamefont {F.}~\bibnamefont {Arute}},
  \bibinfo {author} {\bibfnamefont {K.}~\bibnamefont {Arya}}, \bibinfo {author}
  {\bibfnamefont {J.}~\bibnamefont {Atalaya}}, \bibinfo {author} {\bibfnamefont
  {R.}~\bibnamefont {Babbush}}, \bibinfo {author} {\bibfnamefont {J.~C.}\
  \bibnamefont {Bardin}}, \bibinfo {author} {\bibfnamefont {R.}~\bibnamefont
  {Barends}}, \bibinfo {author} {\bibfnamefont {A.}~\bibnamefont {Bengtsson}},
  \bibinfo {author} {\bibfnamefont {S.}~\bibnamefont {Boixo}}, \bibinfo
  {author} {\bibfnamefont {A.}~\bibnamefont {Bourassa}}, \bibinfo {author}
  {\bibfnamefont {M.}~\bibnamefont {Broughton}}, \bibinfo {author}
  {\bibfnamefont {B.~B.}\ \bibnamefont {Buckley}}, \bibinfo {author}
  {\bibfnamefont {D.~A.}\ \bibnamefont {Buell}}, \bibinfo {author}
  {\bibfnamefont {B.}~\bibnamefont {Burkett}}, \bibinfo {author} {\bibfnamefont
  {N.}~\bibnamefont {Bushnell}}, \bibinfo {author} {\bibfnamefont
  {Y.}~\bibnamefont {Chen}}, \bibinfo {author} {\bibfnamefont {Z.}~\bibnamefont
  {Chen}}, \bibinfo {author} {\bibfnamefont {B.}~\bibnamefont {Chiaro}},
  \bibinfo {author} {\bibfnamefont {R.}~\bibnamefont {Collins}}, \bibinfo
  {author} {\bibfnamefont {W.}~\bibnamefont {Courtney}}, \bibinfo {author}
  {\bibfnamefont {S.}~\bibnamefont {Demura}}, \bibinfo {author} {\bibfnamefont
  {A.~R.}\ \bibnamefont {Derk}}, \bibinfo {author} {\bibfnamefont
  {A.}~\bibnamefont {Dunsworth}}, \bibinfo {author} {\bibfnamefont
  {D.}~\bibnamefont {Eppens}}, \bibinfo {author} {\bibfnamefont
  {C.}~\bibnamefont {Erickson}}, \bibinfo {author} {\bibfnamefont
  {E.}~\bibnamefont {Farhi}}, \bibinfo {author} {\bibfnamefont {A.~G.}\
  \bibnamefont {Fowler}}, \bibinfo {author} {\bibfnamefont {B.}~\bibnamefont
  {Foxen}}, \bibinfo {author} {\bibfnamefont {C.}~\bibnamefont {Gidney}},
  \bibinfo {author} {\bibfnamefont {M.}~\bibnamefont {Giustina}}, \bibinfo
  {author} {\bibfnamefont {J.~A.}\ \bibnamefont {Gross}}, \bibinfo {author}
  {\bibfnamefont {M.~P.}\ \bibnamefont {Harrigan}}, \bibinfo {author}
  {\bibfnamefont {S.~D.}\ \bibnamefont {Harrington}}, \bibinfo {author}
  {\bibfnamefont {J.}~\bibnamefont {Hilton}}, \bibinfo {author} {\bibfnamefont
  {A.}~\bibnamefont {Ho}}, \bibinfo {author} {\bibfnamefont {S.}~\bibnamefont
  {Hong}}, \bibinfo {author} {\bibfnamefont {T.}~\bibnamefont {Huang}},
  \bibinfo {author} {\bibfnamefont {W.~J.}\ \bibnamefont {Huggins}}, \bibinfo
  {author} {\bibfnamefont {L.~B.}\ \bibnamefont {Ioffe}}, \bibinfo {author}
  {\bibfnamefont {S.~V.}\ \bibnamefont {Isakov}}, \bibinfo {author}
  {\bibfnamefont {E.}~\bibnamefont {Jeffrey}}, \bibinfo {author} {\bibfnamefont
  {Z.}~\bibnamefont {Jiang}}, \bibinfo {author} {\bibfnamefont
  {C.}~\bibnamefont {Jones}}, \bibinfo {author} {\bibfnamefont
  {D.}~\bibnamefont {Kafri}}, \bibinfo {author} {\bibfnamefont
  {K.}~\bibnamefont {Kechedzhi}}, \bibinfo {author} {\bibfnamefont
  {J.}~\bibnamefont {Kelly}}, \bibinfo {author} {\bibfnamefont
  {S.}~\bibnamefont {Kim}}, \bibinfo {author} {\bibfnamefont {P.~V.}\
  \bibnamefont {Klimov}}, \bibinfo {author} {\bibfnamefont {A.~N.}\
  \bibnamefont {Korotkov}}, \bibinfo {author} {\bibfnamefont {F.}~\bibnamefont
  {Kostritsa}}, \bibinfo {author} {\bibfnamefont {D.}~\bibnamefont {Landhuis}},
  \bibinfo {author} {\bibfnamefont {P.}~\bibnamefont {Laptev}}, \bibinfo
  {author} {\bibfnamefont {E.}~\bibnamefont {Lucero}}, \bibinfo {author}
  {\bibfnamefont {O.}~\bibnamefont {Martin}}, \bibinfo {author} {\bibfnamefont
  {J.~R.}\ \bibnamefont {McClean}}, \bibinfo {author} {\bibfnamefont
  {T.}~\bibnamefont {McCourt}}, \bibinfo {author} {\bibfnamefont
  {M.}~\bibnamefont {McEwen}}, \bibinfo {author} {\bibfnamefont
  {A.}~\bibnamefont {Megrant}}, \bibinfo {author} {\bibfnamefont
  {X.}~\bibnamefont {Mi}}, \bibinfo {author} {\bibfnamefont {K.~C.}\
  \bibnamefont {Miao}}, \bibinfo {author} {\bibfnamefont {M.}~\bibnamefont
  {Mohseni}}, \bibinfo {author} {\bibfnamefont {W.}~\bibnamefont
  {Mruczkiewicz}}, \bibinfo {author} {\bibfnamefont {J.}~\bibnamefont {Mutus}},
  \bibinfo {author} {\bibfnamefont {O.}~\bibnamefont {Naaman}}, \bibinfo
  {author} {\bibfnamefont {M.}~\bibnamefont {Neeley}}, \bibinfo {author}
  {\bibfnamefont {C.}~\bibnamefont {Neill}}, \bibinfo {author} {\bibfnamefont
  {H.}~\bibnamefont {Neven}}, \bibinfo {author} {\bibfnamefont
  {M.}~\bibnamefont {Newman}}, \bibinfo {author} {\bibfnamefont {M.~Y.}\
  \bibnamefont {Niu}}, \bibinfo {author} {\bibfnamefont {T.~E.}\ \bibnamefont
  {O'Brien}}, \bibinfo {author} {\bibfnamefont {A.}~\bibnamefont {Opremcak}},
  \bibinfo {author} {\bibfnamefont {E.}~\bibnamefont {Ostby}}, \bibinfo
  {author} {\bibfnamefont {B.}~\bibnamefont {Pat{\'{o}}}}, \bibinfo {author}
  {\bibfnamefont {A.}~\bibnamefont {Petukhov}}, \bibinfo {author}
  {\bibfnamefont {C.}~\bibnamefont {Quintana}}, \bibinfo {author}
  {\bibfnamefont {N.}~\bibnamefont {Redd}}, \bibinfo {author} {\bibfnamefont
  {P.}~\bibnamefont {Roushan}}, \bibinfo {author} {\bibfnamefont {N.~C.}\
  \bibnamefont {Rubin}}, \bibinfo {author} {\bibfnamefont {D.}~\bibnamefont
  {Sank}}, \bibinfo {author} {\bibfnamefont {K.~J.}\ \bibnamefont {Satzinger}},
  \bibinfo {author} {\bibfnamefont {V.}~\bibnamefont {Shvarts}}, \bibinfo
  {author} {\bibfnamefont {V.}~\bibnamefont {Smelyanskiy}}, \bibinfo {author}
  {\bibfnamefont {D.}~\bibnamefont {Strain}}, \bibinfo {author} {\bibfnamefont
  {M.}~\bibnamefont {Szalay}}, \bibinfo {author} {\bibfnamefont {M.~D.}\
  \bibnamefont {Trevithick}}, \bibinfo {author} {\bibfnamefont
  {B.}~\bibnamefont {Villalonga}}, \bibinfo {author} {\bibfnamefont
  {T.}~\bibnamefont {White}}, \bibinfo {author} {\bibfnamefont {Z.~J.}\
  \bibnamefont {Yao}}, \bibinfo {author} {\bibfnamefont {P.}~\bibnamefont
  {Yeh}},\ and\ \bibinfo {author} {\bibfnamefont {A.}~\bibnamefont {Zalcman}},\
  }\bibfield  {title} {\bibinfo {title} {{Accurately computing electronic
  properties of materials using eigenenergies}},\ }\href
  {http://arxiv.org/abs/2012.00921} {\bibfield  {journal} {\bibinfo  {journal}
  {arXiv:2012.00921}\ } (\bibinfo {year} {2020})}\BibitemShut {NoStop}%
\bibitem [{\citenamefont {Kitaev}(1995)}]{Kitaev1995}%
  \BibitemOpen
  \bibfield  {author} {\bibinfo {author} {\bibfnamefont {A.~Y.}\ \bibnamefont
  {Kitaev}},\ }\bibfield  {title} {\bibinfo {title} {{Quantum measurements and
  the Abelian Stabilizer Problem}},\ }\href
  {http://arxiv.org/abs/quant-ph/9511026} {\bibfield  {journal} {\bibinfo
  {journal} {arXiv:quant-ph/9511026}\ } (\bibinfo {year} {1995})}\BibitemShut
  {NoStop}%
\bibitem [{\citenamefont {Whitfield}\ \emph {et~al.}(2011)\citenamefont
  {Whitfield}, \citenamefont {Biamonte},\ and\ \citenamefont
  {Aspuru-Guzik}}]{Whitfield2011}%
  \BibitemOpen
  \bibfield  {author} {\bibinfo {author} {\bibfnamefont {J.~D.}\ \bibnamefont
  {Whitfield}}, \bibinfo {author} {\bibfnamefont {J.}~\bibnamefont
  {Biamonte}},\ and\ \bibinfo {author} {\bibfnamefont {A.}~\bibnamefont
  {Aspuru-Guzik}},\ }\bibfield  {title} {\bibinfo {title} {{Simulation of
  electronic structure Hamiltonians using quantum computers}},\ }\href
  {https://doi.org/10.1080/00268976.2011.552441} {\bibfield  {journal}
  {\bibinfo  {journal} {Molecular Physics}\ }\textbf {\bibinfo {volume}
  {109}},\ \bibinfo {pages} {735} (\bibinfo {year} {2011})}\BibitemShut
  {NoStop}%
\bibitem [{\citenamefont {Lacroix}\ \emph {et~al.}(2020)\citenamefont
  {Lacroix}, \citenamefont {Hellings}, \citenamefont {Andersen}, \citenamefont
  {Di~Paolo}, \citenamefont {Remm}, \citenamefont {Lazar}, \citenamefont
  {Krinner}, \citenamefont {Norris}, \citenamefont {Gabureac}, \citenamefont
  {Blais}, \citenamefont {Eichler},\ and\ \citenamefont
  {Wallraff}}]{Lacroix2020}%
  \BibitemOpen
  \bibfield  {author} {\bibinfo {author} {\bibfnamefont {N.}~\bibnamefont
  {Lacroix}}, \bibinfo {author} {\bibfnamefont {C.}~\bibnamefont {Hellings}},
  \bibinfo {author} {\bibfnamefont {C.~K.}\ \bibnamefont {Andersen}}, \bibinfo
  {author} {\bibfnamefont {A.}~\bibnamefont {Di~Paolo}}, \bibinfo {author}
  {\bibfnamefont {A.}~\bibnamefont {Remm}}, \bibinfo {author} {\bibfnamefont
  {S.}~\bibnamefont {Lazar}}, \bibinfo {author} {\bibfnamefont
  {S.}~\bibnamefont {Krinner}}, \bibinfo {author} {\bibfnamefont {G.~J.}\
  \bibnamefont {Norris}}, \bibinfo {author} {\bibfnamefont {M.}~\bibnamefont
  {Gabureac}}, \bibinfo {author} {\bibfnamefont {A.}~\bibnamefont {Blais}},
  \bibinfo {author} {\bibfnamefont {C.}~\bibnamefont {Eichler}},\ and\ \bibinfo
  {author} {\bibfnamefont {A.}~\bibnamefont {Wallraff}},\ }\bibfield  {title}
  {\bibinfo {title} {{Improving the Performance of Deep Quantum Optimization
  Algorithms with Continuous Gate Sets}},\ }\href
  {http://arxiv.org/abs/2005.05275} {\bibfield  {journal} {\bibinfo  {journal}
  {arXiv:2005.05275}\ } (\bibinfo {year} {2020})}\BibitemShut {NoStop}%
\bibitem [{\citenamefont {Chow}\ \emph {et~al.}(2014)\citenamefont {Chow},
  \citenamefont {Gambetta}, \citenamefont {Magesan}, \citenamefont {Abraham},
  \citenamefont {Cross}, \citenamefont {Johnson}, \citenamefont {Masluk},
  \citenamefont {Ryan}, \citenamefont {Smolin}, \citenamefont {Srinivasan},\
  and\ \citenamefont {Steffen}}]{Chow2013}%
  \BibitemOpen
  \bibfield  {author} {\bibinfo {author} {\bibfnamefont {J.~M.}\ \bibnamefont
  {Chow}}, \bibinfo {author} {\bibfnamefont {J.~M.}\ \bibnamefont {Gambetta}},
  \bibinfo {author} {\bibfnamefont {E.}~\bibnamefont {Magesan}}, \bibinfo
  {author} {\bibfnamefont {D.~W.}\ \bibnamefont {Abraham}}, \bibinfo {author}
  {\bibfnamefont {A.~W.}\ \bibnamefont {Cross}}, \bibinfo {author}
  {\bibfnamefont {B.~R.}\ \bibnamefont {Johnson}}, \bibinfo {author}
  {\bibfnamefont {N.~A.}\ \bibnamefont {Masluk}}, \bibinfo {author}
  {\bibfnamefont {C.~A.}\ \bibnamefont {Ryan}}, \bibinfo {author}
  {\bibfnamefont {J.~A.}\ \bibnamefont {Smolin}}, \bibinfo {author}
  {\bibfnamefont {S.~J.}\ \bibnamefont {Srinivasan}},\ and\ \bibinfo {author}
  {\bibfnamefont {M.}~\bibnamefont {Steffen}},\ }\bibfield  {title} {\bibinfo
  {title} {{Implementing a strand of a scalable fault-tolerant quantum
  computing fabric}},\ }\href {http://arxiv.org/abs/1311.6330} {\bibfield
  {journal} {\bibinfo  {journal} {Nature Communications}\ }\textbf {\bibinfo
  {volume} {5}} (\bibinfo {year} {2014})}\BibitemShut {NoStop}%
\bibitem [{\citenamefont {McKay}\ \emph {et~al.}(2017)\citenamefont {McKay},
  \citenamefont {Wood}, \citenamefont {Sheldon}, \citenamefont {Chow},\ and\
  \citenamefont {Gambetta}}]{McKay2017}%
  \BibitemOpen
  \bibfield  {author} {\bibinfo {author} {\bibfnamefont {D.~C.}\ \bibnamefont
  {McKay}}, \bibinfo {author} {\bibfnamefont {C.~J.}\ \bibnamefont {Wood}},
  \bibinfo {author} {\bibfnamefont {S.}~\bibnamefont {Sheldon}}, \bibinfo
  {author} {\bibfnamefont {J.~M.}\ \bibnamefont {Chow}},\ and\ \bibinfo
  {author} {\bibfnamefont {J.~M.}\ \bibnamefont {Gambetta}},\ }\bibfield
  {title} {\bibinfo {title} {{Efficient Z gates for quantum computing}},\
  }\href {https://doi.org/10.1103/PhysRevA.96.022330} {\bibfield  {journal}
  {\bibinfo  {journal} {Physical Review A}\ }\textbf {\bibinfo {volume} {96}},\
  \bibinfo {pages} {022330} (\bibinfo {year} {2017})}\BibitemShut {NoStop}%
\bibitem [{\citenamefont {Motzoi}\ \emph {et~al.}(2009)\citenamefont {Motzoi},
  \citenamefont {Gambetta}, \citenamefont {Rebentrost},\ and\ \citenamefont
  {Wilhelm}}]{Motzoi2009}%
  \BibitemOpen
  \bibfield  {author} {\bibinfo {author} {\bibfnamefont {F.}~\bibnamefont
  {Motzoi}}, \bibinfo {author} {\bibfnamefont {J.~M.}\ \bibnamefont
  {Gambetta}}, \bibinfo {author} {\bibfnamefont {P.}~\bibnamefont
  {Rebentrost}},\ and\ \bibinfo {author} {\bibfnamefont {F.~K.}\ \bibnamefont
  {Wilhelm}},\ }\bibfield  {title} {\bibinfo {title} {{Simple Pulses for
  Elimination of Leakage in Weakly Nonlinear Qubits}},\ }\href
  {https://doi.org/10.1103/PhysRevLett.103.110501} {\bibfield  {journal}
  {\bibinfo  {journal} {Physical Review Letters}\ }\textbf {\bibinfo {volume}
  {103}},\ \bibinfo {pages} {110501} (\bibinfo {year} {2009})}\BibitemShut
  {NoStop}%
\bibitem [{\citenamefont {Gambetta}\ \emph {et~al.}(2011)\citenamefont
  {Gambetta}, \citenamefont {Motzoi}, \citenamefont {Merkel},\ and\
  \citenamefont {Wilhelm}}]{Gambetta2011}%
  \BibitemOpen
  \bibfield  {author} {\bibinfo {author} {\bibfnamefont {J.~M.}\ \bibnamefont
  {Gambetta}}, \bibinfo {author} {\bibfnamefont {F.}~\bibnamefont {Motzoi}},
  \bibinfo {author} {\bibfnamefont {S.~T.}\ \bibnamefont {Merkel}},\ and\
  \bibinfo {author} {\bibfnamefont {F.~K.}\ \bibnamefont {Wilhelm}},\
  }\bibfield  {title} {\bibinfo {title} {{Analytic control methods for
  high-fidelity unitary operations in a weakly nonlinear oscillator}},\
  }\href@noop {} {\bibfield  {journal} {\bibinfo  {journal} {Physical Review A
  - Atomic, Molecular, and Optical Physics}\ }\textbf {\bibinfo {volume} {83}}
  (\bibinfo {year} {2011})}\BibitemShut {NoStop}%
\bibitem [{\citenamefont {Nielsen}\ and\ \citenamefont
  {Chuang}(2010)}]{Nielsen2000}%
  \BibitemOpen
  \bibfield  {author} {\bibinfo {author} {\bibfnamefont {M.~A.}\ \bibnamefont
  {Nielsen}}\ and\ \bibinfo {author} {\bibfnamefont {I.~L.}\ \bibnamefont
  {Chuang}},\ }\href {https://doi.org/10.1017/cbo9780511976667} {\emph
  {\bibinfo {title} {Quantum Computation and Quantum Information}}}\ (\bibinfo
  {year} {2010})\BibitemShut {NoStop}%
\bibitem [{\citenamefont {C{\'{o}}rcoles}\ \emph {et~al.}(2013)\citenamefont
  {C{\'{o}}rcoles}, \citenamefont {Gambetta}, \citenamefont {Chow},
  \citenamefont {Smolin}, \citenamefont {Ware}, \citenamefont {Strand},
  \citenamefont {Plourde},\ and\ \citenamefont {Steffen}}]{Corcoles2013}%
  \BibitemOpen
  \bibfield  {author} {\bibinfo {author} {\bibfnamefont {A.~D.}\ \bibnamefont
  {C{\'{o}}rcoles}}, \bibinfo {author} {\bibfnamefont {J.~M.}\ \bibnamefont
  {Gambetta}}, \bibinfo {author} {\bibfnamefont {J.~M.}\ \bibnamefont {Chow}},
  \bibinfo {author} {\bibfnamefont {J.~A.}\ \bibnamefont {Smolin}}, \bibinfo
  {author} {\bibfnamefont {M.}~\bibnamefont {Ware}}, \bibinfo {author}
  {\bibfnamefont {J.}~\bibnamefont {Strand}}, \bibinfo {author} {\bibfnamefont
  {B.~L.}\ \bibnamefont {Plourde}},\ and\ \bibinfo {author} {\bibfnamefont
  {M.}~\bibnamefont {Steffen}},\ }\bibfield  {title} {\bibinfo {title}
  {{Process verification of two-qubit quantum gates by randomized
  benchmarking}},\ }\href {https://doi.org/10.1103/PhysRevA.87.030301}
  {\bibfield  {journal} {\bibinfo  {journal} {Physical Review A - Atomic,
  Molecular, and Optical Physics}\ }\textbf {\bibinfo {volume} {87}},\ \bibinfo
  {pages} {030301} (\bibinfo {year} {2013})}\BibitemShut {NoStop}%
\bibitem [{\citenamefont {Chow}\ \emph {et~al.}(2009)\citenamefont {Chow},
  \citenamefont {Gambetta}, \citenamefont {Tornberg}, \citenamefont {Koch},
  \citenamefont {Bishop}, \citenamefont {Houck}, \citenamefont {Johnson},
  \citenamefont {Frunzio}, \citenamefont {Girvin},\ and\ \citenamefont
  {Schoelkopf}}]{Chow2009}%
  \BibitemOpen
  \bibfield  {author} {\bibinfo {author} {\bibfnamefont {J.~M.}\ \bibnamefont
  {Chow}}, \bibinfo {author} {\bibfnamefont {J.~M.}\ \bibnamefont {Gambetta}},
  \bibinfo {author} {\bibfnamefont {L.}~\bibnamefont {Tornberg}}, \bibinfo
  {author} {\bibfnamefont {J.}~\bibnamefont {Koch}}, \bibinfo {author}
  {\bibfnamefont {L.~S.}\ \bibnamefont {Bishop}}, \bibinfo {author}
  {\bibfnamefont {A.~A.}\ \bibnamefont {Houck}}, \bibinfo {author}
  {\bibfnamefont {B.~R.}\ \bibnamefont {Johnson}}, \bibinfo {author}
  {\bibfnamefont {L.}~\bibnamefont {Frunzio}}, \bibinfo {author} {\bibfnamefont
  {S.~M.}\ \bibnamefont {Girvin}},\ and\ \bibinfo {author} {\bibfnamefont
  {R.~J.}\ \bibnamefont {Schoelkopf}},\ }\bibfield  {title} {\bibinfo {title}
  {{Randomized benchmarking and process tomography for gate errors in a
  solid-state qubit}},\ }\href {https://doi.org/10.1103/PhysRevLett.102.090502}
  {\bibfield  {journal} {\bibinfo  {journal} {Physical Review Letters}\
  }\textbf {\bibinfo {volume} {102}},\ \bibinfo {pages} {090502} (\bibinfo
  {year} {2009})}\BibitemShut {NoStop}%
\bibitem [{\citenamefont {Cross}\ and\ \citenamefont
  {Gambetta}(2015)}]{Cross2015}%
  \BibitemOpen
  \bibfield  {author} {\bibinfo {author} {\bibfnamefont {A.~W.}\ \bibnamefont
  {Cross}}\ and\ \bibinfo {author} {\bibfnamefont {J.~M.}\ \bibnamefont
  {Gambetta}},\ }\bibfield  {title} {\bibinfo {title} {{Optimized pulse shapes
  for a resonator-induced phase gate}},\ }\href@noop {} {\bibfield  {journal}
  {\bibinfo  {journal} {Physical Review A - Atomic, Molecular, and Optical
  Physics}\ }\textbf {\bibinfo {volume} {91}} (\bibinfo {year}
  {2015})}\BibitemShut {NoStop}%
\bibitem [{\citenamefont {Puri}\ and\ \citenamefont {Blais}(2016)}]{Puri2016}%
  \BibitemOpen
  \bibfield  {author} {\bibinfo {author} {\bibfnamefont {S.}~\bibnamefont
  {Puri}}\ and\ \bibinfo {author} {\bibfnamefont {A.}~\bibnamefont {Blais}},\
  }\bibfield  {title} {\bibinfo {title} {{High-Fidelity Resonator-Induced Phase
  Gate with Single-Mode Squeezing}},\ }\href@noop {} {\bibfield  {journal}
  {\bibinfo  {journal} {Physical Review Letters}\ }\textbf {\bibinfo {volume}
  {116}} (\bibinfo {year} {2016})}\BibitemShut {NoStop}%
\bibitem [{\citenamefont {Filipp}\ \emph {et~al.}(2009)\citenamefont {Filipp},
  \citenamefont {Maurer}, \citenamefont {Leek}, \citenamefont {Baur},
  \citenamefont {Bianchetti}, \citenamefont {Fink}, \citenamefont
  {G{\"{o}}ppl}, \citenamefont {Steffen}, \citenamefont {Gambetta},
  \citenamefont {Blais},\ and\ \citenamefont {Wallraff}}]{Filipp2009}%
  \BibitemOpen
  \bibfield  {author} {\bibinfo {author} {\bibfnamefont {S.}~\bibnamefont
  {Filipp}}, \bibinfo {author} {\bibfnamefont {P.}~\bibnamefont {Maurer}},
  \bibinfo {author} {\bibfnamefont {P.~J.}\ \bibnamefont {Leek}}, \bibinfo
  {author} {\bibfnamefont {M.}~\bibnamefont {Baur}}, \bibinfo {author}
  {\bibfnamefont {R.}~\bibnamefont {Bianchetti}}, \bibinfo {author}
  {\bibfnamefont {J.~M.}\ \bibnamefont {Fink}}, \bibinfo {author}
  {\bibfnamefont {M.}~\bibnamefont {G{\"{o}}ppl}}, \bibinfo {author}
  {\bibfnamefont {L.}~\bibnamefont {Steffen}}, \bibinfo {author} {\bibfnamefont
  {J.~M.}\ \bibnamefont {Gambetta}}, \bibinfo {author} {\bibfnamefont
  {A.}~\bibnamefont {Blais}},\ and\ \bibinfo {author} {\bibfnamefont
  {A.}~\bibnamefont {Wallraff}},\ }\bibfield  {title} {\bibinfo {title}
  {{Two-qubit state tomography using a joint dispersive readout}},\ }\href
  {https://doi.org/10.1103/PhysRevLett.102.200402} {\bibfield  {journal}
  {\bibinfo  {journal} {Physical Review Letters}\ }\textbf {\bibinfo {volume}
  {102}},\ \bibinfo {pages} {200402} (\bibinfo {year} {2009})}\BibitemShut
  {NoStop}%
\bibitem [{\citenamefont {Macklin}\ \emph {et~al.}(2015)\citenamefont
  {Macklin}, \citenamefont {Hover}, \citenamefont {Schwartz}, \citenamefont
  {Zhang}, \citenamefont {Oliver},\ and\ \citenamefont
  {Siddiqi}}]{Macklin2015}%
  \BibitemOpen
  \bibfield  {author} {\bibinfo {author} {\bibfnamefont {C.}~\bibnamefont
  {Macklin}}, \bibinfo {author} {\bibfnamefont {D.}~\bibnamefont {Hover}},
  \bibinfo {author} {\bibfnamefont {M.~E.}\ \bibnamefont {Schwartz}}, \bibinfo
  {author} {\bibfnamefont {X.}~\bibnamefont {Zhang}}, \bibinfo {author}
  {\bibfnamefont {W.~D.}\ \bibnamefont {Oliver}},\ and\ \bibinfo {author}
  {\bibfnamefont {I.}~\bibnamefont {Siddiqi}},\ }\bibfield  {title} {\bibinfo
  {title} {{A near – quantum-limited Josephson traveling-wave parametric
  amplifier}},\ }\href@noop {} {\bibfield  {journal} {\bibinfo  {journal}
  {Science}\ }\textbf {\bibinfo {volume} {350}},\ \bibinfo {pages} {307}
  (\bibinfo {year} {2015})}\BibitemShut {NoStop}%
\end{thebibliography}

%apsrev4-2.bst 2019-01-14 (MD) hand-edited version of apsrev4-1.bst
%Control: key (0)
%Control: author (8) initials jnrlst
%Control: editor formatted (1) identically to author
%Control: production of article title (0) allowed
%Control: page (0) single
%Control: year (1) truncated
%Control: production of eprint (0) enabled
%

\end{document}

\title{Supplemental Materials for\\
`Arbitrary controlled-phase gate on fluxonium qubits using differential ac-Stark shifts'}

\author{H. Xiong et al.}

\date{\today}

\maketitle

\section{Supplementary note 1: Experimental Setup \label{sec:setup}}

\begin{figure}
    \centering
    \includegraphics{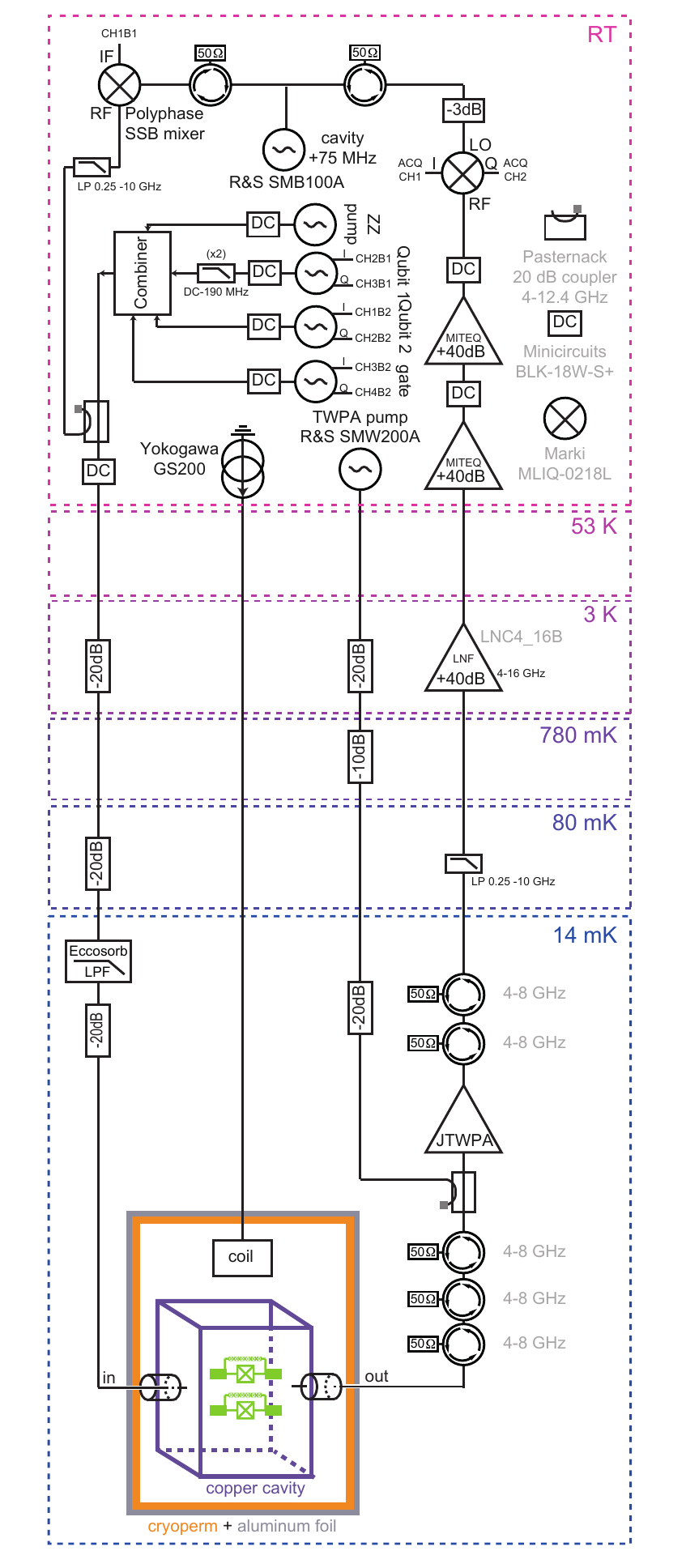}
    \caption{\textbf{Schematics of the experimental setup.} See labeling on the plot for details.}
    \label{fig:wiring}
\end{figure}
Our device geometry and fabrication procedure follow the one described in Ref.~\cite{Ficheux2020}. A schematic of the experimental setup is depicted in Supplementary Figure~\ref{fig:wiring}. We use standard modulation and demodulation techniques of microwave signals. Intermediate frequency pulses are generated by two M3202A PXIe Arbitrary Waveform Generators (not represented) and the readout signal is digitized by a M3102A PXIe Digitizer (not represented) in the same M9010A PXIe Chassis. The signal departing the cavity is amplified by a Josephson Traveling Wave Parametric Amplifier (JTWPA) provided by Lincoln labs \cite{Macklin2015}.

\section{Supplementary note 2: Spectroscopy and time-domain measurements}

\begin{figure}
    \centering
    \includegraphics[width=0.6\linewidth]{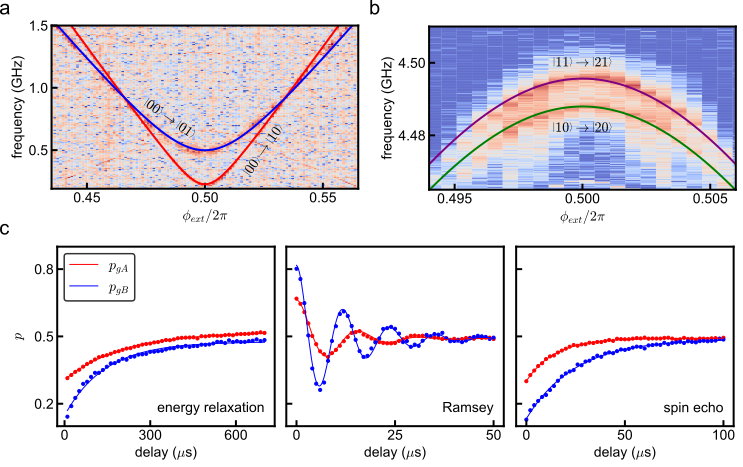}
    \caption{\textbf{Two-tone spectroscopy and time-domain characterization.} \textbf{a} Two-tone spectroscopy of the fluxonium transitions as a function of the external flux threading both fluxonium loops. The $|00\rangle - |10\rangle$ and $|00\rangle - |01\rangle$ transitions show the qubit frequencies at the sweet spot of $217.2 \mathrm{~MHz}$ and $488.9 \mathrm{~MHz}$, respectively. \textbf{b} Two-tone spectroscopy of the $|11\rangle - |21\rangle$ and $|10\rangle - |20\rangle$ used to induce the $\textrm{ZZ}$-interaction in the main text. \textbf{c} Energy relaxation time $T_1$, Ramsey coherence time $T_2^{\mathrm{Ramsey}}$, and spin-echo coherence time $T_2^{\mathrm{Echo}}$ characterization of the qubit transitions.}
    \label{fig:spectrum}
\end{figure}

The two-fluxonium spectrum is obtained with the standard two-tone experiment technique where we readout the cavity after exciting the system with a probe tone. The qubit parameters are extracted by fitting the spectrum at different external flux points with the numerical diagonalization of the Hamiltonian given by Eq.~\ref{eq:Hamiltonian}
as shown in Supplementary Figure~\ref{fig:spectrum}a, b. The $|1\rangle - |2\rangle$ transition of qubit A splits into two transitions $|11\rangle - |21\rangle$ and $|10\rangle - |20\rangle$ due to its capacitive coupling to qubit B. Off-resonantly driving these two transitions induces the $\textrm{ZZ}$-interaction  required to cancel the static $\textrm{ZZ}$-interaction and to perform for CP gates as described in the main text.

The coherence time of the fluxonium transitions are extracted using time-domain measurements preceded by an initialization pulse described in Methods. Measuring the energy decay of the qubits from the initialized state yields $T_1^A = (207 \pm 3.7) \mathrm{~\mu s}$ and $T_1^B = (141 \pm 5.2) \mathrm{~\mu s}$, while standard Ramsey and spin-echo sequences give the coherence times $T_{2,R}^A = (9.69\pm 0.33) \mathrm{~\mu s}$, $T_{2,R}^B = (13 \pm 0.4) \mathrm{~\mu s}$, $T_{2,E}^A = (14.3 \pm 0.27) \mathrm{~\mu s}$, $T_{2,E}^B = (24.6 \pm 0.47) \mathrm{~\mu s}$ (see Fig.\ref{fig:spectrum}c). The fluctuation range of these numbers are reported in Table~I in the main text.

\section{Supplementary note 3: Readout errors}

\begin{figure}
    \centering
    \includegraphics[width=0.6\linewidth]{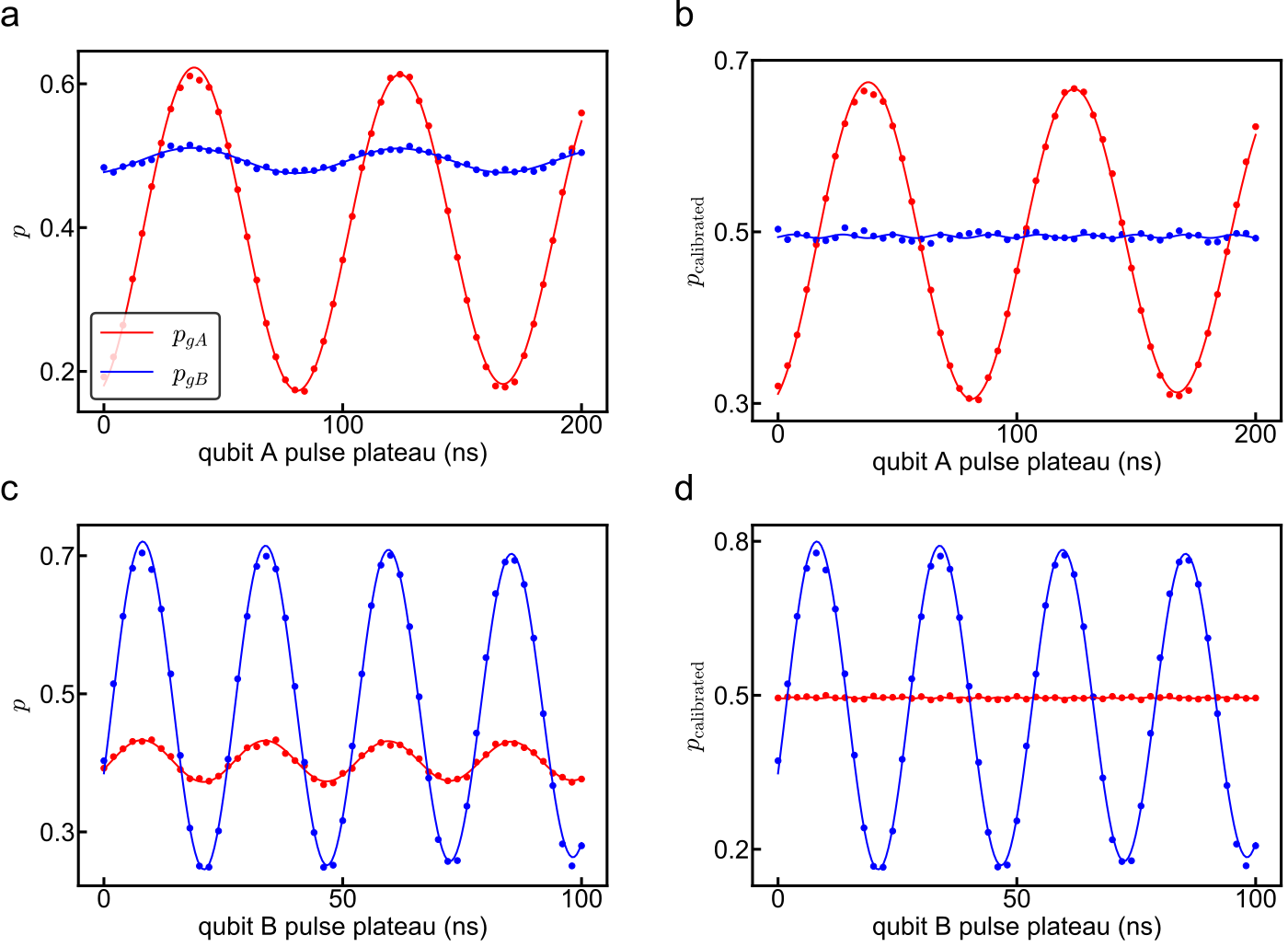}
    \caption{\textbf{Rabi oscillations with and without readout errors correction.} Oscillations of qubit A (Fig.~\textbf{a}, \textbf{b}) and qubit B (Fig.~\textbf{c}, \textbf{d}). Before correction (left), the Rabi oscillations are not centered around $0.5$ because of qubit transitions occuring during the readout. We also correct for readout cross-talk: qubit B readout signal oscillates when qubit A is rotated before correction, and vice-versa.}
    \label{fig:AB_Rabi}
\end{figure}

We adopt an empirical method to compensate for readout errors similar to the one in Ref.~\cite{Ficheux2020}. We assume the measured populations $p_{ij}'$ are linked to populations $p_{ij}$ at the beginning of the readout by a linear transformation $\hat{M}$,
\begin{equation}
    \begin{pmatrix}
    p_{gg}'\\
    p_{ge}'\\
    p_{eg}'\\
    p_{ee}'
    \end{pmatrix} =
    \hat{M}
    \begin{pmatrix}
    p_{gg}\\
    p_{ge}\\
    p_{eg}\\
    p_{ee}
    \end{pmatrix}.
\end{equation}
The error matrix $\hat{M}$ has 6 parameters that account for the incorrect mapping $\left| 0 \right> \rightarrow \left| 1 \right>$ and $\left| 1 \right> \rightarrow \left| 0 \right>$ for both qubits and transitions between the two qubits during the readout. The error matrix reads
\begin{equation}
    M = 
    \begin{pmatrix}
    1 - a_1 - b_1 & b_2 & a_2 & 0\\
    b_1 & 1 - a_1 - b_2 - c_1 & c_2 & a_2\\
    a_1 & c_1 & 1 - a_2 - b_1 - c_2 & b_2\\
    0 & a_1 & b_1 & 1 - a_2 - b_2
    \end{pmatrix}.
\label{eq:Tmatrix}
\end{equation}
Among the 6 parameters, $a_1$ and $a_2$ represent incorrect mappings $\left| 0 \right>_A \rightarrow \left| 1 \right>_A$ and $\left| 1 \right>_A \rightarrow \left| 0 \right>_A$. $b_1$ and $b_2$ represent incorrect mappings $\left| 0 \right>_A \rightarrow \left| 1 \right>_B$ and $\left| 1 \right>_B \rightarrow \left| 0 \right>_A$. $c_1$ and $c_2$ describe the readout cross-talk where the two qubits swap excitations. In addition to these six parameters, the initial populations for the two qubits $p_{gA}^0$, $p_{gB}^0$ are also two unknown variables that need to be 
determined. To calibrate all the parameters, we perform Rabi measurements on each qubit while the other qubit is left in the $\left| + \right>$ state, as shown in Supplementary Figure~\ref{fig:AB_Rabi}. Assuming that the system starts with the populations
\begin{equation}
    \begin{pmatrix}
    p_{gg}\\
    p_{ge}\\
    p_{eg}\\
    p_{ee}
    \end{pmatrix} =
    \begin{pmatrix}
    p_{gA}^0p_{gB}^0\\
    p_{gA}^0(1-p_{gB}^0)\\
    (1-p_{gA}^0)p_{gB}^0\\
    (1-p_{gA}^0)(1-p_{gB}^0)
    \end{pmatrix},
\end{equation}
one can predict the Rabi oscillation amplitudes and offsets in the two Rabi measurements of Supplementary Figure~\ref{fig:AB_Rabi} as a function of the 6 parameters of the error matrix -- and then deduce $\hat{M}$. The corrected populations are obtained by applying $\hat{M}^{-1}$ to the measured populations $(p_{gg}', p_{ge}', p_{eg}', p_{ee}')^T$. With this calibration technique, the populations for both qubits oscillate around 0.5 and only the rotated qubit displays oscillations during the Rabi experiment.

\begin{figure}
    \centering
    \includegraphics[width=0.6\linewidth]{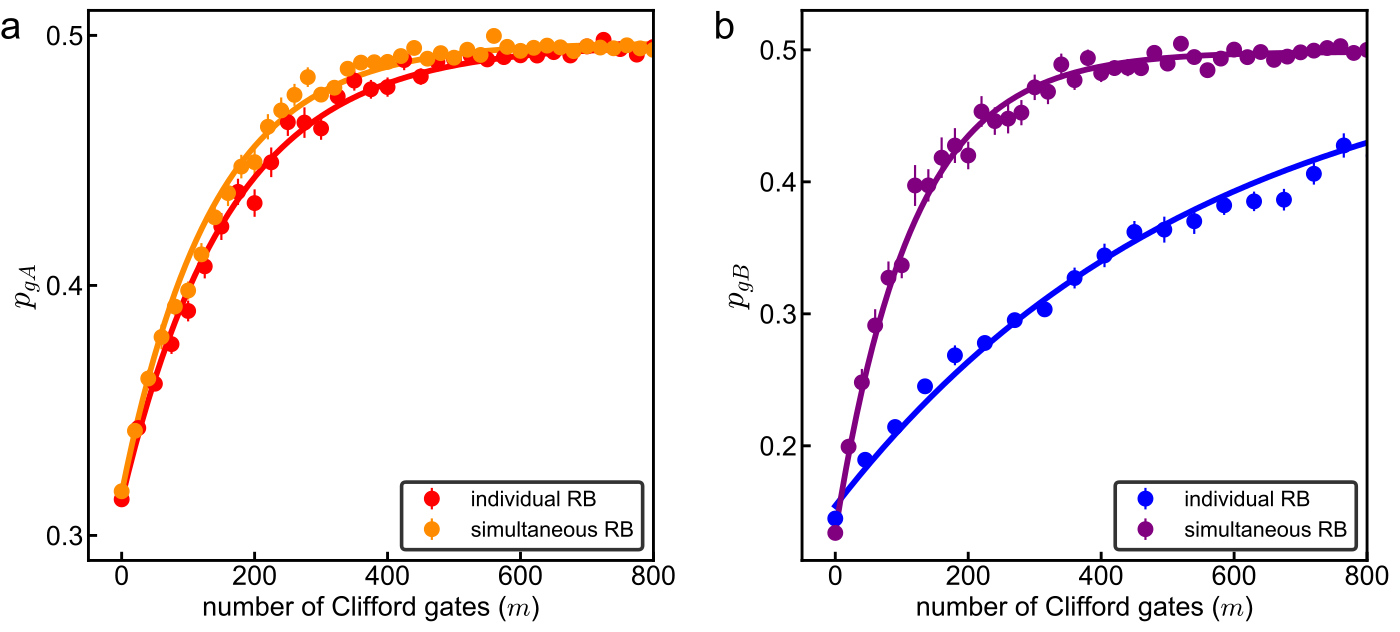}
    \caption{\textbf{Randomized benchmarking of single qubit gates.} \textbf{a} Population of qubit A and \textbf{b} population of qubit B as a function of the sequence length. A list of $51$ randomly chosen sequence of Clifford gates are applied to the individually or simultaneously to the qubits before applying a recovery gate. The average single qubit fidelity for qubit A (resp. qubit B) for a gate duration of $45\mathrm{~ns}$ (resp. $26\mathrm{~ns}$) is $(99.69\pm 0.01) \%$ (resp. $(99.91\pm 0.003) \%$) in the individual case and $(99.63\pm 0.01) \%$ (resp. $(99.57\pm 0.01) \%$) in the simultaneous case. The duration of a simultaneous operation is given by longest gate time (gate time of qubit A or gate time of qubit B is a zero-duration pulse is applied to A). The fidelities extracted from the simultaneous RB are used to calculate the CP gate fidelity in Fig. 4 and Fig. 5 of the main text.}
    \label{fig:AB_RB}
\end{figure}

\section{Supplementary note 4: Single qubit Randomized Benchmarking\label{sec:singleRB}}

\begin{center}
\begin{table}[h!]
\begin{tabular}{ | c | } \hline
Single qubit Clifford group\\ \hline
$I$, $X_\pi$, $Y_\pi$, $Z_\pi$, $X_{\pi/2}$, $X_{-\pi/2}$, $Y_{\pi/2}$, $Y_{-\pi/2}$, $Z_{\pi/2}$, $Z_{-\pi/2}$, \\
$Y_{\pi/2}Z_{\pi/2}$, $Y_{-\pi/2}Z_{-\pi/2}$, $Y_{\pi/2}Z_{-\pi/2}$, $Y_{-\pi/2}Z_{\pi/2}$, $X_{\pi/2}Z_{-\pi/2}$, $X_{-\pi/2}Z_{\pi/2}$, $X_{\pi/2}Z_{\pi/2}$, $X_{-\pi/2}Z_{-\pi/2}$, \\
$Z_{\pi/2}X_{\pi/2}Z_{\pi/2}$, $Z_{\pi/2}X_{-\pi/2}Z_{\pi/2}$, $Z_{-\pi/2}Y_{\pi/2}Z_{-\pi/2}$, $Z_{-\pi/2}Y_{-\pi/2}Z_{-\pi/2}$, $Z_{-\pi/2}X_{\pi}Z_{-\pi}$, $Z_{\pi/2}X_{\pi}Z_{-\pi}$\\ \hline
\end{tabular}
\caption{\label{table:Clifford1} Single-qubit Clifford gates used in the RB sequences. Z rotations are realized with virtual Z gates. Only X and Y pulses have a non-zero duration.}
\end{table}
\end{center}

Single-qubit randomized benchmarking (RB) is performed on each qubit -- individually and simultaneously -- by applying a series of randomly chosen Clifford gates listed in Table~\ref{table:Clifford1}. Z rotations are performed with virtual Z gates \cite{McKay2017} -- which simply change the phase of X, Y pulses in subsequent gates. Therefore, our Z rotations and identity gates have a zero duration. In the individual single-qubit RB sequence, one Clifford gate contains on average 0.83 physical pulses. As shown in Supplementary Figure~\ref{fig:AB_RB}, the average individual single-qubit fidelity is $(99.69\pm 0.01) \%$ for qubit A with $45 \mathrm{~ns}$-long pulses and $(99.91\pm 0.003) \%$ for qubit B with $26 \mathrm{~ns}$-long pulses.

In our simultaneous RB sequence, there is a possible idle time for one qubit when the other qubit is rotated on the Bloch sphere. So the simultaneous RB sequence is longer than the individual RB sequences with the same number of Clifford gates. The change of fidelity when operating the qubits simultaneously cannot be directly associated to cross-talks. With such sequences, the simultaneous single qubit fidelity is $(99.63\pm 0.01) \%$ for qubit A and $(99.57\pm 0.01) \%$ for qubit B. These numbers are used to extract the CP gate fidelity in the XEB measurements.

\section{Supplementary note 5: Quantum Process Tomography \label{sec:QPT}}

We perform a process tomography of the gate operation using a standard procedure described in \cite{Nielsen2000,Corcoles2013}. The quantum process to be characterized is interleaved between one preparation pulse and one tomography pulse chosen in Table~\ref{table:QPTpulses}. The process matrix $\chi$ characterizes fully the quantum process $\mathcal{E}$. We prepare 16 input states $\left\{ \rho_j \right\}$ and measure the output states $\mathcal{E} (\rho_j) = \sum_{m ,n} \chi_{m n} P_m \rho_j P_n^\dagger$ where $P_i$ are the 16 two-qubit Pauli operators.

\subsection{State tomography}

The density matrix of each output state is reconstructed by maximizing the likelihood of a list of $36$ measurements obtained by applying the tomography pulses given in Table~\ref{table:QPTpulses}. The Maximum Likelihood Estimation (MLE) technique \cite{Hradil1997} searches the space of density matrices to find the one that is the most likely to reproduce the observations. To calculate the expected output signal for an arbitrary density matrix, we first calibrate the measurement operator
\begin{equation}
\hat{M} = \beta_{II} \mathbb{I} \otimes \mathbb{I} + \beta_{IZ} \mathbb{I} \otimes \sigma_Z+\beta_{ZI} \sigma_Z \otimes \mathbb{I}+\beta_{ZZ} \sigma_Z \otimes \sigma_Z
\label{eq:MeasOp}
\end{equation}
where $\beta_{ij}$ are complex coefficients and $\sigma_i$ are Pauli matrices. $\beta_{ij}$ are measured by applying the pulses $\{II, IX_\pi, X_\pi I , X_\pi X_\pi\}$ to the initial state before measuring the output signal. After calibrating the measurement operator, we can predict the output signal for a state $\rho$ after any of the $36$ tomography pulses. We parametrize the density matrix with 16 real parameters as $\rho = \mathcal{T}^\dagger\mathcal{T} / \mathrm{Tr}(\mathcal{T}^\dagger\mathcal{T})$, where $\mathcal{T}$ is a lower triangular matrix given by
\begin{equation}
\mathcal{T}=
    \begin{pmatrix}
    t_1 & 0 & 0 & 0\\
    t_5 + i t_6 & t_2 & 0 & 0\\
    t_{11} + i t_{12} & t_7 + i t_8 & t_3 & 0\\
    t_{15} + i t_{16} & t_{13} + i t_{14} & t_9 + i t_{10} & t_4
    \end{pmatrix}.
\label{eq:Tmatrix}
\end{equation}
This Cholesky decomposition ensures that $\rho$ corresponds to a physical quantum state (see section V of the supplementary material of Ref. \cite{Roy2017}). The state $\rho$ is then obtained by maximizing the likelihood of the 36 output measurements over the 16 parameters of $\mathcal{T}$.

\subsection{Process matrix}

The process matrix $\chi$ is then calculated by linear inversion following the procedure described in Ref.~\cite{Nielsen2000}. This procedure is valid in our case even if the input states $\rho_j$ are not pure states because they are still linearly independent. The gate fidelity \cite{Chow2009} compared to the ideal process matrix $\chi_{ideal}$ is $ F(\chi, \chi_{ideal}) = (\mathrm{Tr}(\chi^\dagger\chi_{ideal}) d + 1) / (d + 1)$, where $d$ is the dimension of the Hilbert space. 

\begin{center}
\begin{table}[h!]
\begin{tabular}{ | c | c | } \hline
Preparation pulses & Tomography Pulses\\ \hline

$II$, $IX_{\pi/2}$, & $II$, $X_{\pi}I$, $X_{\pi/2}I$, $X_{-\pi/2}I$,\\
$IY_{\pi/2}$, $IX_{-\pi/2}$, & $Y_{\pi/2}I$, $Y_{-\pi/2}I$, $IX_{\pi}$, $X_{\pi}X_{\pi}$,\\
$X_{\pi}I$, $X_{\pi}X_{\pi}$, & $X_{\pi/2}X_{\pi}$, $X_{-\pi/2}X_{\pi}$, $Y_{\pi/2}X_{\pi}$, $Y_{-\pi/2}X_{\pi}$,\\
$X_{\pi}Y_{\pi/2}$, $X_{\pi}X_{-\pi/2}$, & $IX_{\pi/2}$, $X_{\pi}X_{\pi/2}$, $X_{\pi/2}X_{\pi/2}$, $X_{-\pi/2}X_{\pi/2}$,\\
$Y_{\pi/2}I$, $Y_{\pi/2}X_{\pi}$, & $Y_{\pi/2}Y_{\pi/2}$, $Y_{-\pi/2}Y_{\pi/2}$, $IX_{-\pi/2}$, $X_{\pi}X_{-\pi/2}$,\\
$Y_{\pi/2}Y_{\pi/2}$, $Y_{\pi/2}X_{-\pi/2}$, & $X_{\pi/2}X_{-\pi/2}$, $X_{-\pi/2}X_{-\pi/2}$, \\
$X_{-\pi/2}I$, $X_{-\pi/2}X_{\pi}$, & $Y_{\pi/2}X_{-\pi/2}$, $Y_{-\pi/2}X_{-\pi/2}$, \\
$X_{-\pi/2}Y_{\pi/2}$, $X_{-\pi/2}X_{-\pi/2}$ & $IY_{\pi/2}$, $X_{\pi}Y_{\pi/2}$, $X_{\pi/2}Y_{\pi/2}$, $X_{-\pi/2}Y_{\pi/2}$,\\
& $Y_{\pi/2}Y_{\pi/2}$, $Y_{-\pi/2}Y_{\pi/2}$, $IY_{-\pi/2}$, $X_{\pi}Y_{-\pi/2}$,\\
& $X_{\pi/2}Y_{-\pi/2}$, $X_{-\pi/2}Y_{-\pi/2}$, \\ 
& $Y_{\pi/2}Y_{-\pi/2}$, $Y_{-\pi/2}Y_{-\pi/2}$ \\ \hline
\end{tabular}
\caption{\label{table:QPTpulses} %Table of the
Preparation and tomography pulses used for Quantum Process Tomography (QPT). The first letter refers to the rotation axis and the subscript is the rotation angle. $I$ is the identity. We prepare $d^2=16$ input states (where $d=4$ is the dimension of the Hilbert space). An overcomplete set of $36$ pulses is used for the state tomography -- combined with a maximum likelihood estimation -- to reduce the sensitivity of our tomography to pulse imperfections.}
\end{table}
\end{center}

\section{Supplementary note 6: ZZ-interactions induced when driving around 6.5~GHz}

\begin{figure}
    \centering
    \includegraphics[width=0.6\linewidth]{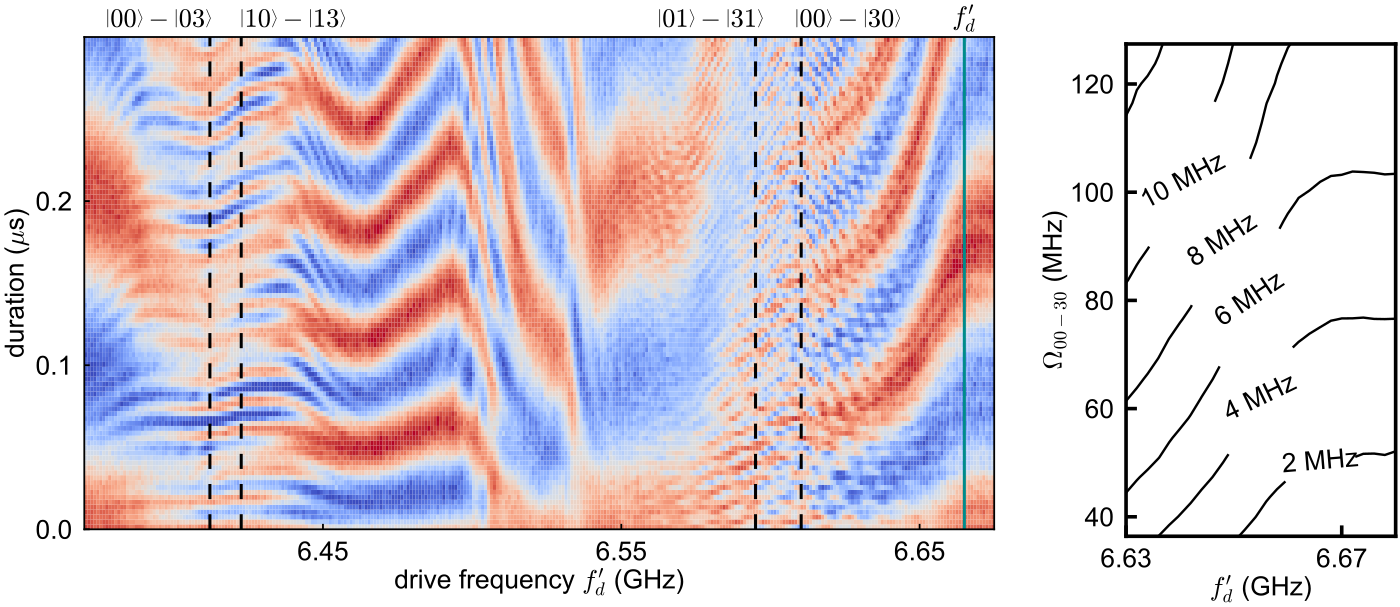}
    \caption{\textbf{$\textrm{ZZ}$-interactions induced by driving around $6.5 \mathrm{~GHz}$.} We use the pulse protocol  depicted in Fig.~2a of the main text to measure $\textrm{ZZ}$-oscillations (left panel). At this drive frequency, the four transitions $|00 \rangle-| 03 \rangle$, $|10 \rangle-| 13 \rangle$, $|01 \rangle-| 31 \rangle$, and $|00 \rangle-| 30 \rangle$ contribute to the ac-Stark shift in the computational subspace. We achieve interaction rates exceeding $10 \mathrm{~MHz}$ (right panel).
    }
    \label{fig:ZZ6p5}
\end{figure}

As stated in the main text, our gate scheme is applicable by driving near any transition from the computational subspace to higher states with a large transition dipole. Indeed, $\textrm{ZZ}$-interactions can be induced using the four transitions $| 00 \rangle- |03 \rangle$, $| 10 \rangle - |13 \rangle$, $| 01 \rangle-| 31 \rangle$, and $| 00 \rangle- | 30 \rangle$. Supplementary Figure~\ref{fig:ZZ6p5} shows ZZ-oscillations similar to the one of Fig.~2 of the main text by driving around $6.5 \mathrm{~GHz}$. We find that the induced interaction rate in this frequency range can exceed the one achievable in the vicinity of the $|10 \rangle - |20 \rangle$ and $|11 \rangle - |21 \rangle$ transitions -- and reach values greater than $10\mathrm{~MHz}$.

Our optimal CZ gate in this frequency window is obtained at $f_d'=6.665$ GHz (blue line in Supplementary Figure~\ref{fig:ZZ6p5}) with a drive amplitude $\Omega_{00-30} = 91 \mathrm{~MHz}$. Because the detuning to the four transitions are on the same order of magnitude compared to the drive amplitude, all four transitions contribute to the induced $\textrm{ZZ}$-interaction during the gate. Even though more transitions are involved when performing a CZ gate at $f_d'$ compared to the gate at $f_d=4.545$ GHz, the gate fidelity is still higher ($99.1\%$ at $f_d'$ compared to $98.9 \%$ at $f_d$) due to the larger $\textrm{ZZ}$-interaction rate.

\section{Supplementary note 7: Over 20 MHz ZZ-interactions\label{sec:faster}}
We realize the same type of CP gate at a faster speed (40 ns for a CZ gate) in the second two-fluxonium device measured in a different laboratory. This device has a similar design but with different parameters shown in Supplementary Table~\ref{table:device_parameters_2}. The coupling constant $J_C$ is comparable with that in the main text, which does not require large coupling capacitors. Despite abnormally low coherence times in this particular device, we measured a CZ gate fidelity of $(97.6\pm 0.3) \%$ using RB.

\begin{center}
\begin{table}[h!]
\begin{tabular}{ | c | c | c | c | c |} \hline
  &$E_{C,i} \mathrm{~(GHz)}$ & $E_{L,i} \mathrm{~(GHz)}$ & $E_{J,i} \mathrm{~(GHz)}$ & $J_C \mathrm{~(GHz)}$\\ \hline
  Qubit A & 1.1 & 0.84 & 3.5 & \multirow{2}{*}{0.33}  \\
  Qubit B & 1.0 & 1.7 & 4.0 &  \\ \hline
\end{tabular}
\caption{\label{table:device_parameters_2}Charging, inductive, Josephson energy and coupling constant of the device discussed in Supplementary Note \ref{sec:faster}.}
\end{table}
\end{center}

This device has a stronger static ZZ interaction $\xi_{ZZ}^\mathrm{static}=-2.1$ MHz. We use an always-on tone (detuned from the $| 11 \rangle- | 21 \rangle$ transition by $\delta = 80$ MHz, Rabi rate $\Omega = 43$ MHz) to cancel the ZZ interaction at all times.  We perform simultaneous randomized benchmarking to characterize our single-qubit gates and obtain gate fidelities of $(99.35\pm 0.03) \%$ and $(99.16\pm 0.03) \%$ for qubits A and B, respectively.  

The single-qubit gate fidelity is predominantly limited by the low coherence times of the computational transitions compared with the single-qubit gate time of 24 ns. Qubit A (707 MHz) has $T_1=11~\mu s$, $T_2^R=4.2~\mu s$, $T_2^E=5.3~\mu s$, Qubit B (1310 MHz) has $T_1=8.1~\mu s$, $T_2^R=1.6~\mu s$, $T_2^E=2.4~\mu s$. The $\ket{1}-\ket{2}$ transitions for both qubits have $T_1=0.7~\mu s$, $T_2^R=1.0~\mu s$, $T_2^E=1.3~\mu s$ on average. The low coherence time is suspected to be associated with dielectric loss deterioration during the mailing or packaging.

By increasing the amplitude of the same tone above $\Omega = 190$ MHz, we can obtain an induced ZZ coupling strength $\xi_{ZZ}^\mathrm{ind}=23$ MHz to implement a fast CP gate. Using a flat-topped Gaussian microwave pulse, we achieve a CZ gate time of 40 ns.  We characterize this two-qubit gate with interleaved RB and show the results in Supplementary Figure~\ref{fig:ChenRB}. The data is produced from 45 randomized gate sequences and the two-qubit Clifford group is constructed with on average 5.583 physical single-qubit gates (incorporating virtual Z gates for the single-qubit Z rotations) and 1.5 CZ gates per Clifford gate. The CP gate fidelity extracted from the interleaved RB measurement is $(97.6\pm 0.3) \%$. This value is slightly lower than the simulated error 98.7 \% from Supplementary Note \ref{sec:leakage}, which should come from the insufficient optimization of the pulse parameters or the temporal fluctuation of the relaxation and dephasing time.

 In all measurements of this device, we did not optimize our readout parameters sufficiently to perform single-shot joint readout of the two-qubit states. As an alternative, to measure the population distribution of any two-qubit state $\vec{P} = [P_{gg}, P_{ge}, P_{eg}, P_{ee}]^T$, we repeatedly prepare the same state and measure the cavity transmission following four different single-qubit rotations: no rotation, $\pi$ rotation on qubit B,  $\pi$ rotation on qubit A, and $\pi$ rotation on both qubits.  These measurements yield a 4-vector of complex transmission coefficients $\vec{V}=[V_{II}, V_{I \pi}, V_{\pi I}, V_{\pi\pi}]^T$.  This measurement signal can be converted to the population vector via a calibrated measurement matrix $M$: $\vec{P}=M\vec{V}$,
 \begin{equation}
\begin{pmatrix}
p_{gg}\\
p_{ge}\\
p_{eg}\\
p_{ee}\\
\end{pmatrix} = 
\begin{pmatrix}
M_{gg} & M_{ge} & M_{eg} & M_{ee}\\
M_{ge} & M_{gg} & M_{ee} & M_{eg}\\
M_{eg} & M_{ee} & M_{gg} & M_{ge}\\
M_{ee} & M_{eg} & M_{ge} & M_{gg}\\
\end{pmatrix}^{-1}
\begin{pmatrix}
V_{II}\\
V_{I\pi}\\
V_{\pi I}\\
V_{\pi\pi}\\
\end{pmatrix}.
\end{equation}
The $M$ matrix is calibrated and updated throughout all experiments by measuring $\vec{V}$ of the initial state of the system $\vec{P_i}$ (which typically is about $[0.87, 0.04, 0.09, 0.00]^T$ following a driven reset protocol).  $\vec{P_i}$ is determined self-consistently with additional calibration routines. 

\begin{figure}
    \centering
    \includegraphics[width=0.3\linewidth]{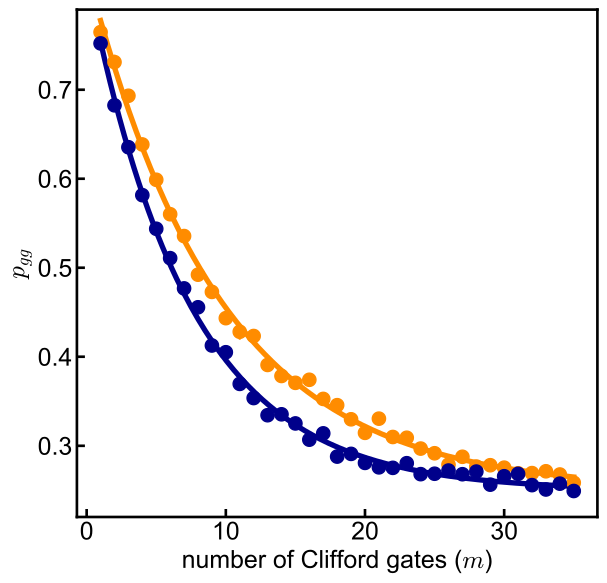}
    \caption{\textbf{Interleaved randomized benchmarking of the CZ gate for the second device.} The blue and yellow circles depict the interleaved and reference sequence. We extract the CZ gate fidelity $(97.6\pm 0.3) \%$.}
    \label{fig:ChenRB}
\end{figure}

\section{Supplementary note 8: Theoretical analysis\label{sec:leakage}}

In this Supplementary Note, we analyze the controlled-phase (CP) gate with fluxonium qubits theoretically. First, we simulate the unitary dynamics during the gate operation, calculate the unitary error, and discuss coherent leakage to higher noncomputational levels. Second, we estimate the incoherent error caused by relaxation and dephasing, which are the dominant sources of error in our devices. Our estimate of this error is in good agreement with experimental gate infidelities in Figure~5 of the main text. Finally, we discuss effects of the $\textrm{ZZ}$-cancellation tone.

\subsection{Unitary dynamics and coherent leakage to higher levels}\label{sec:simulations-unitary}

We describe the unitary dynamics of the driven system of two capacitively coupled fluxonium qubits by the Hamiltonian
\begin{equation}\label{eq:Hamiltonian_total}
    \hat{H} = \hat{H}_{\rm static} + \hat{H}_{\rm drive}(t)\,.
\end{equation}
Here the static part of the Hamiltonian reads ($\alpha$ labels two circuits $A$ and $B$)
\begin{equation}
\begin{aligned}
\frac{\hat{H}_{\rm static}}{h} = &  \sum_{\alpha = A,B} \left[4 E_{C,\alpha} \hat{n}_\alpha^2 + \frac{1}{2}E_{L,\alpha} \hat{\varphi}_\alpha^2 - E_{J,\alpha} \cos(\hat{\varphi}_\alpha-\phi_{\mathrm{ext},\alpha})
\right]
+ J_C \hat{n}_A \hat{n}_B\,,
\label{eq:Hamiltonian}
\end{aligned}
\end{equation}
where $E_{C,\alpha}$ is the charging energy (antenna) of fluxonium $\alpha$, $E_{L,\alpha}$ is its inductive energy (junction array), $E_{J,\alpha}$ is Josephson energy (small junction), and $\phi_{\rm ext, \alpha}$ is proportional to the external magnetic flux threading the loop formed by the junction array and the small junction of qubit $\alpha$. Here we operate at the flux sweet spot where $\phi_{\rm ext, \alpha} = \pi$.  The operators $\hat{\varphi}_\alpha$ and $\hat{n}_\alpha$ are the generalized flux and Cooper-pair number operators of fluxonium $\alpha$, respectively, which satisfy the commutation relation $[\hat{\varphi}_\alpha,\hat{n}_{\alpha'}] = i\delta_{\alpha \alpha'}$. Supplementary Table \ref{table:device_parameters} gives the experimental values of the parameters in the Hamiltonian~(\ref{eq:Hamiltonian}) obtained by fitting two-tone spectroscopy measurements to the numerical diagonalization of the Hamiltonian and calculated values of single-qubit transition frequencies and charge matrix elements. The notation $f^\alpha_{kl}$ stands for the frequency of the $\ket{k}_\alpha-\ket{l}_\alpha$ transition. Our qubits are coupled by an interaction term $J_C \hat{n}_A \hat{n}_B$ with a coupling constant $J_C = 0.248 \mathrm{~GHz}$. This affects two-qubit transition frequencies, which are shown in Figure~1c of the main text.

\begin{center}
\begin{table}[h!]
\begin{tabular}{ | c | c | c | c | c | c | c | c | c |} \hline
  &$E_{C,\alpha} \mathrm{~(GHz)}$ & $E_{L,\alpha} \mathrm{~(GHz)}$ & $E_{J,\alpha} \mathrm{~(GHz)}$ & $J_C \mathrm{~(GHz)}$ & $f^\alpha_{01}$ (GHz) & $f^\alpha_{12}$ (GHz) & $|\bra{0}\hat{n}_\alpha\ket{1}_\alpha|$ & $|\bra{1}\hat{n}_\alpha\ket{2}_\alpha|$\\ \hline
  Qubit A & 1.051 & 0.753 & 5.263 & \multirow{2}{*}{0.248} & 0.217 & 4.489 & 0.066 & 0.576 \\
  Qubit B & 1.069 & 0.771 & 3.870 & & 0.489 & 3.510 & 0.131 & 0.559 \\ \hline
\end{tabular}
\caption{\label{table:device_parameters}Parameters of the Hamiltonian~(\ref{eq:Hamiltonian}) extracted from the two-tone spectroscopy measurements and predicted qubit transition frequencies and charge matrix elements. }
\end{table}
\end{center}

We model the time-dependent drive term in the Hamiltonian~\eqref{eq:Hamiltonian_total} by 
\begin{equation}\label{eq:H_drive}
    \frac{\hat{H}_{\rm drive}(t)}{h} = (\epsilon_A \hat{n}_A + \epsilon_B \hat{n}_B) \left[g_x(t) \cos(2\pi f_d t) + g_y(t) \sin(2\pi f_d t)\right]\,.
\end{equation}
Here the envelope functions $g_x(t)$ and $g_y(t)$ describe two independent quadrature controls. Assuming that they are normalized as $\sqrt{g_x^2 + g_y^2}=1$ for a continuous microwave tone, we find that the on-resonance Rabi frequency for the two-qubit transition $|kl\rangle -|k'l'\rangle$ is given by
\begin{equation}\label{eq:Rabi_freq}
    \Omega_{kl-k'l'} = |\langle kl |(\epsilon_A \hat{n}_A + \epsilon_B \hat{n}_B)|k'l'\rangle|\,.
\end{equation}
Here $|kl\rangle$ is the eigenstate of the interacting Hamiltonian~\eqref{eq:Hamiltonian} that is connected adiabatically to the noninteracting state $|k\rangle_A \otimes |l\rangle_B$. To match the experimentally measured induced $\textrm{ZZ}$-rate of $\xi_{ZZ}^{\rm ind} = 2.9$ MHz at $f_d = 4.545$ GHz at the drive power corresponding to $\Omega_{11-21} = 52.4$ MHz, we choose $\epsilon_B / \epsilon_A = 1.3$ in the drive term~\eqref{eq:H_drive}. Eq.~\eqref{eq:Rabi_freq} has been used in simulating Rabi frequencies shown in Figure~1c in the main text.

For the control $g_x(t)$, we use a Gaussian flat-topped pulse  with the length of each of the rising and lowering edges $t_{\rm rise}$ and the duration of the flat part $t_{\rm flat}$. Up to a normalization constant, the rising edge of the pulse at $0<t<t_{\rm rise}$ is given by
\begin{equation}\label{eq:x_control_flattop}
    g_x(t) \propto \exp\left[-\frac{(t-t_{\rm rise})^2}{2\sigma^2}\right] - \exp\left[-\frac{t_{\rm rise}^2}{2\sigma^2}\right]\,,
\end{equation}
where $\sigma = t_{\rm rise} / \sqrt{2\pi}$, and the lowering edge is given by a similar expression at $t_{\rm gate} - t_{\rm rise} < t< t_{\rm gate}$, where $t_{\rm gate} = 2t_{\rm rise} + t_{\rm flat}$. For the orthogonal quadrature, we use the DRAG approach~\cite{Motzoi2009}, which gives
\begin{equation}\label{eq:y_control_DRAG}
    g_y(t) = \alpha_{\rm DRAG} \frac{dg_x}{dt}\,,
\end{equation}
where $\alpha_{\rm DRAG}$ is the DRAG coefficient adjusted to avoid leakage errors during the gate. 

To calculate the unitary gate errors, we first simulate the operator of evolution between $t=0$ and $t=t_{\rm gate}$ for a system described by the Hamiltonian~\eqref{eq:Hamiltonian_total} and project it into the computational subspace to obtain $\hat{U}$. To properly compare $\hat{U}$ with the target CP gate operator $\hat{U}_{\rm CP}(\phi)$, where
\begin{equation}\label{eq:U_CP}
    \hat{U}_{\rm CP}(\phi) = {\rm diag}\left(1, 1, 1, e^{-i\phi}\right)\,,
\end{equation}
we adjust $\hat{U}$ with single-qubit $Z$ rotations as follows. We define the phase accumulation for operator $\hat{U}$ as
\begin{equation}\label{eq:phase_accumulation}
    \phi_{U} = \phi_{00} + \phi_{11} - \phi_{10} - \phi_{01}\,,
\end{equation}
where $\phi_{kl} = -{\rm arg}\langle kl |\hat{U}|kl\rangle$ is the opposite of the phase of the corresponding diagonal matrix element. The combination \eqref{eq:phase_accumulation} is invariant under single-qubit $Z$ rotations. It results in the phase mismatch 
\begin{equation}
    \delta\phi = \phi_{U} - \phi
\end{equation}
between the accumulated and target phases. We then calculate $\hat{U}' = \hat{U}_Z \hat{U}$, where
\begin{equation}\label{eq:U_Z}
    \hat{U}_Z = {\rm diag}\left[e^{i(\phi_{00} - \delta\phi/4)}, e^{i(\phi_{10} + \delta\phi/4)}, e^{i(\phi_{01} + \delta\phi/4)},
    e^{i(\phi_{11} - \delta\phi/4 - \phi)}\right]
\end{equation}
is the product of two single-qubit $Z$ rotations and an overall phase factor. Using the operator $\hat{U}'$, we use the standard expression for the two-qubit gate fidelity~\cite{Pedersen2007}:
\begin{equation}\label{eq:fidelity}
    F = \frac{{\rm Tr}\left[\left(\hat{U}'\right)^\dagger \hat{U}'\right] + \left|{\rm Tr} \left[\hat{U}^\dagger_{\rm CP}(\phi) \hat{U}'\right]\right|^2}{20}\,.
\end{equation}

There are two major sources of coherent errors: the phase error and the error due to leakage to noncomputational levels. To calculate the phase error, we take $\hat{U} = {\rm diag} \left(e^{-i\phi_{00}}, e^{-i\phi_{01}}, e^{-i\phi_{10}}, e^{-i\phi_{11}}\right)$ and find using Eq.~\eqref{eq:U_Z} that $\hat{U}' = {\rm diag}\left(e^{-i\delta\phi/4}, e^{i\delta\phi/4}, e^{i\delta\phi/4}, e^{-i\delta\phi/4 -i\phi}\right)$. The infidelity $1-F$ calculated using Eq.~\eqref{eq:fidelity}  then gives the phase error
\begin{equation}\label{eq:phase_error}
    {\cal E}_\phi = \frac 45 \sin^2\left(\frac{\delta \phi}{4}\right)\,.
\end{equation}
The coherent leakage error is the average leakage probability, which is given by 
\begin{equation}\label{eq:leakage_error}
    P_{\rm leak} = 1 - \frac 14{\rm Tr}\left(\hat{U}^\dagger \hat{U}\right)\,.
\end{equation}
Since $\hat{U}$ is obtained following the projection into the computational subspace, it is generally nonunitary, so $P_{\rm leak}$ can be nonzero.

\begin{figure}
    \centering
    \includegraphics[width=0.8\textwidth]{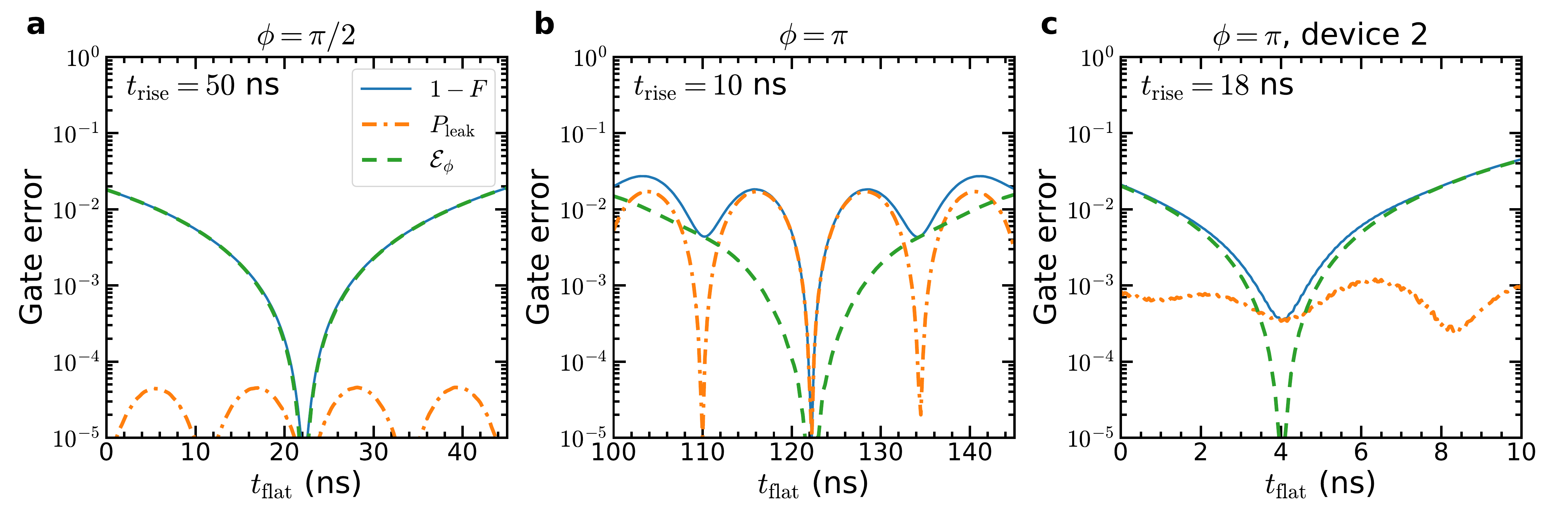}
    \caption{\textbf{Simulated coherent gate errors.} The total unitary error $1-F$ (solid lines), leakage error $P_{\rm leak}$ (dash-dotted lines), and the phase error ${\cal E}_\phi$ (dashed lines) are shown vs $t_{\rm flat}$, the duration of the flat part of the pulse, for the target phases $\phi=\pi/2$ ({\bf a}) and $\pi$ ({\bf b}) and for the second device that is described in Supplementary Note~\ref{sec:faster} with the target phase of $\pi$ ({\bf c}).  Each gate is first optimized for a fixed Gaussian flat-topped envelope $g_x(t)$ with experimental values of $t_{\rm flat}$, $t_{\rm rise}$, and $\sigma$, see Eq.~\eqref{eq:x_control_flattop}, and then simulated at other values of $t_{\rm flat}$. 
     }
    \label{fig:fid_tflatdep}
\end{figure}

Using this procedure, we calculated coherent errors for 16 target phases $\phi = \pi/16, \pi/8, \ldots, \pi$, discussed in the main text. For each target phase, we used experimental pulse duration parameters $t_{\rm rise}$ and $t_{\rm flat}$ and optimized $1-F$ numerically over the DRAG parameter $\alpha_{\rm DRAG}$, the overall drive amplitude, and the drive frequency, which we allowed to vary within a $\pm 5$ MHz window around its experimental value $f_d = 4.545$ GHz. For each of the 16 CP gates optimized this way, we found that $1-F < 10^{-4}$, $P_{\rm leak} < 10^{-4}$, and ${\cal E}_\phi < 10^{-5}$.  We show these errors as a function of  $t_{\rm flat}$ in Supplementary Figures~\ref{fig:fid_tflatdep}a and \ref{fig:fid_tflatdep}b for $\phi=\pi/2$ and $\phi=\pi$, respectively, where we keep other parameters fixed at their optimal values. For $\phi=\pi/2$, Supplementary Figure~\ref{fig:fid_tflatdep}a, we notice that $P_{\rm leak} < 10^{-4}$ for any value of $t_{\rm flat}$, which is explained by long edges of the pulse with $t_{\rm rise} = 50$ ns. Therefore, for this set of gate parameters, we can obtain a complete family of CP gates with the total unitary error $1-F < 10^{-4}$ for any target phase $\phi$ by simply varying $t_{\rm flat}$. We have checked that in this case, the accumulated phase \eqref{eq:phase_accumulation} is a linear function of $t_{\rm flat}$ and is given by  $\phi_{U} = 2\pi \xi_{ZZ} t_{\rm flat}$ up to a constant with $\xi_{ZZ} = 4.3$ MHz. In comparison, a shorter $t_{\rm rise} = 10$ ns has been used in simulations and experiment for the target phase $\phi=\pi$, see Supplementary Figure~\ref{fig:fid_tflatdep}b, which resulted in maxima of $P_{\rm leak}(t_{\rm flat})$ around $10^{-2}$. In this case, the optimized gate requires simultaneous elimination of leakage  in both $\ket{10}-\ket{20}$ and $\ket{11}-\ket{21}$ transitions at the end of the pulse, which happens at sharp minima of $P_{\rm leak}(t_{\rm flat})$ in Supplementary Figure~\ref{fig:fid_tflatdep}b.

In Supplementary Figure~\ref{fig:fid_tflatdep}c, we show simulations for the target phase $\phi=\pi$ for the second device, which is discussed in detail in Supplementary Note~\ref{sec:faster}. This device has a stronger repulsion between $|12\rangle$ and $|21\rangle$, which results in a larger detuning between frequencies of the $|10\rangle-|20\rangle$ and $|11\rangle$ - $|21\rangle$ transitions and, therefore, in a faster CP gate. In simulations, we have chosen $\epsilon_A=\epsilon_B$ in the drive term~\eqref{eq:H_drive} to match the experimental value of $\Omega_{11-21}/\Omega_{10-20}$. To match experimental time parameters,  we have used $t_{\rm rise} = 18$ ns and $\sigma = 30 / \sqrt{2\pi}$ ns in the Gaussian edge~\eqref{eq:x_control_flattop}. As is evident from Supplementary Figure~\ref{fig:fid_tflatdep}c, this device allows a fast CP, while maintaining a small leakage $P_{\rm leak} < 10^{-3}$. Here, it is a specific device spectrum rather than a short gate duration that prevents us from easily achieving leakage error below $10^{-4}$ as in Supplementary Figures~\ref{fig:fid_tflatdep}a and \ref{fig:fid_tflatdep}b. For this device, the  $|01\rangle - |02\rangle$ and $|11\rangle - |12\rangle$ transitions are additional important leakage channels as their frequencies are only 100-200 MHz below those of the $|10\rangle - |20\rangle$ and $|11\rangle - |21\rangle$ transitions. This makes it harder to match off-resonant Rabi oscillations in all the leaking transitions to achieve sharp minima as in Supplementary Figure~\ref{fig:fid_tflatdep}b. Thus, a longer Gaussian edge as in Supplementary Figure~\ref{fig:fid_tflatdep}a or more advanced pulse shaping is required to obtain $1-F < 10^{-4}$.

\subsection{Incoherent error}

In the present experiment, the actual gate fidelity is limited by decoherence rather than coherent errors. To estimate this limit, we use a simple model in the rotating-wave approximation with only six levels in the Hilbert space: four computational states  as well as states $|20\rangle$ and $|21\rangle$. Thus, we use
\begin{multline}\label{Hamiltonian_6levels}
    \frac{\hat{H}_{\rm RWA}}{h} = -\delta |20\rangle \langle 20| - (\delta - \Delta)|21\rangle \langle 21| + \frac{g_x(t)}{2} \left[\Omega_{10-20}\left(|10\rangle \langle 20| + |20\rangle \langle 10|\right) + \Omega_{11-21}\left(|11\rangle \langle 21| + |21\rangle \langle 11|\right)\right] \\
    -  \frac{ig_y(t)}{2} \left[\Omega_{10-20}\left(|10\rangle \langle 20| - |20\rangle \langle 10|\right) + \Omega_{11-21}\left(|11\rangle \langle 21| - |21\rangle \langle 11|\right)\right]\,,
\end{multline}
where $\delta = f_d - f_{10-20} \approx 57$ MHz and $\Delta = f_{11-21} - f_{10-20}\approx 8 $ MHz, see Figure~1c in the main text. In addition, here $\Omega_{11-21}/ \Omega_{10-20} = 1.114$ to match the ratio $\epsilon_B/\epsilon_A$ in Eq.~\eqref{eq:H_drive} and we use the same pulse-shaping envelopes as in Eqs.~\eqref{eq:x_control_flattop} and \eqref{eq:y_control_DRAG}. We find that contributions due to decoherence of both computational and noncomputational transitions are important in our device. The relevant time parameter for the first contribution is the total gate duration $2t_{\rm rise} + t_{\rm flat}$, while the relevant time parameter for the second contribution is the effective duration $t_{\rm rise} + t_{\rm flat}$.  Similarly to the calculation of unitary errors, we use the experimental values of $t_{\rm rise}$ and $t_{\rm flat}$ to better match our estimates with measurements.

For given target phase accumulation $\phi$ and experimental $t_{\rm rise}$ and $t_{\rm flat}$, we  optimize the unitary gate error over the  drive amplitude  and the DRAG coefficient $\alpha_{\rm DRAG}$. We calculate this error following the same procedure as in Supplementary Note~\ref{sec:simulations-unitary}, but for the Hamiltonian \eqref{Hamiltonian_6levels}. Once the optimal parameters are found, we perform simulations of nonunitary dynamics using Lindblad master equation. For the $6\times 6$ density matrix $\rho$, it has the form
\begin{equation}\label{eq:Lindblad}
    \frac{d\hat{\rho}}{dt} = -\frac{i}{\hbar}\left[\hat{H}_{\rm RWA}, \hat{\rho}\right] + \sum_k \left[\hat{L}_k \hat{\rho}\hat{L}_k^\dagger - \frac 12 \left(\hat{L}_k^\dagger \hat{L}_k \hat{\rho} + \hat{\rho} \hat{L}_k^\dagger \hat{L}_k\right)\right]\,,
\end{equation}
where we use six collapse operators $\hat{L}_k$. We form them by continuing tensor products of single-qubit collapse and identity operators into interacting states. E.g., we take $|0\rangle_A \langle 1|_A \otimes (|0\rangle_B \langle 0|_B + |1\rangle_B \langle 1|_B)$ to obtain $|00\rangle \langle 10| + |01\rangle \langle 11|$,  which is exact for an uncoupled system. Thus, for the relaxation in the main qubit transitions, we use the following collapse operators:
\begin{subequations}
\begin{gather}
\hat{L}_1^{A}  = \sqrt{\Gamma_1^A} \left(|00\rangle \langle 10| + |01\rangle \langle 11|\right)\,, \\
\hat{L}_1^{B}  = \sqrt{\Gamma_1^B} \left(|00\rangle \langle 01| + |10\rangle \langle 11| + |20\rangle \langle 21|\right)\,.
\end{gather}
\end{subequations}
We notice that while the unitary dynamics of states $\ket{00}$ and $\ket{01}$ is independent of that of other levels in our simplified model~\eqref{Hamiltonian_6levels}, these states are no longer uncoupled from the rest of the Hilbert space once incoherent channels are accounted for. For the pure dephasing of the main qubit transitions, we use
\begin{subequations}
\begin{gather}
\hat{L}_\varphi^{A}  = \sqrt{2\Gamma_\varphi^A} \left(|00\rangle \langle 00| + |01\rangle \langle 01|\right)\,, \\
\hat{L}_\varphi^{B}  = \sqrt{2\Gamma_\varphi^B} \left(|00\rangle \langle 00| + |10\rangle \langle 10| + |20\rangle \langle 20|\right)\,.
\end{gather}
\end{subequations}
For relaxation and pure dephasing of the $|1\rangle_A \to |2\rangle_A$ transition of qubit A, we use
\begin{subequations}
\begin{gather}
\label{eq:L_1_12}
\hat{L}_1^{1- 2, A}  = \sqrt{\Gamma_1^{1- 2, A}} \left(|10\rangle \langle 20| + |11\rangle \langle 21|\right)\,, \\
\label{eq:L_phi_12}
\hat{L}_\varphi^{1- 2, A}  = \sqrt{2\Gamma_\varphi^{1- 2, A}} \left(|20\rangle \langle 20| + |21\rangle \langle 21|\right)\,.
\end{gather}
\end{subequations}
In these equations, the relaxation and dephasing rates are given by $\Gamma_1 = 1/T_1$ and $\Gamma_\varphi = 1/T_2^E - 1/2T_1$, where $T_1$ and $T_2^E$ are the relaxation and $T_2$ echo times of the corresponding transition. In simulations, we use average values of $T_1$ and $T_2^E$ for the ranges shown in Table I in the main text.

After solving the master equation \eqref{eq:Lindblad} for an initial state $|\psi_0\rangle$ at $t=0$, we find the density matrix $\hat{\rho}$ describing the system at time $t=t_{\rm gate}$. We calculate the state fidelity as  $F_\rho = {\rm Tr}(\hat{\rho}\hat{\rho}_{\rm ideal})$, where $\rho_{\rm ideal} = \ket{\psi_{\rm ideal}}\bra{\psi_{\rm ideal}}$ with $\ket{\psi_{\rm ideal}}= \hat{U}_Z^\dagger\hat{U}_{\rm CP}(\phi) \ket{\psi_0}$ and $\hat{U}_Z$ describing virtual $Z$ rotations used to calculate unitary gate error, see Eq.~(\ref{eq:U_Z}). We then estimate the gate error $1-F$ by averaging $1-F_\rho$ over 36 initial two-qubit states generated from the set of six initial single-qubit states $\{|0\rangle, |1\rangle, (|0\rangle \pm |1\rangle)/\sqrt{2}, (|0\rangle \pm i |1\rangle)/\sqrt{2}\}$. Gate errors calculated this way are shown by squares in Figure~5 of the main text. 

The simulated gate error for the accumulated phase $\phi = \pi$ is $1 - F \approx 1.1\times 10^{-2}$, which agrees well with the experimental value. A natural question to ask is how small this error can be in devices with longer coherence times. For $T_1 = T_2 = 500$ $\mu$s of single-qubit transitions and $T_1 = T_2 = 50$ $\mu$s of the $|1\rangle_A \to |2\rangle_A$ transition, we find that this error reduces to $7\times 10^{-4}$. When these relaxation and coherence times are further increased by a factor of two to $1000$ and $100$ $\mu$s, we estimate $1-F \approx 4\times 10^{-4}$. To obtain $1-F < 10^{-4}$, we additionally need to change the device parameters of Supplementary Table~\ref{table:device_parameters} to allow for a shorter gate. For example, this can be achieved by increasing $J_C$ by a factor of 2, which increases $\Delta$ by a factor of 4 and thus reduces the shortest possible gate duration by the same factor.

A similar approach to estimate gate error due to relaxation and dephasing applied to the second device, which is discussed in Supplementary Figure~\ref{fig:fid_tflatdep}c and in Supplementary Note~\ref{sec:faster}, gives the error  of about $1.3\%$. Since in that device, the frequencies of the $\ket{01}-\ket{02}$ and $\ket{11}-\ket{12}$ transitions are only 100-200 MHz away from those of the $\ket{10}-\ket{20}$ and $\ket{11}-\ket{21}$ transitions, we extended the model~\eqref{Hamiltonian_6levels} to add levels $\ket{02}$ and $\ket{12}$ and modified the set of collapse operators accordingly with the extra relaxation and dephasing channels.

\subsection{Effect of the $\textrm{ZZ}$-cancellation tone}

We have seen in our simulations that the relaxation and dephasing of the $\ket{1}_A-\ket{2}_A$ transition of qubit A is responsible for about half of the CP gate error because the population of state $\ket{2}_A$ can become as large as $20\%$ during gate operation. Similarly, the microwave tone that is used to cancel static $\textrm{ZZ}$-interaction results in modified relaxation and dephasing channels within the computational subspace as expected from the hybridization of $\ket{10}$ with $\ket{20}$ and $\ket{11}$ with $\ket{21}$. However, in comparison to a fast CP gate, where $\Omega_{11\to 21} \sim \delta-\Delta$, the $\textrm{ZZ}$-cancellation tone has a lower power of $\Omega_{11-21} \approx 30$ MHz and a larger detuning $|f_d - f_{11-21}| \approx 150$ MHz. In the rotating frame [the frame of the Hamiltonian~\eqref{Hamiltonian_6levels}], this leads to a weak hybridization between computational and higher levels. Keeping only terms that are linear in $\lambda = \Omega_{11-21}/|f_d - f_{11-21}|\approx \Omega_{10-20}/|f_d - f_{10-20}|\approx 0.2$, we find the following dressed states [eigenstates of the Hamiltonian~\eqref{Hamiltonian_6levels}]:
\begin{subequations}
\begin{gather}\label{eq:11_dressed}
    \ket{{11}}_\lambda = \ket{11} + \frac{\lambda}{2}\ket{21}\,,
    \\
\label{eq:21_dressed}
    \ket{{21}}_\lambda = -\frac{\lambda}{2}\ket{11} + \ket{21}\,,
\end{gather}
\end{subequations}
and similarly for $\ket{10}_\lambda$ and $\ket{20}_\lambda$. The set of states $\{\ket{00}, \ket{01}, \ket{10}_\lambda, \ket{11}_\lambda\}$ is the computational subspace in the presence of the cancellation tone (the dressed computational subspace). When the system is in one of the dressed states $\ket{{10}}_\lambda$ or $\ket{{11}}_\lambda$, the population of the bare state $\ket{20}$ or $\ket{21}$ is only $\lambda^2/4 \approx 0.01$, so we anticipate only a weak reduction of the relaxation and dephasing rates of qubit transitions in comparison to their values at $\lambda=0$.

To calculate these rates accurately, one needs to apply the transformation given by Eqs.~\eqref{eq:11_dressed} and \eqref{eq:21_dressed} to the Hamiltonian describing system-bath interaction, trace out bath degrees of freedom, and obtain a master equation in the new basis~\cite{Boissonneault2009}. Here we only estimate the effect of the interplay of the cancellation tone and relaxation and dephasing of the $\ket{1}_A - \ket{2}_A$ transition to demonstrate that such an effect is small. To this end, we apply the transformation given by Eqs.~\eqref{eq:11_dressed} and \eqref{eq:21_dressed} to the collapse operators \eqref{eq:L_1_12} and \eqref{eq:L_phi_12}. To the zeroth order in $\lambda$, we find the collapse operators having the same forms as Eqs.~\eqref{eq:L_1_12} and \eqref{eq:L_phi_12} but written in terms of the dressed states. The corrections to them are given by
\begin{subequations}
\begin{gather}
\label{eq:L_1_12_dressed}
    \delta\hat{L}_1^{1- 2, A}  = \frac{\lambda}{2} \sqrt{\Gamma_1^{1- 2, A}} \left( \ket{{10}}_\lambda\bra{{10}}_\lambda - \ket{{20}}_\lambda\bra{{20}}_\lambda + \ket{{11}}_\lambda\bra{{11}}_\lambda - \ket{{21}}_\lambda\bra{{21}}_\lambda \right) + O(\lambda^2)\,, \\
\label{eq:L_phi_12_dressed}
\delta\hat{L}_\varphi^{1- 2, A}  = \frac{\lambda}{2} \sqrt{2\Gamma_\varphi^{1- 2, A}} \left(\ket{{10}}_\lambda\bra{{20}}_\lambda + \ket{{20}}_\lambda\bra{{10}}_\lambda + \ket{{11}}_\lambda\bra{{21}}_\lambda + \ket{{21}}_\lambda\bra{{11}}_\lambda\right) + O(\lambda^2)\,.
\end{gather}
\end{subequations}
We observe that the relaxation channel introduces a small dephasing of the $\ket{1}_A-\ket{2}_A$ transition and vice versa. In addition, the terms with $\ket{{10}}\bra{{10}}$ and $\ket{{11}}\bra{{11}}$ in the r.h.s. of Eq.~\eqref{eq:L_1_12_dressed} yield additional dephasing of computational transitions. This additional dephasing can be characterized by the rate $\Gamma_\phi'$ defined as $\sqrt{2\Gamma'_\phi} \approx (\lambda/2) \sqrt{\Gamma_1^{1-2, A}}$. For our parameters, we find $1/\Gamma'_\phi \approx 8 / \left(\lambda^2 \Gamma_1^{1-2, A}\right) \approx 1 ~{\rm ms}$, which is much longer than qubit dephasing times in our device. This number is comparable with longest fluxonium coherence times reported so far~\cite{Nguyen2018}. As we anticipate a significant improvement in $T_1$ of the $\ket{1}\to \ket{2}$ transitions of future devices, $1/\Gamma_\phi'$ is expected to become much longer. Therefore, we do not predict any significant increase in the dephasing rate caused by the cancellation tone in both current and future devices. 

We notice that Eqs.~\eqref{eq:L_1_12_dressed} and \eqref{eq:L_phi_12_dressed} do not predict directly any relaxation within the computational subspace. Cancellation-tone-induced relaxation is possible in a more detailed model \eqref{eq:Hamiltonian}, which properly accounts for the capacitive interaction. With the interaction-induced hybridization of states $\ket{10}$ and $\ket{01}$ taken into account, an additional relaxation channel is possible with the rate $\Gamma_1'$, where $\sqrt{\Gamma_1'} \sim \lambda \nu \sqrt{\Gamma_1^{1\to 2, A}}$, $\nu = J_C n^A_{01} n^B_{01} / |f^A_{01} - f^B_{01}|\approx 0.008$ describes the hybridization strength, and $n^\alpha_{kl} = \left|\bra{k}\hat{n}_\alpha \ket{l} \right|$. Alternatively, qubit relaxation can be facilitated by a stronger hybridization of states $\ket{21}$ and $\ket{12}$ with the strength $J_C n^A_{12} n^B_{12} / |f^A_{12} - f^B_{12}|\approx 0.08$. For both of these cases, we find that the theory predicts a correction to the qubit relaxation rate negligible compared to the energy relaxation rates of the qubit transitions of our devices.

%\bibliography{references2}

%apsrev4-2.bst 2019-01-14 (MD) hand-edited version of apsrev4-1.bst
%Control: key (0)
%Control: author (8) initials jnrlst
%Control: editor formatted (1) identically to author
%Control: production of article title (0) allowed
%Control: page (0) single
%Control: year (1) truncated
%Control: production of eprint (0) enabled
%